\documentclass[manuscript,screen]{acmart}
\AtBeginDocument{%
  }

\setcopyright{acmlicensed}
\copyrightyear{2025}
\acmYear{2025}
\acmDOI{XXXXXXX.XXXXXXX}

\usepackage{svg}
\usepackage{multirow}
\usepackage{multicol}
\usepackage{fontawesome5}

\begin{document}

\title{Generative AI for Testing of Autonomous Driving Systems: A Survey}

\author{Qunying Song}
\email{qunying.song@ucl.ac.uk}
\orcid{0000-0002-8653-0250}
\affiliation{%
  \institution{University College London}
  \city{London}
  \country{United Kingdom}
}

\author{He Ye}
\affiliation{%
  \institution{University College London}
  \city{London}
  \country{United Kingdom}}
\email{he.ye@ucl.ac.uk}

\author{Mark Harman}
\affiliation{%
  \institution{University College London}
  \city{London}
  \country{United Kingdom}}
\email{mark.harman@ucl.ac.uk}

\author{Federica Sarro}
\affiliation{%
  \institution{University College London}
  \city{London}
  \country{United Kingdom}}
\email{f.sarro@ucl.ac.uk}

\renewcommand{\shortauthors}{Song et al.}

\begin{abstract}
  Autonomous driving systems (ADS) have been an active area of research, with the potential to deliver significant benefits to society. However, before large-scale deployment on public roads, extensive testing is necessary to validate their functionality and safety under diverse driving conditions. Therefore, different testing approaches are required, and achieving effective and efficient testing of ADS remains an open challenge. Recently, generative AI has emerged as a powerful tool across many domains, and it is increasingly being applied to ADS testing due to its ability to interpret context, reason about complex tasks, and generate diverse outputs. To gain a deeper understanding of its role in ADS testing, we systematically analyzed 91 relevant studies and synthesized their findings into six major application categories, primarily centered on scenario-based testing of ADS. We also reviewed their effectiveness and compiled a wide range of datasets, simulators, ADS, metrics, and benchmarks used for evaluation, while identifying 27 limitations. This survey provides an overview and practical insights into the use of generative AI for testing ADS, highlights existing challenges, and outlines directions for future research in this rapidly evolving field.
\end{abstract}

\ccsdesc[500]{General and reference~Surveys and overviews}
\ccsdesc[500]{Hardware~Safety critical systems}
\ccsdesc[500]{Software and its engineering~Software verification and validation}
\ccsdesc[500]{Computing methodologies~Artificial intelligence}

\keywords{Literature Survey, Autonomous Driving Systems, Testing, Generative AI}

\received{20 February 2007}
\received[revised]{12 March 2009}
\received[accepted]{5 June 2009}

\maketitle

\section{Introduction}
\label{sec:introduction}

Autonomous driving systems (ADS) have gained increasing attentions and are expected to enhance future mobility and road safety~\cite{favaro2017examining, neurohr2020fundamental}. However, before large-scale deployment, rigorous testing is required to ensure safe and reliable operation across a wide range of real-world scenarios~\cite{riedmaier2020survey, song2024empirically}. This testing process is complex and involves multiple testing environments and approaches~\cite{lou2022testing, tang2023survey}. According to the latest EU regulation on type approval of AVs~\cite{eu2022}, manufacturers must demonstrate an acceptable level of functional and operational safety for their ADS, including defining performance requirements, conducting assessments via testing, providing detailed documentation, maintaining safety management systems, and validating any simulation tools used. Eventually, the objective of testing, as articulated by the SOTIF (Safety of the Intended Functionality) standard~\cite{sotif2022}, is to test all relevant scenarios, especially the critical ones that are challenging and hazardous for the systems and other road users~\cite{simens2022whitepaper}.

Testing ADS is inherently complex, as these systems integrate multiple functional modules and advanced technologies, and are expected to manage an extensive range of possible scenarios in open operational environments~\cite{song2021concepts, song2024empirically, karunakaran2022challenges}.  Effectively identifying all relevant scenarios and ensuring adequate coverage in testing remains a major challenge~\cite{lou2022testing}. The adoption of machine learning related techniques in ADS further heightens concerns regarding their explainability, safety, and robustness~\cite{adigun2022collaborative, klas2022integrating}. Furthermore, significant gaps remain across multiple facets of ADS engineering, including standardizing requirements, design, and testing processes for ADS~\cite{hacohen2022autonomous}. As a result, \emph{current testing practices seem still insufficient to fully address the complexities of ADS} and fatal accidents occur, such as the Uber's AV struck a pedestrian in Arizona, US~\cite{yang2020lessons}, and Tesla Autopilot incident in which the system failed to detect a truck against a bright sky~\cite{Abbas2017driver}.

Generative AI refers to a class of AI models -- such as ChatGPT~\cite{openai2024gpt4technicalreport, hurst2024gpt}, DALL-E~\cite{ramesh2022hierarchicaltextconditionalimagegeneration, pmlr-v139-ramesh21a}, and GitHub Copilot~\cite{dakhel2023github} -- capable of producing new content, including text, images, audio, video, code, and other formats, based on patterns learned from training data~\cite{feuerriegel2024generative}. Generative AI has gained tremendous interest in recent years and has been applied across diverse domains, such as creative content production, software development, and healthcare~\cite{sengar2024generative, bengesi2024advancements, cao2023comprehensive}. Recently, there has been a growing trend in applying generative AI to testing ADS~\cite{gao2025foundation, tian2024large}. Leveraging its ability to understand context, reason about complex tasks, and generate outputs in various forms, \emph{generative AI is an emerging approach and shows strong potential to support a range of testing-related activities}, including test scenario generation and analysis. 

Despite its potential, the use of generative AI in ADS testing is still in its early stages, with questions about its effectiveness, limitations, and best practices remaining largely unanswered. To \emph{gain a deeper understanding of the role of generative AI in ADS testing}, we identified 91 relevant studies from several major literature databases. We \emph{analyzed these papers and synthesized findings on their applications, effectiveness, and limitations, using a thematic analysis}~\cite{cruzes2011recommended, huang2018synthesizing} approach. In particular, this study focuses on three research questions:

\begin{itemize}
    \item \textbf{RQ1: How is generative AI used for testing ADS?} $\rightarrow$ This question examines which generative AI models are used, how they are applied in ADS testing, and what types of ADS are targeted.
    \item \textbf{RQ2: How effective is generative AI for testing ADS?} $\rightarrow$ This question focuses on the evaluation, including the datasets, simulators, ADS, metrics, and benchmarks employed, as well as the evaluation results.
    \item \textbf{RQ3: What limitations do exist in the use of generative AI for testing ADS? } $\rightarrow$ This question identifies the limitations of generative AI and potential solutions that have been proposed or discussed.
\end{itemize}

We carefully extracted relevant findings from the 91 papers, with particular attention to the mechanisms through which generative AI is applied in testing ADS, its demonstrated effectiveness (as evidenced by evaluations), and the limitations that were observed or discussed in these papers. Following that, we consolidated these findings into a unified model focused on applying generative AI to ADS testing, using a thematic analysis approach~\cite{cruzes2011recommended, huang2018synthesizing}. 

Our analysis reveals a clear upward trend in publications exploring the use of generative AI for ADS testing. Various generative models are employed, such as large language models (LLMs), vision-language models (VLMs), and diffusion-based models. These models are predominantly used in the context of scenario-based testing, encompassing \emph{(critical) scenario generation, transformation, reconstruction, augmentation, and understanding}. While they demonstrate remarkable performance in producing desired scenarios for testing and improving ADS, several limitations persist. Common challenges include \emph{generating suboptimal outputs, limited generalization to underrepresented data, difficulties in processing complex and domain-specific tasks, and high computational costs}. Overall, this survey \emph{provides an overview and practical insights into the use of generative AI for testing ADS, highlights current challenges, and serves as a reference for future research} aimed at improving ADS testing. While several survey studies have been conducted on ADS testing and on the use of generative AI in autonomous driving, very few provide an in-depth examination of generative AI specifically for testing ADS, which to some extent underscores the novelty and value of this study.

The remainder of this article is organized as follows: Section~\ref{sec:preliminaries} introduces the key terms and concepts used in this study; Section~\ref{sec:methodology} outlines the methodology, particularly the literature selection, analysis, and synthesis; and Section~\ref{sec:results} presents the results and analysis of the selected papers in relation to the research questions. Section~\ref{sec:discussion} discusses our findings, and Section~\ref{sec:related_work} reviews related literature and compares it with our work. Finally, Section~\ref{sec:conclusion} concludes the article.

\section{Preliminaries}
\label{sec:preliminaries}

In this study, we primarily focus on three key terms: \emph{ADS}, \emph{testing} (in the context of autonomous driving), and \emph{generative AI}. In this section, we briefly present the conceptualizations we adopt for each term for clarity. Related work discussing studies connected to these concepts is further detailed in Section~\ref{sec:related_work} (titled Related Work).

\subsection{ADS}

ADS are designed to \emph{perform dynamic driving tasks on a sustained basis and manage diverse driving scenarios in road traffic independently}~\cite{j30162021}. Two primary architectures are commonly employed for ADS~\cite{tang2023survey}. The first is the modular ADS, which divides the system into several functional modules, such as sensing, perception, planning, and control, each responsible for a specific aspect of autonomous driving~\cite{tang2023survey}. Specifically, the sensing module collects environmental information using various sensors (e.g., cameras, radar, LiDAR, and GPS). The perception module processes these sensory inputs to identify objects and road events. The planning module predicts and determines the trajectory and maneuvers, while the control module executes the planned actions to operate the vehicle. The second architecture is the end-to-end ADS, which directly processes environmental and traffic sensory inputs to control the AV~\cite{tang2023survey}.

ADS are categorized into six levels of automation by SAE International~\cite{j30162021}, ranging from Level 0 to Level 5. Level 0 represents no automation, with the human driver fully responsible for vehicle operation. Levels 1 and 2 correspond to Advanced Driver Assistance Systems (ADAS): Level 1 provides either longitudinal or lateral control to assist the driver, while Level 2 offers assistance for both. Levels 3 to 5 are considered true ADS. Specifically, Level 3 enables conditional autonomous driving, requiring human intervention when necessary. Level 4 supports high automation within specific constraints, such as designated regions or conditions, without the need for human control in those contexts. Finally, Level 5 represents full automation, enabling autonomous driving under all conditions without human involvement. 

\subsection{ADS Testing} 

As described earlier, testing ADS is complex and remains an open challenge that has not yet been fully addressed. Different environments and approaches are employed for testing ADS, including: simulation testing, which tests ADS in virtual simulated driving environments; closed-track testing, which tests ADS in the real world but on controlled testing tracks; and public road testing, which tests ADS in the real world and public roads~\cite{lou2022testing}. Particularly, \emph{scenario-based testing approach is widely adopted}, which aims to evaluate the functionality and safety of ADS across diverse driving scenarios and environmental conditions, typically executed within (virtual) simulation platforms. 

However, scenario generation (a.k.a., creation, identification, synthesis) remains a key challenge~\cite{riedmaier2020survey, lou2022testing, beringhoff2022thirty}, as it raises the question of how to \emph{obtain realistic and relevant test scenarios, especially the critical ones} that challenge the system~\cite{zhang2022finding, ding2023survey, song2024industry}. Existing approaches have been explored and categorized into taxonomies, including: knowledge-based approaches, which rely on domain experts and established ontologies to derive test scenarios; data-driven approaches, which extract and extrapolate scenarios from collected (real and synthetic) data; and search-based approaches, which explore and optimize desired scenarios within the scenario space using search algorithms~\cite{riedmaier2020survey}. Specifically, search-based approaches are commonly adopted for testing both ADS modules and full systems, aiming to effectively identify potentially critical scenarios, as highlighted in several recent surveys on ADS testing~\cite{lou2022testing, tang2023survey}.

\subsection{Generative AI}

Generative AI (or GenAI) refers to a class of \emph{AI models capable of producing new content}, such as text, images, audio, video, code, and even 3D models, \emph{based on patterns learned from training data}~\cite{feuerriegel2024generative, sengar2024generative, bengesi2024advancements, cao2023comprehensive}. Unlike traditional AI models, which are primarily designed for classification or prediction tasks, generative AI focuses on creating novel outputs that resemble the underlying data distribution without duplicating the original data. Notable examples include ChatGPT~\cite{openai2024gpt4technicalreport, hurst2024gpt}, Claude~\cite{TheC3, caruccio2024claude}, and Google Gemini~\cite{geminiteam2025geminifamilyhighlycapable}, commonly used for text generation; DALL-E~\cite{ramesh2022hierarchicaltextconditionalimagegeneration, pmlr-v139-ramesh21a}, Stable Diffusion~\cite{Rombach_2022_CVPR}, and MidJourney~\cite{hanna2023use} for image creation; and GitHub Copilot~\cite{dakhel2023github} and Code Llama~\cite{touvron2023llamaopenefficientfoundation, roziere2023code} for code generation. Generative AI has rapidly emerged and found applications across diverse domains, including creative content production, software development, healthcare, gaming, and others. We briefly discuss a few commonly used terms, models, frameworks, and architectures for generative AI to provide an overview of the techniques involved.

\begin{itemize}
    \item \textbf{Transformer} is a deep learning architecture widely adopted in generative AI models for tasks such as text generation, image synthesis, and even music composition~\cite{feuerriegel2024generative, sengar2024generative, bengesi2024advancements, cao2023comprehensive}. It is built on self-attention mechanisms, which enable the model to weigh the importance of different parts of the input data and capture their relationships when generating output~\cite{vaswani2017attention}. Unlike sequential processing, the self-attention mechanism considers all input elements simultaneously and determines which components are most relevant for each output element~\cite{vaswani2017attention}. The Transformer architecture typically employs an encoder-decoder structure, where the encoder processes the input sequence (e.g., a sentence) and the decoder generates the output sequence (e.g., translated text). Examples of models based on this architecture include GPT~\cite{radford2018improving, brown2020language}, BERT~\cite{devlin2019bert}, and T5~\cite{raffel2020exploring}.

    \vspace{1mm}

    \item \textbf{Large Language Models (LLMs)} are deep learning models designed to process and generate human-like text~\cite{chang2024survey, feuerriegel2024generative, sengar2024generative, bengesi2024advancements, cao2023comprehensive}. The models are trained on massive amounts of general text data, and can understand, generate, and manipulate text in ways that mimic human conversations. Many LLMs adopt the Transformer architecture~\cite{vaswani2017attention}, which enables efficient handling and generation of long sequences of words. Given a carefully crafted input (or prompt), LLMs generate task-specific outputs. Rather than retraining the model from scratch, prompt design focuses on formulating or optimizing inputs to elicit the most relevant and accurate responses~\cite{liu2023summary}, which is referred to as prompt engineering~\cite{sahoo2024systematic}. For example, zero-shot learning provides only contextual information of the task, while few-shot learning supplements prompts with a few illustrative examples of the task, enabling the model to infer patterns and generate accurate outputs~\cite{sahoo2024systematic}. Examples of LLMs include examples include GPT~\cite{openai2024gpt4technicalreport, hurst2024gpt}, Claude~\cite{TheC3, caruccio2024claude}, and Google Gemini~\cite{geminiteam2025geminifamilyhighlycapable}.

    \vspace{1mm}

    \item \textbf{Vision Language Models (VLMs)} are a class of generative models that integrate computer vision and natural language processing to jointly understand and process both visual and textual information~\cite{dosovitskiy2020image, lu2019vilbert, su2019vl}. These models are typically pre-trained on large collections of image-text pairs, enabling them to perform a variety of visual language tasks without requiring task-specific training~\cite{gao2025foundation, cao2023comprehensive}. VLMs are widely applied to multimodal tasks such as visual question answering, image captioning, image tagging, and image generation. By combining visual machine learning techniques with LLMs, VLMs can capture semantic relationships within images and between objects, facilitating contextual understanding of visual data. Examples include OpenAI’s GPT-4V~\cite{openai2023gpt4v}, CLIP~\cite{radford2021learning}, Flamingo~\cite{alayrac2022flamingo}, and the open-source Large Language and Vision Assistant (LLaVA)~\cite{liu2023visual}.

    \vspace{1mm}

    \item \textbf{Diffusion probabilistic models} are a class of generative models that gradually transform random noise into meaningful content, and are widely used for generating images, videos, and other types of data~\cite{ho2020denoising}. These models operate by first adding noise to training data (e.g., images) and then learning to reverse this process through a step-by-step denoising procedure, effectively reconstructing realistic outputs from noise. Diffusion models have been extensively applied in AI art, e.g., text-to-image generation (e.g., GLIDE~\cite{nichol2021glide}, Stable Diffusion~\cite{Rombach_2022_CVPR}), video creation (e.g., Imagen Video~\cite{ho2022imagen}), and other scientific domains. Their ability to produce high-quality, diverse, and original content has established them as a powerful tool in generative AI.

    \vspace{1mm}

    \item \textbf{Generative Adversarial Network (GAN)} is a class of artificial intelligence frameworks and one of the most prominent architectures in generative AI, capable of producing realistic synthetic content such as images, music, and videos~\cite{goodfellow2014generative, goodfellow2020generative}. A GAN consists of two neural networks trained simultaneously: a generator, which creates synthetic data, and a discriminator, which evaluates whether the data is real or generated~\cite{goodfellow2014generative, goodfellow2020generative}. The two networks are trained in a competitive process: the generator improves its ability to produce data that can fool the discriminator, while the discriminator becomes increasingly adept at distinguishing between real and synthetic inputs. Over time, this adversarial training drives the generator to produce highly realistic outputs. Examples of GAN architectures include DCGAN~\cite{radford2015unsupervised}, CycleGAN~\cite{zhu2017unpaired}, and StyleGAN~\cite{karras2019style, Karras_2020_CVPR}.

    \vspace{1mm}

    \item \textbf{Variational Autoencoder (VAE)} is a type of generative model that learns to represent data (such as images or text) in a lower-dimensional latent space and then reconstruct it~\cite{kingma2013auto}. A VAE operates by encoding input data into a compact latent representation and then decoding it back to approximate the original data. It is trained to maximize the likelihood of the data while enforcing a structured and smooth latent space through variational inference\cite{blei2017variational}. This property allows VAEs to generate new, similar data by sampling from the latent space. Examples of VAE architectures include Vanilla VAE~\cite{kingma2013auto}, Beta-VAE~\cite{higgins2017beta}, and VQ-VAE~\cite{van2017neural}.

    \vspace{1mm}

    \item \textbf{Foundation models} are general-purpose models trained on large-scale, heterogeneous datasets that span multiple modalities, domains, and tasks~\cite{bommasani2021opportunities}. As the term gains popularity, we introduce it to provide a general understanding. Rather than referring to a specific category of generative models, "foundation model" is an umbrella term that encompasses a range of models, including, but not limited to, LLMs, VLMs, and diffusion-based models, which have been described earlier. These models have demonstrated impressive capabilities in contextual understanding, reasoning, and generation, often across tasks and domains they were not explicitly trained for~\cite{wu2025foundation, gao2024survey, gao2025foundation}. Their adaptability allows them to be fine-tuned or prompted for a wide array of downstream applications, making them a powerful and versatile component of generative AI systems.
\end{itemize}

\section{Methodology}
\label{sec:methodology}

We referred to the guidelines from Kitchenham et al.~\cite{kitchenham2009systematic} and Keele et al.~\cite{keele2007guidelines} for performing systematic literature reviews in software engineering to design our study, and our methodology is also inspired by two recent literature surveys~\cite{hort2024bias, chen2024fairness}. Specifically, our methodology consists of four main stages: literature selection, data extraction, data synthesis, and data validation, which we describe in Section~\ref{sec:methodology:literature_selection}-\ref{sec:methodology:data_validation}. In Section~\ref{sec:methodology:threats_to_validity}, we describe potential threats to validity of our study and how we incorporated the mitigation of them as part of our methodology design.

\begin{figure}[tbp]
    \centering
    \includegraphics[trim=0 15mm 0 10mm, clip, width=\textwidth]{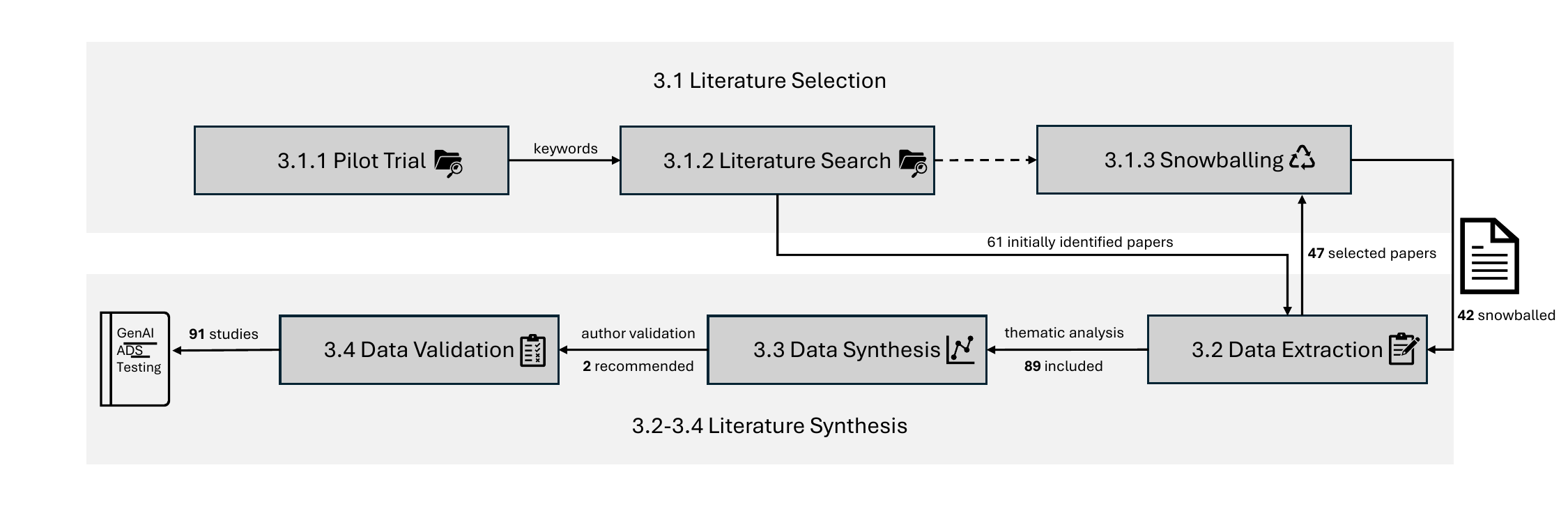}
    \caption{Methodology of the survey includes four main stages: literature search, data extraction, data synthesis, and data validation. The literature search begins with a pilot trial to identify relevant keywords, which are then used for database searches, followed by snowballing based on the initially selected papers. The identified papers are analyzed and synthesized into findings on GenAI for ADS testing, which are subsequently validated by the original authors of those studies.}
    \label{fig:methodology}
\end{figure}

\subsection{Literature Selection}
\label{sec:methodology:literature_selection}
The literature selection stage aims to search for and identify studies relevant to the focus of our study. To start with, we performed a pilot trial to obtain a preliminary view of the existing research landscape and to refine the search string. Following that, we conducted a literature search across several databases using the refined search string. We screened through all searched papers and selected those that align with the objectives of our study. Finally, we performed both backward and forward snowballing~\cite{badampudi2015experiences, jalali2012systematic, wohlin2014guidelines} on selected papers to identify additional relevant studies. 

\subsubsection{Pilot Trial}
\label{sec:methodology:literature_selection:pilot_search}

\begin{table}[hbp]
\centering
\caption{Keywords used in the search string for literature search. The asterisk sign (*) denotes a wildcard match in the search process. For example, ``Validat*'' captures variations such as ``Validate'', ``Validates'', ``Validated'', ``Validating'', and ``Validation''. }
\begin{tabular}{|l|l|l|}
    \hline 
    \textbf{Generative AI} & \textbf{Test} & \textbf{Autonomous Driving} \\
    \hline
    Generative & Test & Autonomous Driving \\
    GenAI & Validat* & Autonomous Vehicle \\
    Language Model & Verif* & Autonomous Car \\
    LLM & Evaluat* & Automated Driving \\
    Foundation Model & Assur* & Automated Vehicle \\
    GAN & Assess & Automated Car \\
    Diffusion & Analys* & Self-driving\\
    Variational Autoencoder & Generat* & Driverless \\
    VAE &  & \\
    Transformer &  & \\
    \hline
\end{tabular}
\label{table:methodology:keywords}
\end{table}

We performed a pilot trial (on March 28, 2025) using Scopus to gain an initial sense of the studies published in this area -- using generative AI for testing ADS. Scopus is an extensive abstract and citation database that covers 330 disciplines and more than 7 000 publishers~\cite{scopus}. Our search string was built around three primary concepts, and consisted of multiple keywords in relation to ``Generative AI'', ``test'', and ``autonomous driving''. Using the search string on Scopus, we retrieved 228 journal and conference papers in English. We carefully analysed the titles, abstracts, and when necessary, the full texts to identify additional keywords and refine our search string. To maximize coverage, we intentionally used broad and general keywords, despite the potential for retrieving a significant amount of irrelevant literature. The final keywords in our search string are presented in Table~\ref{table:methodology:keywords}, grouped according to the three primary concepts that we focus on in this study. 

\subsubsection{Literature Search}
\label{sec:methodology:literature_selection:literature_search}

We searched relevant literature across several comprehensive and widely used databases (on April 7, 2025) using the search query obtained in Section~\ref{sec:methodology:literature_selection:pilot_search}. Apart from Scopus, we used IEEE Xplore~\cite{ieee} and ACM Digital Library~\cite{acm}, both are digital library that provides comprehensive access to publications in several disciplines such as computer science and information technology. We also used arXiv~\cite{arxiv} to capture relevant literature that are not published yet. Notably, we considered another database ScienceDirect~\cite{sciencedirect}, but it did not support wildcard match and allowed a maximum of eight AND/OR operators in the search string; thus, it can be extremely challenging to account for all possible combinations of the identified keywords and is not used in this study.

\begin{table}[hbp]
\centering
\caption{Search results for each repository used in this study. Number of Results indicates the results from applying the search query in the repository, and Number of Results (Limited) limits the results to conference papers and journal articles in English only.}
\begin{tabular}{|c|c|c|}
    \hline 
    \textbf{Repository} & \textbf{Number of Results} & \textbf{Number of Results (Limited)} \\
    \hline
    Scopus & 2,746 & 2,214 \\
    IEEE & 3,460 & 3,174 \\
    ACM & 799 & 591 \\
    arXiv & 1,901 & 1,881 \\
    \hline
    Total & 8,906 & 7,860 \\
    \hline
\end{tabular}
\label{table:methodology:search_results}
\end{table}

As a result, we retrieved a total of 8,906 documents, as shown in Table~\ref{table:methodology:search_results}. We limited the search results to conference papers and journal articles published in English, and found 7,860 remained. After that, we removed duplicates and obtained 6,355 unique entries. We examined their titles, abstracts, and introductions if necessary, and identified 61 relevant studies. We adopted a rigorous selection process, including only studies that explicitly satisfy the following four inclusion criteria: (1) the paper presents a primary study, (2) it targets ADS, (3) it is framed or grounded in the context of testing, and (4) it involves the use of generative AI models. In addition to substantial studies that focused on non-ADS or were unrelated to generative AI, a considerable number of papers were also excluded because their primary focus was on the development of autonomous driving systems or enabling technologies, rather than on testing.

\subsubsection{Snowballing}
\label{sec:methodology:literature_selection:snowballing}

We performed snowballing~\cite{wohlin2022successful} on the selected papers (on May 15, 2025) to identify additional relevant studies. Prior to the snowballing process, we analysed and extracted key findings from the 61 identified studies, which is detailed in Section~\ref{sec:methodology:data_extraction}. We further excluded 14 papers that did not meet all the criteria outlined in Section~\ref{sec:methodology:literature_selection:literature_search}, resulting in a final set of \textbf{47} studies. The reason for conducting snowballing after analyzing the initially retrieved papers is, as already noted, that some of the papers could be irrelevant and excluded following full-text analysis. For backward snowballing, we examined the reference lists of each included paper, particularly the studies discussed in their Related Work sections. For forward snowballing, we used Google Scholar~\cite{googlescholar} to review their citations, and selected those relevant ones for further analysis. As a result of the snowballing process, we identified an additional \textbf{42} papers. In total, we obtained \textbf{89} papers, comprising both the initially searched and the snowballed studies.

\subsection{Data Extraction}
\label{sec:methodology:data_extraction}

The data extraction stage aims to analyse the included papers and extract key information from them. Our analysis was anchored on the three research questions of this study, as described in Section~\ref{sec:introduction}. Specifically, for \textbf{RQ1}, we extracted the generative AI models used, the ADS targeted, and the overall approach of how generative AI is applied in the testing of ADS in each study. For \textbf{RQ2}, we gathered the datasets, simulators, systems under test, metrics, and benchmark methods that were employed to assess and demonstrate the effectiveness of generative AI for testing ADS in each study. For \textbf{RQ3}, we collected all reported limitations of using generative AI for testing ADS, including not only specific challenges encountered in each individual study, but also general limitations that are discussed in these papers. During the analysis, we relied entirely on the descriptions provided in the papers, without making any additional inferences or assumptions, and we prepared a brief summary for each research question for each paper. In cases where information was ambiguous or missing (i.e., 7 studies), we reached out to the first/corresponding authors for clarification and additional details.

\subsection{Data Synthesis}
\label{sec:methodology:data_synthesis}

Following the data extraction stage described in Section~\ref{sec:methodology:data_extraction}, the data synthesis stage aims to synthesize the extracted information into a coherent model using a thematic analysis approach~\cite{cruzes2011recommended, huang2018synthesizing}. First, we coded the summaries prepared in Section~\ref{sec:methodology:data_extraction} with appropriate codes -- short labels that accurately capture key information presented in the summaries. Second, similar codes were grouped under a common theme, and similar themes were grouped under a higher-level theme. This process was repeated iteratively until all codes and themes formed a unified model, meaning each code and theme was distinct, with no overlaps, and accurately positioned at the appropriate level of the thematic structure. To ensure accuracy and consistency, we held a workshop among the authors (on July 4, 2025) to review all extracted data and the thematic model, resolving any ambiguities and disagreements through discussion.

\subsection{Data Validation}
\label{sec:methodology:data_validation}

Finally, after completing data synthesis (Section~\ref{sec:methodology:data_synthesis}) and the initial draft of this survey, we sent the manuscript on August 13, 2025, to the 81 unique first or corresponding authors of all 89 papers identified through our search and snowballing process. We invited their feedback on our analysis of their work and on the survey as a whole. We received 18 responses. While most of them confirmed the accuracy of our analysis, some provided updates on the publication status of their (arXiv) pre-print papers, suggested additional relevant work, or offered minor feedback on our interpretations. We revised the manuscript accordingly and incorporated two additional papers they recommended. The inclusion criteria remained the same as those described in Section~\ref{sec:methodology:literature_selection}: only primary studies that explicitly focused on testing ADS with generative AI were considered. As a result, this survey now includes a total of \textbf{91} studies.

\subsection{Threats to Validity}
\label{sec:methodology:threats_to_validity}

As a literature review aiming to collect and synthesize existing research, our primary focus is on ensuring the construct validity and reliability of the study~\cite{verdecchia2023threats, lago2024threats}. Construct validity refers to how accurately the study measures the concepts it intends to investigate~\cite{wohlin2012experimentation}. A potential threat to construct validity lies in the definitions of the three key concepts central to our topic -- generative AI, testing, and ADS. To mitigate this concern, we took several steps. Before conducting the literature search, we explored existing definitions in the field and adopted the most commonly accepted and clearly defined ones, as outlined in Section~\ref{sec:preliminaries}. We also conducted a pilot trial to refine our search strings, detailed in Section~\ref{sec:methodology:literature_selection:pilot_search}. During the literature analysis phase, we reached out to the authors when clarification was needed regarding the content. After completing the analysis, we held internal workshops among the authors to resolve any disputes during data synthesis and ensured full consensus was reached on the final results, as described in~\ref{sec:methodology:data_synthesis}. Furthermore, as described in Section~\ref{sec:methodology:data_validation}, we invited the authors of the included studies to review our manuscript and provide feedback on our analysis. Reliability refers to the repeatability and consistency of the research procedures~\cite{wohlin2012experimentation}. To enhance reliability, we ensured that each step of our study was thoroughly documented throughout the study and can be traced.

\section{Results and Analysis}
\label{sec:results}

In this section, we present the results of our analysis of the included studies. We first present the demographics of the included studies in Section~\ref{sec:results:demographics} to understand some meta characteristics. Following that, we present the results for the three research questions in this study, in Section~\ref{sec:results:RQ1}-\ref{sec:results:RQ3}, respectively, to understand how generative AI is used in testing ADS, how effective they are, and what are their limitations.

\subsection{Demographics}
\label{sec:results:demographics}

\begin{figure}[hbp]
    \centering
    \includegraphics[trim=0 5mm 0 0, clip, width=0.85\textwidth]{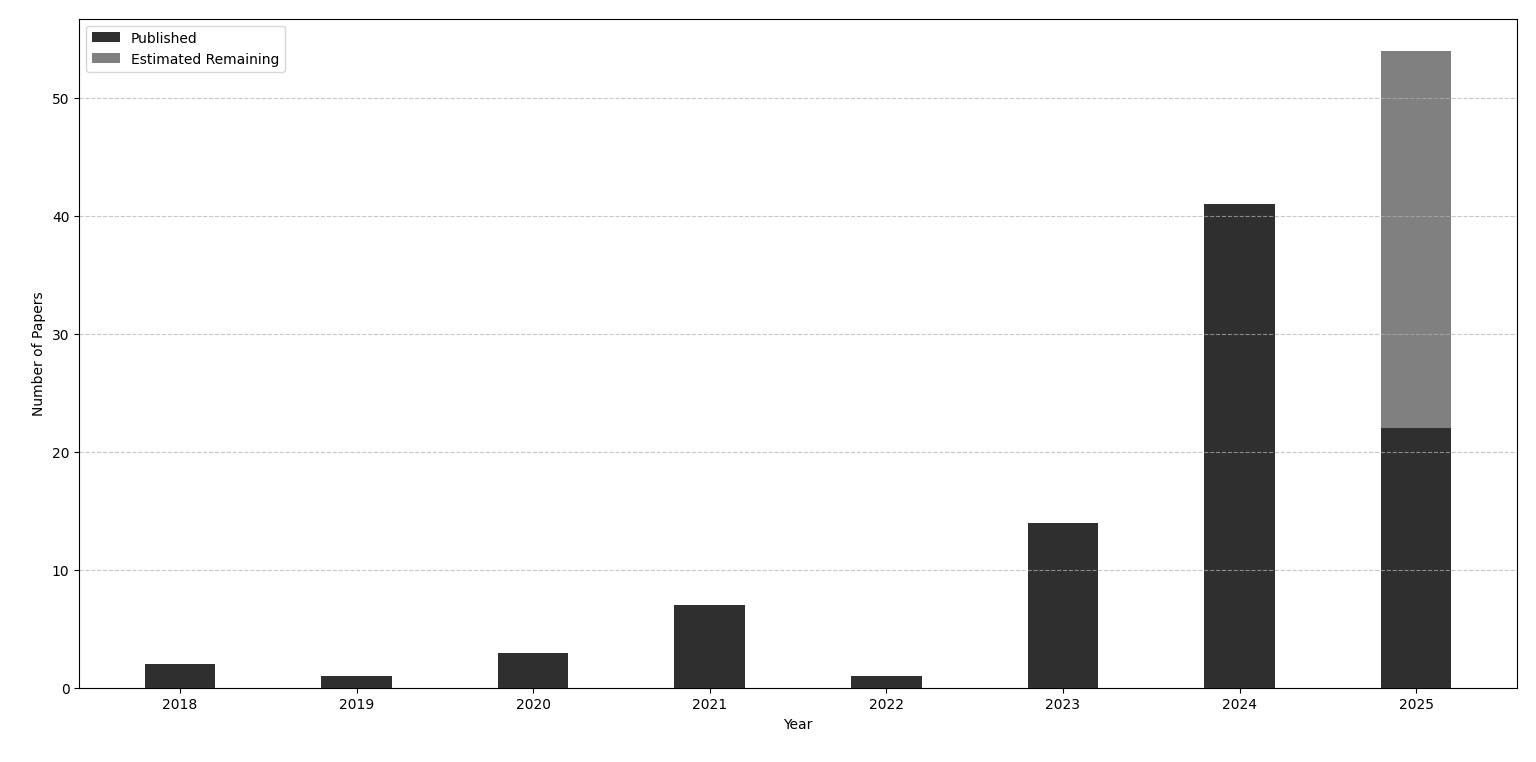}
    \caption{Number of papers published per year. Particularly for 2025, since the snowballing was conducted on May 15, the estimated number of publications for the remainder of the year was projected based on the number of papers identified up to that date.}
    \label{fig:results:demographics:papers}
\end{figure}

We observe a clear upward trend in the number of publications over the years, particularly from 2023 onward, as shown in Figure~\ref{fig:results:demographics:papers}. Although the included studies span from 2018 to 2025, the vast majority of them -- 77 out of 91 papers (85\%) -- were published between 2023 and 2025. This sharp increase in recent years suggests that this topic represents \emph{an emerging and promising research area that has received increasing attention}. However, it is important to highlight that the 91 included studies consists of 28 (31\%) pre-prints from arXiv, which have not yet been published through peer review. We contacted the authors of these preprint papers. 12 of them had already been accepted for publication, and the authors confirmed that the camera-ready versions were exactly, or did not differ substantively from, the preprints uploaded on arXiv. 12 out of the remaining 16 preprints, while still under peer review, were expected to receive decisions soon and some even entered the rebuttal stage with rather positive feedback. 

\begin{figure}[tbp]
    \centering
    \includegraphics[trim=0 70mm 0 20mm, clip, width=0.9\textwidth]{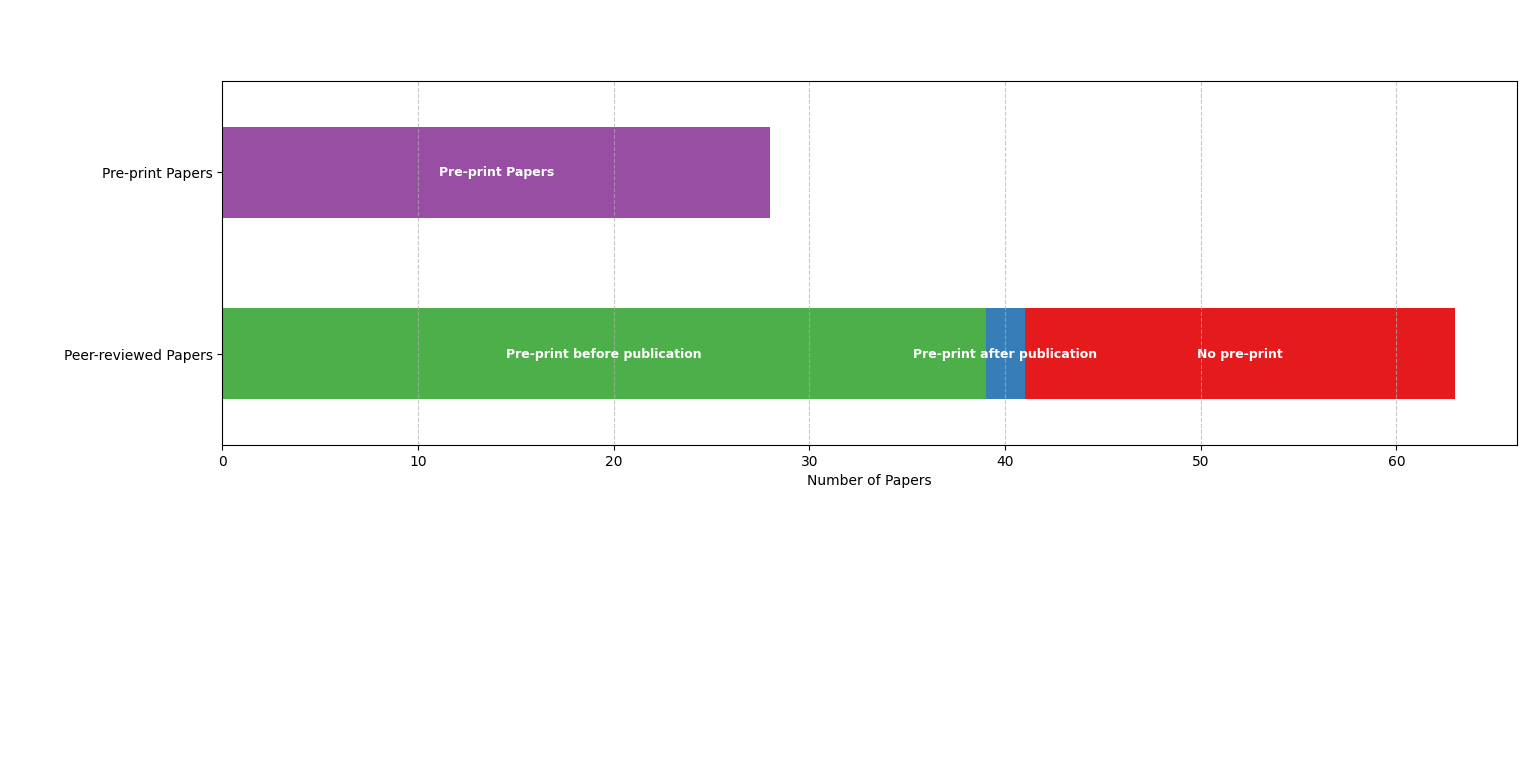}
    \caption{Number of pre-prints and peer-reviewed papers accompanied by pre-prints released before or after publication.}
    \label{fig:results:demographics:pre-prints}
\end{figure}

We observe that a significant proportion of peer-reviewed papers (41 out of 63, or 65\%, as shown in Table~\ref{fig:results:demographics:pre-prints}) have a pre-print available, either on arXiv, OpenReview, or other platforms. For this analysis, we verified each published paper on Google Scholar to determine the existence of a pre-print version. Notably, 39 of these papers released their pre-print prior to publication, while 2 released a pre-print afterward. In contrast, only 22 papers (35\% of the peer-reviewed set) were published without any pre-print. This indicates that \emph{the majority of work in this area is shared as a pre-print before formal publication}, highlighting the widespread adoption of this practice and underscoring the importance of considering pre-prints in our research.

\begin{table}[hbp]
\centering
\caption{Publication venues for the published and accepted papers among the included studies. For commonly occurring venues, the number in parentheses indicates the frequency of publication. All other venues appeared only once.}
\begin{tabular}{|c|c|}
    \hline 
    \textbf{Common Publication Venues} & \textbf{Other Publication Venues} \\
    \hline
    CVPR(12), ICRA(8) & ELECO, ICTAI, ITSSC, LRA, ML4AD \\
    IROS(4), ASE(4), ICSE(2) & CICTP, AS, AML, ASPDAC, TrustCom, FLLM, TSMC \\
    IV(3), ECCV(3), NeurIPS(3) & JICV, ICMLA, AAP, AIAT, PMLR, AAAI, IFAC-PapersOnLine \\
    ITSC(3), Automotive Innovation(2) & DeepTest, QRS-C, WACV, SN Computer Science \\
    T-ITS(2), T-IV(2), CoRL(2) & CASE, ICVES, ICCV \\
    \hline
\end{tabular}
\label{table:results:demographics:publication_venues}
\end{table}

Among the 75 published or accepted (informed by authors) papers, the majority are conference papers (62 out of 75, or 83\%), with the remaining 13 (18\%) being journal articles. These papers are distributed across 39 unique publication venues, as detailed in Table~\ref{table:results:demographics:publication_venues}, spanning a diverse set of domains. This includes AI/ML venues (e.g., CVPR, ECCV, NeurIPS, AAAI, ICMLA, ICCV), software engineering related venues (e.g., ASE, ICSE, QRS-C, SN Computer Science), robotics and automation related venues (e.g., ICRA, IROS, CoRL, LRA), automotive venues (e.g., IV, T-IV, Automotive Innovation, JICV, ICVES), and intelligent transportation venues (e.g., ITSC, T-ITS, ITSSC). This wide distribution of publication venues indicates that the use of generative AI for testing ADS is \emph{a multidisciplinary research topic, drawing interest from researchers and practitioners specialized in several different domains}.

\subsection{RQ1 -- How Generative AI is used for Testing ADS}
\label{sec:results:RQ1}
To address RQ1, we primarily focus on understanding the approaches and mechanisms through how generative AI is applied in the testing of ADS (see Section~\ref{sec:results:RQ1:testing}). Before delving into these details, we first present the distribution of generative AI models used and the types of ADS targeted in the selected studies (in Section~\ref{sec:results:RQ1:models} and \ref{sec:results:RQ1:ADS}), in order to provide a basic overview of the models and systems involved, without specifics on the models or their usage.

\subsubsection{Generative AI Models}
\label{sec:results:RQ1:models}

\begin{table}[tbp]
\centering
\small
\caption{RQ1 -- LLMs used in the selected studies.}
\begin{tabular}{|c|c|l|l|}
    \hline 
    \textbf{Category} & \textbf{Training} & \textbf{Model} & \textbf{Studies} \\
    \hline
    \multirow{25}{*}{LLM} & \multirow{24}{*}{Pre-trained} & \multirow{2}{*}{GPT-4} & \cite{P3}, \cite{P10}, \cite{P34}, \cite{P35}, \cite{P39}, \cite{P45}, \cite{P60}, \cite{P74}, \cite{P75}, \cite{P81} \\
    & & & \cite{P83}, \cite{P90} \\
    \cline{3-4} & & GPT-4o & \cite{P5}, \cite{P22}, \cite{P27}, \cite{P32}, \cite{P56}, \cite{P79}, \cite{P94} \\
    \cline{3-4} & & gpt-4-1106-preview & \cite{P9}, \cite{P46}, \cite{P85} \\
    \cline{3-4} & & GPT-4o-mini	& \cite{P21}, \cite{P79} \\
    \cline{3-4} & & gpt-4-0613 & \cite{P33} \\
    \cline{3-4} & & gpt-4o-0513	& \cite{P37} \\
    \cline{3-4} & & gpt-3.5-turbo-0613 & \cite{P46} \\
    \cline{3-4} & & Llama 3.1 8B & \cite{P58} \\
    \cline{3-4} & & Llama 3.1 & \cite{P59} \\
    \cline{3-4} & & Llama 3 8B & \cite{P34} \\
    \cline{3-4} & & Llama-2-7B & \cite{P35} \\
    \cline{3-4} & & Llama-2 13B & \cite{P70} \\
    \cline{3-4} & & CodeLlama 7B & \cite{P26} \\
    \cline{3-4} & & DeepSeek-V3	& \cite{P57} \\
    \cline{3-4} & & DeepSeek-R1	& \cite{P57} \\
    \cline{3-4} & & Genimi-1.5 Pro & \cite{P79} \\
    \cline{3-4} & & Gemini V1.0 Ultra & \cite{P43} \\
    \cline{3-4} & & Claude-3-opus & \cite{P17} \\
    \cline{3-4} & & Mistral 7B & \cite{P26} \\
    \cline{3-4} & & GLM-4-Air & \cite{P21} \\
    \cline{3-4} & & yi-34B-v1 & \cite{P46} \\
    \cline{3-4} & & Gemma 7B & \cite{P26} \\
    \cline{3-4} & & OpenChat-3.5-7B & \cite{P35} \\
    \cline{2-4} & Fine-tuned & DistilBERT & \cite{P82} \\
    \hline
\end{tabular}
\label{table:results:RQ1:models:llm}
\end{table}

Language Language Models (\textbf{LLMs}), as listed in Table~\ref{table:results:RQ1:models:llm}, emerged as the most commonly used type of generative models among the selected studies. Specifically, 33 out of 91 papers (36\%) utilized LLMs, with the majority of them (24 out of 32, or 75\%) employing GPT models for ADS testing tasks such as scenario understanding, generation, and augmentation, which we detail in Section~\ref{sec:results:RQ1:testing}. GPT-based models, especially GPT-4~\cite{openai2024gpt4technicalreport} and GPT-4o~\cite{hurst2024gpt}, dominated due to their widespread attention since release and their strong capabilities in understanding, reasoning, and generating intended responses based on complex contexts and natural language instructions across various domains. Notably, while GPT-4o was used as an LLM in these studies, it is technically a large multi-modal model (LMM) capable of processing not only natural language but also inputs in other formats such as images and videos, thus broadening its applicability across diverse tasks and contexts. In addition to GPT models, several studies also employed LLaMA~\cite{touvron2023llamaopenefficientfoundation}, DeepSeek~\cite{deepseekai2025deepseekv3technicalreport, deepseekai2025deepseekr1incentivizingreasoningcapability}, Gemini~\cite{geminiteam2025geminifamilyhighlycapable}, Claude~\cite{TheC3}, Mistral~\cite{jiang2023mistral7b}, and a few other LLMs. Except Chang et al.~\cite{P82} who finetuned a DistilBERT model~\cite{sanh2020distilbertdistilledversionbert}, all LLMs used in included studies are pre-trained models applied directly without training or any additional fine-tuning. 

\begin{table}[hbp]
\centering
\small
\caption{RQ1 -- VLMs used in the selected studies.}
\begin{tabular}{|c|c|l|l|}
    \hline 
    \textbf{Category} & \textbf{Training} & \textbf{Model} & \textbf{Studies} \\
    \hline
    \multirow{7}{*}{VLM} & \multirow{7}{*}{Pre-trained} & GPT-4V & \cite{P32}, \cite{P36}, \cite{P77} \\
    \cline{3-4} & & GPT-4o & \cite{P24}, \cite{P93}\\
    \cline{3-4} & & GPT-4-turbo & \cite{P17}\\
    \cline{3-4} & & DALL-E 2 & \cite{P78}\\
    \cline{3-4} & & Qwen2-VL-7B & \cite{P6}\\
    \cline{3-4} & & Gemini-1.5-flash & \cite{P93}\\
    \cline{3-4} & & Claude 3.7 & \cite{P93}\\
    \hline
\end{tabular}
\label{table:results:RQ1:models:vlm}
\end{table}

While Vision Language Models (\textbf{VLMs}) were not frequently used to support ADS testing, GPT-based models such as GPT-4V~\cite{openai2023gpt4v} continue to dominate among the VLMs used, as shown in Table~\ref{table:results:RQ1:models:vlm}. These models are capable of processing multi-modal inputs such as natural language instructions and visual data, e.g., images, for tasks including scenario generation, reconstruction, and augmentation, which is detailed in Section~\ref{sec:results:RQ1:testing}. Other VLMs used included DALL-E 2~\cite{ramesh2022hierarchicaltextconditionalimagegeneration, pmlr-v139-ramesh21a}, Qwen2-VL-7B~\cite{wang2024qwen2vlenhancingvisionlanguagemodels}, Gemini-1.5-flash~\cite{team2024gemini}, and Claude 3.7~\cite{TheC3}, and they are all pre-trained models directly applied in several studies we surveyed.

\begin{table}[hbp]
\centering
\small
\caption{RQ1 -- Diffusion-based models used in the selected studies.}
\begin{tabular}{|c|c|l|l|}
    \hline 
    \textbf{Category} & \textbf{Training} & \textbf{Model} & \textbf{Studies} \\
    \hline
    \multirow{13}{*}{Diffusion} & \multirow{3}{*}{Custom-trained} & \multirow{2}{*}{DDPM} & \cite{P14}, \cite{P18}, \cite{P42}, \cite{P50}, \cite{P51}, \cite{P53}, \cite{P60}, \cite{P61}, \cite{P64}, \\
    & & & \cite{P81}, \cite{P82}, \cite{P84}, \cite{P88} \\
    \cline{3-4} & & LDM & \cite{P4}, \cite{P41}, \cite{P43}, \cite{P52}, \cite{P68}, \cite{P69}, \cite{P94}\\
    \cline{2-4} & \multirow{5}{*}{Pre-trained} &  stable-diffusion-v1-5 & \cite{P20}\\
    \cline{3-4} & & Reliberate-v2 & \cite{P20}\\
    \cline{3-4} & & Deliberate-v5 & \cite{P20}\\
    \cline{3-4} & & RealisticVision & \cite{P23}\\
    \cline{3-4} & & ControlNet & \cite{P70}\\
    \cline{2-4} & \multirow{5}{*}{Fine-tuned} & Stable Diffusion & \cite{P19}, \cite{P62}\\
    \cline{3-4} & & stable-diffusion-v1-4 & \cite{P12}\\
    \cline{3-4} & & stable-diffusion-v1-5 & \cite{P30}\\
    \cline{3-4} & & InstructPix2Pix & \cite{P19}\\
    \cline{3-4} & & ControlNet & \cite{P73}, \cite{P19}\\
    \hline
\end{tabular}
\label{table:results:RQ1:models:diffusion}
\end{table}

\textbf{Diffusion}-based generative models are used to a considerable extent across the included studies, with 28 out of 91 papers (31\%) employing them for tasks such as scenario generation, augmentation, and transformation. Among Custom-trained models, the standard Denoising Diffusion Probabilistic Model (DDPM)~\cite{NEURIPS2020_4c5bcfec} and its variant, the Latent Diffusion Model (LDM)~\cite{Rombach_2022_CVPR}, are the most commonly adopted. While DDPM operates directly in the original data space (e.g., pixel space) and and learns to denoise noisy inputs into realistic data (e.g., driving scenarios represented in images), LDM performs the diffusion process in a compressed latent space, thereby improving computational efficiency. Among pre-trained or fine-tuned diffusion models, Stable Diffusion~\cite{Rombach_2022_CVPR} and its different checkpoints (e.g., stable-diffusion-v1-4, stable-diffusion-v1-5, Reliberate-v2, Deliberate-v5) and variants (e.g., InstructPix2Pix~\cite{Brooks_2023_CVPR}) are the most frequently used. Other than that, a few studies also utilized the ControlNet model~\cite{Zhang_2023_ICCV} and its variants (e.g., ControlNet with edge conditioning~\cite{4767851}), as shown in Table~\ref{table:results:RQ1:models:diffusion}, to enable more precise control over the generated outputs.

\begin{table}[tbp]
\centering
\small
\caption{RQ1 -- GAN-based models used in the selected studies.}
\begin{tabular}{|c|c|l|l|}
    \hline 
    \textbf{Category} & \textbf{Training} & \textbf{Model} & \textbf{Studies} \\
    \hline
    \multirow{13}{*}{GAN} & \multirow{11}{*}{Custom-trained} & Wasserstein GAN & \cite{P29}, \cite{P40}, \cite{P48} \\
    \cline{3-4} & & AE-GAN & \cite{P28}, \cite{P80}, \cite{P55} \\
    \cline{3-4} & & RC-GAN & \cite{P28}, \cite{P80} \\
    \cline{3-4} & & TimeGAN & \cite{P49}, \cite{P76} \\
    \cline{3-4} & & SurfelGAN & \cite{P63} \\
    \cline{3-4} & & StyleGAN & \cite{P72} \\
    \cline{3-4} & & Style-controlled GAN & \cite{P15} \\
    \cline{3-4} & & SeqGAN & \cite{P25} \\
    \cline{3-4} & & CycleGAN & \cite{P31} \\
    \cline{3-4} & & TTS-GAN & \cite{P13}\\
    \cline{2-4} & Pre-trained & Pix2pixHD & \cite{P73}\\
    \cline{2-4} & \multirow{2}{*}{Fine-tuned} & foggy-CycleGAN & \cite{P71}\\
    \cline{3-4} & & OASIS & \cite{P73}\\
    \hline
\end{tabular}
\label{table:results:RQ1:models:gan}
\end{table}

Generative Adversarial Network (\textbf{GAN})-based models are used to a moderate extent across the included studies (16 out of 91, or 18\%), with the majority being customized GANs implemented and trained from scratch using specific datasets selected in the studies. For instance, Wasserstein GAN~\cite{arjovsky2017wassersteingan} and AE-GAN (combining autoencoders and GANs)~\cite{P28, P55, P80} are employed for generating realistic driving scenarios in several studies. Other models, while generally follow the conventional GAN architecture -- comprising a generator and a discriminator for producing realistic data, are adapted to accommodate specific data modalities such as SurfelGAN~\cite{P63} for surfel data, SeqGAN~\cite{Yu_Zhang_Wang_Yu_2017} for sequence data, and TimeGAN~\cite{NEURIPS2019_c9efe5f2} and TTS-GAN~\cite{ttsgan} to process time-series data, or to incorporate application constraints or specific objectives for more controlled generation of desired outputs, such as style control in StyleGAN~\cite{Karras_2020_CVPR}, length-conditioning in RC-GAN~\cite{P28, P80}, and cycle-consistency in CycleGAN~\cite{P31}. Only a few studies employed or fine-tuned pre-trained GAN-based models, as shown in  Table~\ref{table:results:RQ1:models:gan}, including Pix2PixHD~\cite{Wang_2018_CVPR}, foggy-CycleGAN~\cite{foggy-cyclegan}, and OASIS~\cite{1360017285535413888}, primarily for scenario transformation (e.g., switch weather in existing scenarios) tasks.

\begin{table}[hbp]
\centering
\small
\caption{RQ1 -- Autoencoder-based models used in the selected studies.}
\begin{tabular}{|c|c|l|l|}
    \hline 
    \textbf{Category} & \textbf{Training} & \textbf{Model} & \textbf{Studies} \\
    \hline
    \multirow{6}{*}{Autoencoder (AE)} & \multirow{6}{*}{Custom-trained} & VAE & \cite{P7}, \cite{P41}, \cite{P55}, \cite{P66}, \cite{P72} \\
    \cline{3-4} & & CVAE & \cite{P11}, \cite{P16}, \cite{P54} \\
    \cline{3-4} & & Tree-structured VAE & \cite{P91} \\
    \cline{3-4} & & Dual-VAE & \cite{P65} \\
    \cline{3-4} & & Sparse Autoencoder & \cite{P7} \\
    \cline{3-4} & & Constrastive Autoencoder & \cite{P38} \\
    \hline
\end{tabular}
\label{table:results:RQ1:models:ae}
\end{table}

Autoencoder (\textbf{AE})-based generative models are used to a significantly lesser extent compared to other approaches. Most of them are customized Variational Autoencoders (VAEs)~\cite{kingma2022autoencodingvariationalbayes}, trained from scratch (referred to as Custom-trained) and primarily designed for scenario generation. Notably, many of these models are not standalone generative models capable of fulfilling the generation task independently. Instead, they function as components within a larger generative framework (we refer to these as Hybrid models), which is described later in the same section. The architecture of those AE-based models are often tailored to accommodate specific data modalities such as tree structures representing scenarios~\cite{P91}, or to incorporate particular constraints and objectives in the generation process. For example, Conditional VAEs (CVAEs)~\cite{NIPS2015_8d55a249} are used to condition generation on contextual inputs like historical trajectories, road layouts, or other relevant information~\cite{P11, P16, P54}. Besides, A Dual-VAE~\cite{P65} architecture is designed where two VAEs operate in parallel: one generates the trajectory for the leading vehicle, while the other generates the trajectory for the following vehicle. Together, they produce realistic vehicle-following scenarios for testing ADS.

\begin{table}[hbp]
\centering
\small
\caption{RQ1 -- Other models used in the selected studies.}
\begin{tabular}{|c|c|l|l|}
    \hline 
    \textbf{Category} & \textbf{Training Status} & \textbf{Model} & \textbf{Studies} \\
    \hline
    \multirow{5}{*}{Other} & \multirow{5}{*}{Custom-trained} & Transformer-based Autoregressive model  & \cite{P86}, \cite{P87}, \cite{P89}\\
    \cline{3-4} & & Deep Autoregressive Model & \cite{P44}, \cite{P47}, \cite{P67}\\
    \cline{3-4} & & Flow-based Autoregressive model & \cite{P8}\\
    \cline{3-4} & & Customized Neural Network-based Model & \cite{P90}\\
    \cline{3-4} & & Query-Based Transformer model & \cite{P83} \\
    \hline
\end{tabular}
\label{table:results:RQ1:models:other}
\end{table}

\textbf{Other} generative models in addition to the previously introduced models in the included studies were essentially deep neural networks specifically designed and trained for generative purposes, employing particular architectures (e.g., Transformer~\cite{NIPS2017_3f5ee243}) or generative strategies (e.g., autoregressive~\cite{pmlr-v32-gregor14}, flow-based~\cite{pmlr-v80-huang18d}) to generate desired driving scenarios for testing ADS, as shown in Table~\ref{table:results:RQ1:models:other}. We explain how they are used in details in Section~\ref{sec:results:RQ1:testing}.

\begin{table}[hbp]
\centering
\small
\caption{RQ1 -- Hybrid models used in the selected studies.}
\begin{tabular}{|c|l|l|}
    \hline 
    \textbf{Category} & \textbf{Models} & \textbf{Studies} \\
    \hline
    \multirow{6}{*}{Hybrid} & LLM \& VLM & \cite{P17}, \cite{P32} \\
    \cline{2-3} & LLM \& Diffusion-based & \cite{P43}, \cite{P60}, \cite{P70}, \cite{P81}, \cite{P82}, \cite{P94} \\
    \cline{2-3} & LLM \& Other & \cite{P83}, \cite{P90} \\
    \cline{2-3} & AE-based \& GAN-based & \cite{P28}, \cite{P40}, \cite{P80}, \cite{P55}, \cite{P72}\\
    \cline{2-3} & AE-based \& Diffusion-based & \cite{P41} \\
    \cline{2-3} & AE-based \& AE-based & \cite{P7} \\
    \hline
\end{tabular}
\label{table:results:RQ1:models:hybrid}
\end{table}

\textbf{Hybrid} models refer to generative frameworks that combine two or more distinct generative models (e.g., LLMs and diffusion-based models). In these frameworks, each model is responsible for a specific component, task, or function within the complete generative process. A typical pattern found in the included studies involved integrating LLMs with other models. In such cases, the LLM typically handles natural language instructions and produces such as intermediate representations, supporting functions, or partial scenario elements, which are then processed by another generative model downstream to fulfill the generative tasks. In addition, several studies combined AE-based models with other generative methods. AE-based models were often employed to map data between the original data space and a latent representation, enabling subsequent models to predict or generate new driving scenarios within that latent space.

\begin{figure}[tbp]
    \centering
    \includegraphics[width=\textwidth]{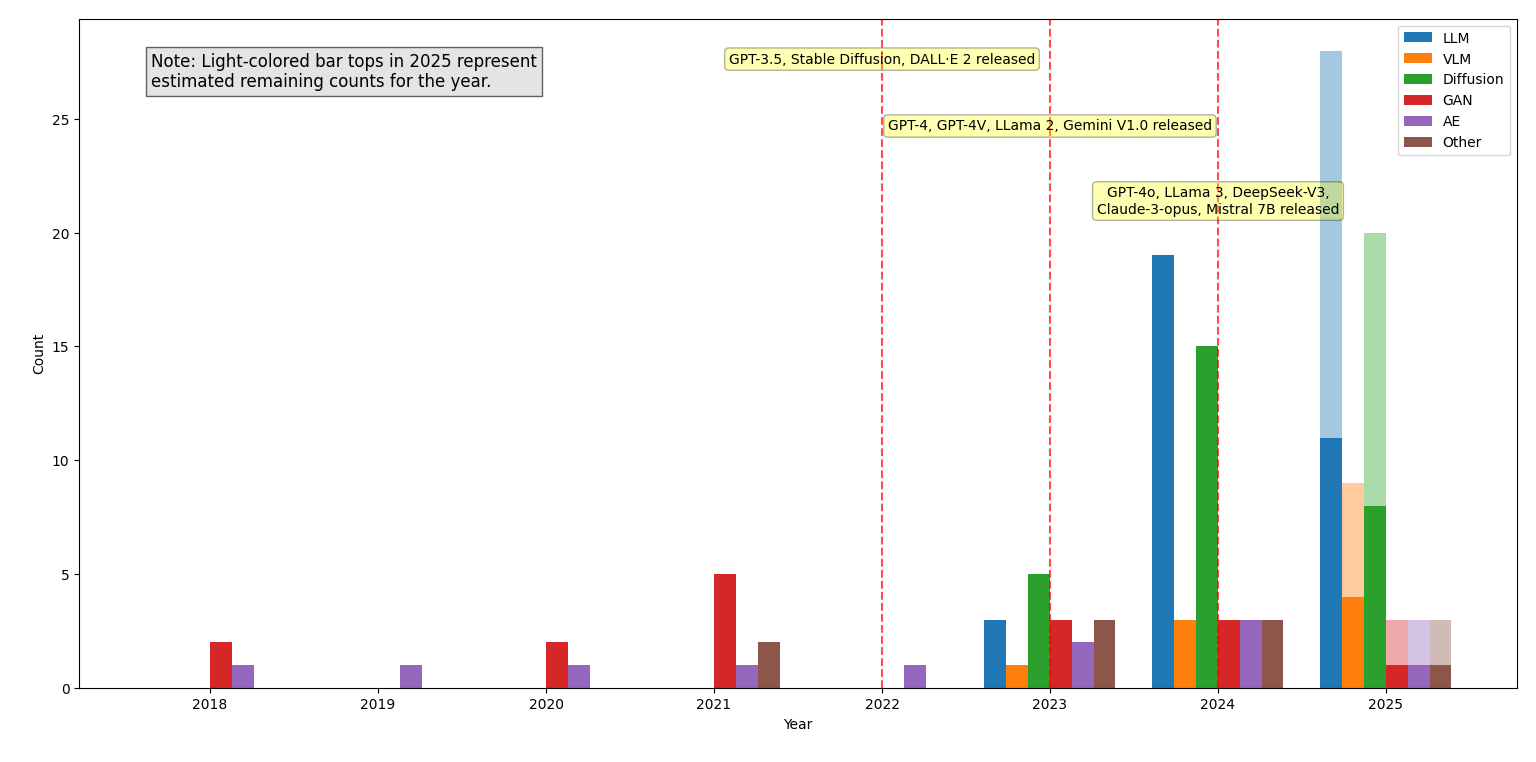}
    \caption{Breakdown of the number of papers published per year based on the types of generative models used. Particularly for 2025, since the snowballing was conducted on May 15, the estimated number of publications (the light-colored bar tops in 2025) for the remainder of the year was projected and rounded based on the number of papers identified up to that date.}
    \label{fig:results:demographics:modelsperyear}
\end{figure}

\par\vspace{2mm}

\faHandPointRight \ Overall, \emph{various types of generative AI models are employed for testing ADS} across the selected studies. Furthermore, we analyzed the yearly distribution of publications based on the types of generative models used in the included studies. As described in Section~\ref{sec:results:demographics}, the number of publications has steadily increased in recent years, with the majority appearing from 2023 onward. Figure~\ref{fig:results:demographics:modelsperyear} further illustrates that \emph{this sharp rise is largely driven by the adoption of LLMs and diffusion-based models} (e.g., GPT and Stable Diffusion models), which account for 58\% (53 of 91) of all included studies. From 2018 to 2022, only GAN-based models, AE-based models, and other customized generative models were utilized, with no studies incorporating LLMs, VLMs, or diffusion-based models during that period. The widespread attention around LLMs, VLMs, and diffusion models has stimulated their use in testing of ADS simultaneously. Therefore, we anticipate that \emph{research leveraging these models will continue to grow in the near future}.  

\subsubsection{Targeted ADS}
\label{sec:results:RQ1:ADS}

\begin{table}[hbp]
\centering
\small
\caption{RQ1 -- Targeted ADS in the selected studies.}
\begin{tabular}{|c|l|l|}
    \hline 
    \textbf{Level} & \textbf{Targeted ADS} & \textbf{Studies} \\
    \hline
    \multirow{2}{*}{System} & \multirow{2}{*}{ADS/AVs} & \textbf{*} All (\textbf{67}) studies that are \textbf{not} listed on the Module \\
    & & and Function levels \\
    \cline{1-3} \multirow{7}{*}{Module} & Planning module & \cite{P4}, \cite{P15}, \cite{P18}, \cite{P41}, \cite{P54}, \cite{P64}, \cite{P68}, \cite{P69}, \cite{P74}, \cite{P79}, \cite{P94} \\
    \cline{2-3} & Decision-making module & \cite{P18}, \cite{P21}, \cite{P41}, \cite{P76} \\
    \cline{2-3} & Perception module & \cite{P30}, \cite{P78} \\
    \cline{2-3} & Large Visual language models & \cite{P6}, \cite{P77} \\
    \cline{2-3} & Deep Learning components & \cite{P70} \\
    \cline{2-3} & Control algorithm & \cite{P76} \\
    \cline{2-3} & Driving algorithm & \cite{P5} \\
    \cline{1-3} \multirow{3}{*}{Function} & Object detection function & \cite{P7}, \cite{P62}, \cite{P63} \\
    \cline{2-3} & Lane-keeping function & \cite{P7}, \cite{P59} \\
    \cline{2-3} & Vehicle-following function & \cite{P65} \\
    \hline
\end{tabular}
\label{table:results:RQ1:models:targetads}
\end{table}

To better understand how generative AI is applied in testing, we also examined which aspects of ADS the included studies aimed to test -- whether ADS/AVs as a whole, or specific modules (i.e., components) or functions (i.e., features) within the system. As shown in Table~\ref{table:results:RQ1:models:targetads}, the majority of studies (67 out of 91, or 74\%) focus on testing ADS/AVs in general, without explicitly targeting or prioritizing any particular module or function. The remaining 24 studies (26\%), however, specifically target individual components such as planning (e.g., motion or trajectory planning), decision-making, perception (e.g., vehicle detection or weather classification), or control and driving algorithms. Notably, two studies focus on testing large vision-language models (LVLMs) used in ADS for vision-language tasks, while another explicitly targets deep learning enabled components within ADS. Additionally, several studies concentrate on specific driving functions, such as object detection, lane-keeping, or vehicle-following.

\par\vspace{2mm}

\faHandPointRight \ Overall, \emph{most included studies refer to ADS or AVs in general terms}, without explicitly specifying which modules or functions are being targeted. As a key takeaway and recommendation, we suggest that \emph{future research should clearly specify the targeted system modules, functions, or testing attributes} that the proposed generative models are designed to test, which facilitates a clearer understanding of their intended application and scope.  

\subsubsection{Generative AI for Testing ADS}
\label{sec:results:RQ1:testing}

\begin{figure}[tbp]
    \centering
    \includegraphics[width=0.9\textwidth]{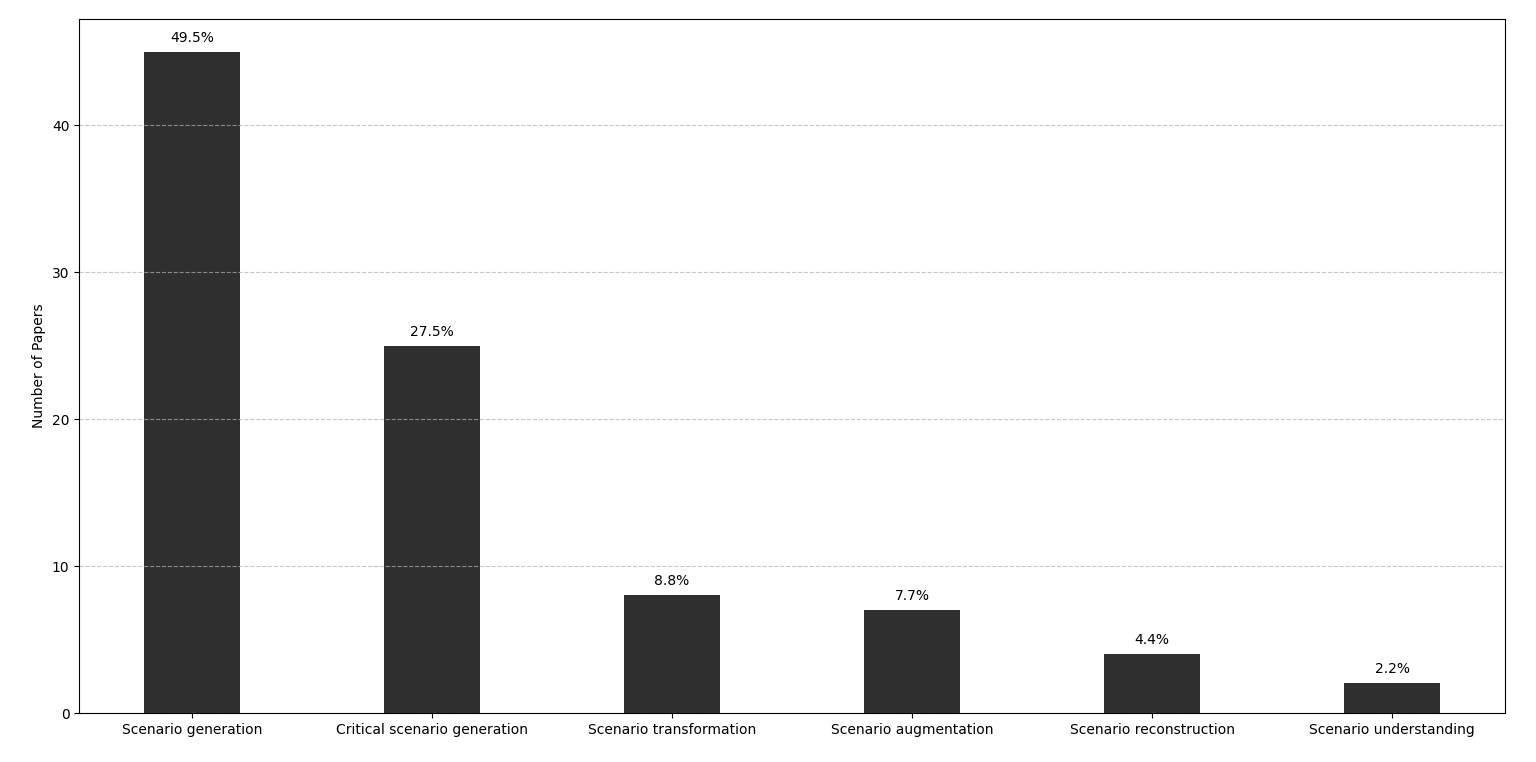}
    \caption{Distribution of papers by task (objective) in ADS testing, including counts and proportions.}
    \label{fig:results:RQ1:testing:numbersandratiospertask}
\end{figure}

Based on our analysis, the included studies applied generative AI to support six key tasks related to testing of ADS: (1) scenario generation, (2) critical scenario generation, (3) scenario transformation, (4) scenario augmentation, (5) scenario reconstruction, and (6) scenario understanding, as in Figure~\ref{fig:results:RQ1:testing:numbersandratiospertask}. Each task leverages different types of generative models and mechanisms tailored to its specific objectives. We present these studies and their corresponding mechanisms to provide an overview of how generative AI is used in the context of ADS testing.

\begin{table}[hbp]
\centering
\small
\caption{RQ1 -- Studies using Generative AI for scenario generation.}
\begin{tabular}{|c|l|p{0.21\textwidth}|l|}
    \hline 
    \textbf{Test objective} & \textbf{Generative model} & \textbf{Highlighted mechanisms} &  \textbf{Studies} \\
    \hline
    \multirow{13}{*}{Scenario generation} & \multirow{2}{*}{LLM} & Prompt engineering & \cite{P3}, \cite{P46}, \cite{P56} \\
    \cline{3-4} & & Prompt engineering with iterative refinement & \cite{P59}, \cite{P85}, \cite{P26}, \cite{P34}, \cite{P37}\\
    \cline{2-4} & VLM & Prompt engineering & \cite{P6}, \cite{P78} \\
    \cline{2-4} & \multirow{2}{*}{Diffusion-based} & Goal conditioned & \cite{P14}, \cite{P51}, \cite{P68}, \cite{P69} \\
    \cline{3-4} & & Context conditioned & \cite{P42}, \cite{P18}, \cite{P53}, \cite{P84}, \cite{P88} \\
    \cline{2-4} & \multirow{2}{*}{GAN-based} & Non-context conditioned & \cite{P13}, \cite{P29}, \cite{P48}, \cite{P49} \\
    \cline{3-4} & & Context conditioned & \cite{P28}, \cite{P63}, \cite{P80} \\
    \cline{2-4} & AE-based & Latent generative modeling, Context conditioned & \cite{P38}, \cite{P66}, \cite{P65} \\
    \cline{2-4} & \multirow{2}{*}{Other} & Autoregressive & \cite{P67} \\
    \cline{3-4} & & Autoregressive, Context conditioned & \cite{P47}, \cite{P44}, \cite{P86}, \cite{P87}, \cite{P89} \\
    \cline{2-4} & \multirow{2}{*}{Hybrid} & Latent generative modeling & \cite{P28}, \cite{P40}, \cite{P80}, \cite{P41}, \cite{P72} \\
    \cline{3-4} & & Multi-stage generation & \cite{P17}, \cite{P43}, \cite{P60}, \cite{P81}, \cite{P82}, \cite{P83}, \cite{P90} \\
    \hline
\end{tabular}
\label{table:results:RQ1:models:testing:scenario_generation}
\end{table}

\noindent \textbf{(1) Scenario generation. } The majority of the included studies employed generative AI to support scenario generation, specifically \emph{to create intended driving scenarios for testing ADS} using different generative models, as shown in Table~\ref{table:results:RQ1:models:testing:scenario_generation}. While critical scenario generation is fundamentally a form of scenario generation, we present it separately due to its distinct objective to create safety-critical scenarios that are likely to challenge or cause failures for ADS. 

\par\vspace{2mm}

\textbf{(1.1) LLMs} are used to create driving scenarios based on natural language instructions. To ensure the scenarios closely match the instructions, advanced \textbf{prompt engineering} techniques~\cite{sahoo2024systematic}, e.g., In-context learning, Contextual prompting, Chain-of-thought, Retrieval-Augmented generation, Self-consistency, are applied~\cite{P3, P46, P56}. Specifically, Cai et al.~\cite{P46} designed a framework (Text2Scenario) to generate scenarios through a two-stage process. In the first stage, an LLM takes a scenario text description and a hierarchical scenario description dictionary as input to identify key elements -- such as the ego vehicle, climate, traffic participants, transportation facilities, and road topology -- and produce a concise scenario representation. The prompt is meticulously constructed and includes several components: a role setting, a task description, a series of input-output examples (scenario descriptions paired with their concise representations), a chain-of-thought process that breaks down reasoning steps, a syntax alignment check to validate the output against the original scenario description, and a self-consistency check that selects the most consistent result via majority voting across multiple reasoning paths. In the second stage, the framework converts the LLM-generated scenario representation into an OpenScenario~\cite{openscenario} specification using a library of predefined descriptive fragments. Ruan et al.~\cite{P56} proposed a framework (TTSG) that generates traffic scenarios from user-provided text prompts through a multi-step process. First, an LLM agent analyzes the input to extract key components such as required objects and agents. Next, another LLM agent queries a pre-constructed road database to identify candidate roads that match the user's specifications. A third LLM agent plans the agents involved based on the user input. The framework then ranks road candidates based on the intended road type and agent plan. Finally, the system generates the scenario by translating the LLM-generated output into a scenario within Carla simulator~\cite{dosovitskiy2017carla} through a rendering interface.

To ensure the generated scenarios accurately align with the user's intent, \textbf{prompt engineering} is often employed \textbf{with an iterative refinement} strategy by iteratively refining the LLM-generated outputs with feedback~\cite{P59, P85, P26, P34, P37}. For instance, Zhou et al.~\cite{P59} introduced a method that leverages an LLM to generate driving scenarios from text prompts. The LLM agent first decomposes the user's scenario description into key components and organizes them into a structured representation with six elements: road, traffic infrastructure, temporary manipulations, objects, environment, and digital information. Users can dynamically modify or extend the scenario through natural language instructions. Once the user confirms the scenario, the LLM agent invokes appropriate functions to convert the structured data into a OpenScenario~\cite{openscenario} specification file. Rubavicius et al.~\cite{P26} proposed a dialogue-based system that generates test scenarios in Scenic code~\cite{fremont2019scenic} through natural language conversation. If an error occurs in the generated scenario (e.g., it is non-executable), the system feeds the description, generated code, and error message back to the LLM for correction. Additionally, the generated scenarios are simulated and presented to the user, enabling feedback to refine the scenario, better align it with user intent, and eliminate undesired behaviors or hallucinations. Lu et al.~\cite{P37} designed a testing pipeline (OmniTester) for generating test scenarios. It uses multiple LLM-enabled components to interpret user requests, construct road networks, and configure vehicle dynamics, to form a complete scenario. A separate LLM-based evaluator then assesses whether the generated scenario aligns with the user's intent. If the resulting vehicle behavior fails to meet the desired criteria, feedback is incorporated into the prompt to iteratively refine the scenario.

\par\vspace{2mm}

\textbf{(1.2) VLMs} are also used to generate test scenarios from user instructions~\cite{P6, P78}, leveraging various \textbf{prompt engineering}~\cite{sahoo2024systematic} techniques. Unlike LLMs, VLMs can incorporate and interpret visual information (e.g., images) as part of the input, enabling more context-rich and visually grounded scenario generation. Instead of generating driving scenarios for testing purposes, Khalili and Smyth~\cite{P6} designed a pipeline (AutoDrive-QA) to evaluate VLMs for autonomous driving. It first generates plausible but incorrect answers using tailored prompts applied to existing open-ended and visual-based question–answer (QA) datasets (e.g., LingoQA~\cite{marcu2024lingoqa}, DriveLM~\cite{sima2024drivelm}, NuScenes-QA~\cite{qian2024nuscenes}). The pipeline then filters these incorrect responses and converts the open-ended QA pairs into multiple-choice questions, facilitating testing of VLMs in driving-related scenarios. Marathe et al.~\cite{P78} provided a dataset (WEDGE), which consists of synthetic scenarios under 16 different adverse weather conditions, generated using a VLM. The dataset is created using structured prompts that follow a pattern: "\textit{objects} on \textit{scenes} when \textit{weather}". Objects (e.g., cars), scenes (e.g., highway), and weather conditions (e.g., sunny) are sampled from predefined option sets. The generated scenarios (i.e., images) are then manually verified and cross validated to correct any errors, mismatches, or inconsistencies.

\par\vspace{2mm}

\textbf{(1.3) Diffusion-based models} are used for controllable generation of driving scenarios, \textbf{conditioned on} various factors such as \textbf{user goals}~\cite{P14, P51, P68, P69} or scenario context (e.g., map information, road layout, and vehicle trajectory history)~\cite{P42, P18, P53, P84, P88}. These conditions are encoded and integrated into the diffusion process, guiding the model to generate realistic driving scenarios that align with the specified constraints or objectives. Wang et al.~\cite{P51} introduced a conditional diffusion-based framework (RADE) for generating realistic, risk-adjustable driving scenarios. RADE employs a multi-agent diffusion model to jointly generate vehicle trajectories based on current vehicle states and a specified risk level. To ensure physical plausibility, it incorporates a tokenized dynamics check module, which refines each predicted motion by aligning it with a pre-built motion token library derived from real-world data. This guarantees that state transitions are consistent with realistic vehicle dynamics. Chitta et al.~\cite{P68} proposed a driving scene synthesis framework that learns from real-world driving logs. It employs a Diffusion Transformer (DiT), trained using the DDPM algorithm~\cite{NEURIPS2020_4c5bcfec}, to generate new RLMs (rasterized latent maps representing driving scenarios) from noise, conditioned on a city label (e.g., Las Vegas). In addition, Zhou et al.~\cite{P14} designed a diffusion model-based framework (DiffRoad) for generating road scenarios in OpenDrive~\cite{opendrive} format. The framework trains a diffusion model to reconstruct road scenarios from random noise, conditioned on road attributes such as road type and road size. Rowe et al.~\cite{P69} used a latent diffusion model, which embeds lanes and objects with factorized attention (lane-to-lane, lane-to-object, and object-to-object) into a latent representation to generate initial driving scenes. Users can control scene density by specifying the number of agents and lanes. The diffusion model iteratively denoises sampled latent to create realistic initial scenes as test scenarios.

\textbf{Scenario context} is frequently used in diffusion-based models~\cite{P42, P18, P53, P69, P84, P88} for \textbf{conditional} scenario generation based \textbf{on given contextual information}, such as map data, road layout, and vehicle trajectory history. Lu et al.~\cite{P42} presents a diffusion-based framework (SceneControl) for controllable scenario generation, conditioned on a high-definition (HD) map and guided by high-level constraints such as spatial region constraints (e.g., region of interest within the scene) and actor attribute constraints (e.g., vehicle speed, bounding box size). The framework consists of three key components: a lane graph neural network (GNN) that encodes HD maps to capture road layouts and lane structures; a Transformer encoder that extracts interactions between vehicles and between vehicles and the map; a diffusion model that learns to denoise random noise into realistic driving scenarios, guided by both the encoded map features and the interaction representations captured by the Transformer decoder. Zhong et al.~\cite{P53} proposed a conditional diffusion-based method (CTG) for generating realistic driving scenarios that adhere to user-specified traffic rules. The model predicts future trajectories for a target agent, conditioned on an agent-centric semantic map, the historical states of the target agent, and information about surrounding agents, and guided by user-defined rules. During the sampling process, multiple trajectory candidates are generated, and the one that best satisfies the specified rules is selected as the final output. Wang et al.~\cite{P88} developed a controllable traffic scenario generation framework (DragTraffic), which generates desired agent behaviors based on the current trajectory, a context description (e.g., vehicle type, size, start and target positions and speeds), and a road map. The framework first employs Multipath++~\cite{varadarajan2022multipath++}, a trajectory prediction model, to generate an initial trajectory from the given context. Then, a conditional diffusion model refines this trajectory by gradually denoising it, guided by the context, the initial trajectory, and the road map.

\par\vspace{2mm}

\textbf{(1.4) GAN-based models} are used to generate new scenarios that closely resemble real-world driving scenarios~\cite{P13, P29, P48, P49}. Specifically, Fang et al.~\cite{P13} developed a generative framework (LC-GAN) to synthesize highway cut-in scenarios. The framework includes a generator that produces a sequence of trajectory points representing the cut-in scenario from a random noise vector, and a discriminator that evaluates whether the generated trajectories appear realistic. It is built on a transformer-based time-series GAN (TTS-GAN)~\cite{ttsgan}, enhanced with a temporal feature extraction module to capture motion patterns over time and an attention mechanism to improve the smoothness and realism of the generated trajectories. Spooner et al.~\cite{P29, P48} proposed a deep generative model (Ped-Cross GAN) to generate pedestrian-crossing scenarios, represented as sequences of human poses. The framework includes a generator that produces new pose sequences based on input human pose data, effectively simulating pedestrian crossing behaviors, and a discriminator that evaluates whether the generated sequences resemble realistic human movements. Li et al.~\cite{P49} presents a generative framework for lane-changing scenarios using TimeGAN~\cite{NEURIPS2019_c9efe5f2}. The framework includes four key components: an embedding function that maps real time-series data into a latent space, a recovery function that reconstructs the original sequences from these embeddings, a generator that produces synthetic latent sequences from random noise, and a discriminator that distinguishes between real and generated sequences. During training, the generator learns to produce realistic latent sequences that are indistinguishable from real data. After training, new lane-changing scenarios are generated by sampling noise, passing it through the generator and recovery function, and producing synthetic trajectories that capture realistic and temporally coherent vehicle behaviors.

\textbf{Scenario context} (e.g., vehicle trajectory history, trajectory length) can also be incorporated as \textbf{conditions} for the GAN-based \textbf{generative models}, enabling generation of scenarios that based on given context information or constraints. Demetriou et al.~\cite{P28, P80} presents a generative model based on Recurrent Conditional GAN (RC-GAN), in which both the generator and discriminator are implemented using LSTM-based recurrent neural networks (RNNs). The model generates complete trajectory sequences directly from random noise, conditioned on the desired sequence length. Yang et al.~\cite{P63} designed a generative framework (SurfelGAN), which produces realistic camera images (representing scenarios) from surfel renderings -- 2D projections of 3D scenes derived from LiDAR and semantic information. The framework first employs a surfel-based rendering pipeline to project scene geometry and semantics into surfel maps from novel viewpoints. These maps are then input to a conditional GAN, where the generator learns to translate them into photorealistic images, while the discriminator evaluates the realism of the generated outputs.

\par\vspace{2mm}

\textbf{(1.5) Autoencoder-based models} offer efficient generation within \textbf{a compressed latent space}, and are used for controllable driving scenario generation \textbf{conditioned on} certain \textbf{context information}~\cite{P38, P66, P65}. Ding et al.~\cite{P38} introduced a retrieval-augmented framework (RealGen), built on a contrastive autoencoder. The autoencoder consists of three encoders that extract latent representations for vehicle behaviors, map layout, and initial positions, provided by users, and a decoder that reconstructs full scenarios from these embeddings. For scenario generation, the framework uses K-nearest neighbors (KNN) to retrieve K most similar behavior embeddings to a given scenario from the database. A combiner module then fuses the retrieved behaviors with the target map and user-defined initial positions to form a new composite latent representation, which is decoded into a complete scenario that satisfies all specified conditions. Ding et al.~\cite{P66} proposed a deep generative model (MTG) for generating multi-vehicle interactive scenarios. MTG is built on a VAE framework with two key enhancements: a bi-directional GRU encoder that captures past trajectories in both forward and backward time directions, and a multi-branch decoder that generates future trajectories for each vehicle separately while sharing hidden states to preserve interaction context. During inference, the model encodes the past trajectories of multiple vehicles into a latent space, samples from it, and generates future trajectories that capture realistic inter-vehicle interactions. Gong et al.~\cite{P65} presents a Dual-VAE-based generative model for generating two-vehicle trajectories in vehicle-following scenarios. The model consists of two parallel VAEs -- one for the leading vehicle and one for the following vehicle. Each VAE encodes the past trajectory of its respective vehicle, samples a latent vector through the reparameterization trick, and generates future trajectory points using an LSTM-based decoder.

\par\vspace{2mm}

\textbf{(1.6) Other models} for scenario generation typically adopt an \textbf{autoregressive} approach, where scenario entities are generated sequentially, \textbf{conditioned on} specific \textbf{context information} such as HD maps and the states of existing entities~\cite{P67, P47, P44, P86, P87, P89}. For example, Feng et al.~\cite{P47} introduced TrafficGen, a method designed to generate driving scenarios for testing ADS. TrafficGen employs an autoregressive encoder-decoder architecture, where the encoder processes HD maps and vehicle states to capture the traffic context, including road layout, vehicle positions, and traffic signals, using multi-context gating and cross-attention to integrate information across different map regions. The decoder then sequentially places vehicles and generates their long-term trajectories using probabilistic modeling. Similarly, Tan et al.~\cite{P44} proposed a generative model (SceneGen) for creating realistic scenarios. SceneGen employs a neural autoregressive model that sequentially inserts actors such as vehicles, pedestrians, and bicyclists into a scene, conditioned on a HD map and ego vehicle’s state. This iterative process continues until a stopping token is generated, signaling the end of the scenario. Several studies~\cite{P86, P87, P89} have proposed transformer-based autoregressive models for scenario generation. Specifically, Mahjourian et al.~\cite{P87} introduced UniGen, which uses a shared scenario encoder to process both static context (e.g., map data, traffic lights) and dynamic elements (e.g., existing agents), producing a dense feature representation. Three specialized decoders are then used to predict the initial positions of new agents, their attributes (e.g., heading, speed), and their future trajectories. UniGen then adds one agent at a time while ensuring that each new agent's behavior accounts for previously generated agents. In contrast, Riegl et al.~\cite{P89} embeds each existing road user and road infrastructure (from HD maps) using a multi-layer perceptron (MLP), and employs a transformer-based model to autoregressively predict the attributes of the next agent. A rule-based scenario planner is then used to generate future trajectories, leveraging physical vehicle models initialized from the predicted starting states.

\par\vspace{2mm}

\textbf{(1.7) Hybrid models} are commonly used for scenario generation, which combine two distinct generative models, each responsible for a specific task within the overall generation process. A typical approach \textbf{integrates autoencoder-based models} with GAN-based or diffusion-based models~\cite{P28, P40, P80, P41, P72}, in which the autoencoder-based model is used to map scenarios between the scenario space and a latent space, while another generative model (e.g., GAN or diffusion model) is employed to generate new latent representations that correspond to newly generated scenarios. For example, Demetriou et al.~\cite{P28, P80} proposed an AE-GAN architecture for scenario generation conditioned on trajectory length. In this framework, an autoencoder first learns to encode trajectories into a latent space. A GAN is then trained to generate new latent vectors, and the decoder from the autoencoder reconstructs full trajectories from the generated latent vectors, conditioned on the specified trajectory length. Yang et al.~\cite{P40} applied a similar mechanism for highway cut-in scenario generation, where an autoencoder and a Transformer-based encoder are used to capture the temporal and spatial patterns of the scenarios. A Wasserstein GAN~\cite{arjovsky2017wassersteingan} is then used to generate new latent vectors, and a decoder reconstructs full scenarios from these latent vectors. Kim et al.~\cite{P72} presents a generative approach (DriveGAN), which learns to generate realistic driving scenarios by observing past driving history. It adopts a VAE-GAN framework with an encoder that extracts disentangled latent codes representing the scenario’s theme (e.g., weather, background color) and content (e.g., spatial layout). These latent codes are fed into a Dynamics Engine, implemented as a RNN, which predicts the next-step latent codes based on the current state and the agent’s action latent. The predicted latent codes are then passed to a modified StyleGAN generator, which decodes them into high-quality scenario images. In contrast, Pronovost et al.~\cite{P41} proposed a novel architecture (Scenario Diffusion), which uses a denoising diffusion model to generate realistic scenarios, conditioned on a Bird’s-Eye View (BEV) map image and a set of descriptive tokens. The framework includes an autoencoder that encodes and decodes data between the input and latent spaces, and a diffusion model that learns to denoise random latent vectors into cleaned latent. These latent are then decoded into final scenario outputs. Conditioning is applied during the denoising process via a cross-attention mechanism, using the BEV map and tokens (e.g., vehicle speed, position, number of agents) to guide generation.

Another common approach \textbf{integrates LLMs} with VLMs, diffusion-based models, or other generative models~\cite{P17, P43, P60, P81, P82, P83, P90} into a \textbf{multi-stage generation} process. The LLM processes natural language instructions to generate intermediate representations, supporting functions, or partial scenario elements and so on. These outputs are then passed to another model in the pipeline, which completes the scenario generation. Specifically, Zhang et al.~\cite{P17} proposed a two-stage scenario generation framework (DriveGen). In the first stage, a LLM, augmented with a retrieval mechanism, is used to generate a map and place vehicles on it, forming the initial scenario setup. In the second stage, a VLM selects goals for each vehicle, and a diffusion-based trajectory planner generates realistic dynamic trajectories for the vehicles based on those goals. Zhong et al.~\cite{P60} proposed a conditional diffusion model (CTG++) for generating controllable multi-agent traffic scenarios based on natural language commands. The model leverages a spatio-temporal transformer architecture to jointly model agent dynamics, capturing temporal patterns in trajectories, inter-agent interactions, and map context. Trajectories for all agents are generated by iteratively denoising Gaussian noise, with each step guided by a differentiable loss function. This loss function is written in code by an LLM, based on the user’s natural language input (e.g., “\textit{Vehicle A should collide with Vehicle B}”), and serves as an optimization objective. The diffusion model then adjusts predicted trajectories to meet the specified behavioral constraints. Liu et al.~\cite{P81} presents a diffusion-based framework for generating controllable driving scenarios, guided by cost functions derived from natural language instructions using an LLM. The LLM decomposes the user’s instructions into individual events, identifies each agent’s intended behavior, and generates corresponding cost functions. These functions guide the diffusion model during the denoising process to produce realistic, goal-aligned multi-agent trajectories. Tan et al.~\cite{P83} introduced a language-conditioned generative model (LCTGen), designed to generate realistic traffic scenarios from natural language descriptions and a library of real-world maps. LCTGen comprises three key modules: An interpreter module that uses an LLM to convert user-provided text into a structured representation, a retrieval module takes this structured representation and selects a matching map from a real-world map dataset, and a transformer-based generator module that encodes both map and agent features, captures agent-agent and agent-map interactions, produces a latent representation, and decodes it into a complete, meaningful scenario. Xia et al.~\cite{P90} proposed a generative model (InteractTraj) that produces interactive traffic trajectories from natural language descriptions. It comprises two main modules: a language-to-code encoder, based on an LLM, which converts natural language commands into numeric codes representing interactions, vehicles, and map elements, and a code-to-trajectory decoder, implemented as a neural network, which generates interactive traffic trajectories by aggregating and decoding these numeric codes, capturing both vehicle behavior and their interactions.

\par\vspace{2mm}

\begin{table}[tbp]
\centering
\small
\caption{RQ1 -- Studies using Generative AI for critical scenario generation.}
\begin{tabular}{|c|l|l|l|}
    \hline 
    \textbf{Test objective} & \textbf{Generative model} & \textbf{Highlighted mechanisms} &  \textbf{Studies} \\
    \hline
    \multirow{9}{*}{Critical scenario generation} & \multirow{2}{*}{LLM} & Prompt engineering & \cite{P5}, \cite{P10}, \cite{P21}, \cite{P35} \\
    \cline{3-4} & & Prompt engineering, multi-stage generation & \cite{P27}, \cite{P39}, \cite{P57}, \cite{P58} \\
    \cline{2-4} & VLM & Prompt engineering, multi-stage generation & \cite{P36}, \cite{P93} \\
    \cline{2-4} & \multirow{2}{*}{Diffusion-based} & Goal conditioned & \cite{P61}, \cite{P62}, \cite{P64} \\
    \cline{3-4} & & Goal conditioned, context conditioned & \cite{P4}, \cite{P50}, \cite{P52} \\
    \cline{2-4} & GAN-based & Context conditioned & \cite{P15}, \cite{P25}, \cite{P76} \\
    \cline{2-4} & AE-based & Goal conditioned, context conditioned  & \cite{P16}, \cite{P54}, \cite{P91} \\
    \cline{2-4} & Other & Causality & \cite{P8} \\
    \cline{2-4} & Hybrid & Prompt Engineering with iterative refinement & \cite{P32}, \cite{P94} \\
    \hline
\end{tabular}
\label{table:results:RQ1:models:testing:critical_scenario_generation}
\end{table}

\noindent\textbf{(2) Critical scenario generation.} A considerable number of studies used generative AI for critical scenario generation, primarily focusing on \emph{creating safety-critical scenarios that are hazardous and may lead to failures in ADS}, such as collisions. Similar to general scenario generation, various types of generative AI models (see Table~\ref{table:results:RQ1:models:testing:critical_scenario_generation}) are employed. In the following, we outline how these models are used and highlight the key mechanisms and strategies behind their designs.

\par\vspace{2mm}

\textbf{(2.1) LLMs} are used to generate critical driving scenarios (e.g., safety-critical scenarios, out-of-distribution scenarios involving rare road events) based on natural language instructions. As with general scenario generation we introduced previously, various \textbf{prompt engineering} techniques~\cite{sahoo2024systematic} (e.g., In-context learning, Contextual prompting, Chain-of-thought, Retrieval-augmented generation, Self-consistency) and generation strategies (e.g., multi-stage generation) are applied to improve the alignment between the generated scenarios and the user's intent~\cite{P5, P10, P21, P35}. Qiu et al.~\cite{P5} proposed a framework (AED) that generates reward functions in code to support vulnerability discovery in ADS. A carefully crafted prompt, incorporating both the environment context and the task description, is used to guide the generation process. The generated reward functions are then used to train adversarial agents that create accident-inducing scenarios, enabling the identification of potential vulnerabilities in ADS. The generated scenarios are evaluated by humans using predefined criteria and accident video playback to determine whether the accident type aligns with the original prompt, and are then used to create new reward functions through preference-based reinforcement learning, enabling the generation of more effective scenarios that uncover a wider range of vulnerabilities. Zhang et al.~\cite{P10} developed an LLM-based agent (ChatScene) for generating safety-critical driving scenarios in Scenic code. ChatScene uses an LLM to generate a natural language description of a critical scenario, then parses the description to extract key characteristics (e.g., entities, behaviors). These are encoded into embeddings and used to retrieve relevant Scenic code snippets from a pre-constructed database. Finally, the retrieved snippets are assembled into a complete and executable Scenic script for simulation in Carla~\cite{dosovitskiy2017carla} and testing of ADS. Xu et al.~\cite{P21} presents a generative framework (LLMTester), designed to generate scenarios for evaluating the decision-making module of ADS. An LLM-based generator transforms seed scenarios selected from a pre-established database and a set of modifications (derived from expert knowledge and testing feedback) into prompts to generate new and challenging scenarios. A scenario evaluation module assesses both the seed and generated scenarios based on criticality and diversity, providing guidance for future seed selection, scenario database updates, and generation of new scenarios. Similarly, Chang et al.~\cite{P35} proposed a generative framework (LLMScenario) for generating corner-case risky scenarios. The framework begins by establishing a scenario database from real-world data. Scenarios are then generated using an LLM with structured prompts that include elements such as role definition, task description, input scenario context, prior generation experience and evaluation feedback (e.g., bad cases and reasons), step-by-step reasoning, and a specified output format. The generated scenarios are evaluated using a scoring function based on realism and rarity. Evaluation results and identified bad cases are fed back into the prompt as experiential feedback, enabling in-context tuning for improved future generation.

\par\vspace{2mm}

A \textbf{multi-stage generation} strategy, involving \textbf{multiple LLM agents}, is commonly applied to LLMs to support more precise and structured scenario generation, building on prompt engineering~\cite{P27, P39, P57, P58}. Aasi et al.~\cite{P27} introduced an LLM-based framework for generating diverse out-of-distribution (OOD) scenarios, which are typically rare and challenging. The process begins with an LLM generating similar examples based on a given OOD scenario, classifying them using predefined patterns, and organizing them into a tree structure (initial tree). This tree is then pruned according to the available assets in Carla simulator~\cite{dosovitskiy2017carla}, resulting in a simulatable tree. Finally, the LLM generates textual descriptions of OOD scenarios based on this tree and augments scenario details to prepare them for simulation in Carla. Lu et al.~\cite{P39} introduced a testing framework (AutoScenario), which generates safety-critical scenarios from multimodal inputs such as user text requests, accident reports, images, and videos. The framework first uses multiple interpreters, including an LLM-based text interpreter and VLM-based image and video interpreters, to extract key components (e.g., road layout, objects, agents, and weather) and produce a standardized scenario description. Next, several LLM-powered generators are used to create the road network, dynamic agents (e.g., pedestrians, cyclists, vehicles), and static elements (e.g., traffic signs, fences, traffic cones). Finally, a scenario generator integrates all elements into a complete scenario. Mei et al.~\cite{P57} proposed a framework for online generation of safety-critical scenarios through a multi-stage process. First, an LLM-based behavior analyzer infers the safety-critical intent of background vehicles based on their observed history. Then, an LLM-based code generator predicts the vehicle’s future endpoint from the inferred intent and generates a specialized trajectory planner that produces a temporally smooth and kinematically feasible trajectory. If the resulting trajectory fails to induce meaningful safety-critical events in simulation, a code modifier refines the planner code. All historical behavior intents and their associated trajectory-generating codes are stored in a memory bank. When a new intent is inferred, the system first searches the memory for the closest matching code; if not found, it adds the new intent-code pair to the memory for future reuse. Mei et al.~\cite{P58} presents a collaborative agentic framework (LLM-Attacker) for generating adversarial scenarios to test ADS. In the first stage, the framework uses multiple LLM-enabled modules to identify optimal attacker vehicles. These modules generate initial attacker identification code, assess its validity, provide targeted suggestions for improvement, and modify the code accordingly to enhance adversarial behavior detection. In the second stage, a pre-trained probabilistic traffic forecasting model (DenseTNT)~\cite{gu2021densetnt} is used to compute realistic trajectories for the identified adversarial vehicles, forming a complete driving scenario.

\par\vspace{2mm}

\textbf{(2.2) VLMs} are used to generate critical driving scenarios based on natural language instructions combined with visual inputs (e.g., driving videos)~\cite{P36}. These models are guided by various \textbf{prompt engineering}~\cite{sahoo2024systematic} techniques to ensure the generated scenarios align with user intent and visual context. Tian et al.~\cite{P36} proposed a generative approach (LEADE) for creating safety-critical scenarios. It begins by analyzing traffic videos, identifying key frames, and extracting static and dynamic elements using a VLM. These elements are parsed into a semi-structured abstract scenario, which is then used to generate a concrete scenario. A dual-layer search is then applied to find safety-critical scenarios with safety violations: the outer layer compares the ego vehicle’s behavior in the generated scenario with human driving from the original video, and uses a feedback-guided fuzzer to find scenarios that possess significant differences. The inner layer checks whether the identified safety violation consistently occurs in the scenario. Li et al.~\cite{P93} proposed a multi-agent framework (CrashAgent) that generates crash scenarios from accident reports using VLMs. The framework consists of three parts: a Sketch Agent, which interprets crash sketches with visual tree-of-thoughts and priority voting, producing a textual description summarizing the sketches; a Road Agent that turns the textual layout descriptions into standardized OpenDrive~\cite{opendrive} road specification; and a Scenario Agent, which builds OpenScenario~\cite{openscenario} specification from 42 predefined scenario elements, then applies a genetic algorithm to tune parameter values so that collisions occur as reported, colliding users stay close, non-colliding users remain separated, and event order is preserved. 

\par\vspace{2mm}

\textbf{(2.3) Diffusion-based models} are used to generate critical driving scenarios by \textbf{incorporating specific objectives}, such as adversarial goals (e.g., collisions) or realism (e.g., rule compliance), \textbf{to guide the diffusion process}~\cite{P61, P62, P64}. Xu et al.~\cite{P61} proposed a framework (DiffScene) to generate safety-critical scenarios by introducing diverse safety-critical objectives. It first uses a goal-agnostic diffusion model, trained on natural driving data, to generate safe scenarios. These scenarios are then optimized during the diffusion process using three types of objectives: adversarial objectives (to increase driving risk), function-based objectives (to prevent task completion by the vehicle), and constraint-based objectives (to enforce traffic rules and physical realism), ensuring both high risk and realism in the final scenarios. Similarly, Chang et al.~\cite{P64} developed a diffusion-based simulation framework (SAFE-SIM) to generate realistic and controllable safety-critical driving scenarios. The diffusion model is guided at each denoising step by a loss function combining several objectives: a collision term (to promote collisions between ego and adversarial agents), two control terms (to manage relative speed and time-to-collision), a regularization term (to prevent collisions among non-adversarial agents), and a route guidance term (to keep adversarial agents on the road). To create specific collision types, the framework applies a partial diffusion process guided by rules: it locates the centerlines of the ego and adversarial agents, identifies possible intersections, and samples initial conditions (e.g., acceleration, lateral offset) likely to trigger the intended collision along the ego’s path. Jiang et al.~\cite{P62} presents a method for generating safety-critical scenarios by combining causal inference with a fine-tuned diffusion model. It first builds a causal graph from observational data and domain knowledge to identify key environmental factors that impact object detection, such as fog, rain, or time of day. These factors are encoded and optimized using a multi-objective genetic algorithm (NSGA-II) to select prompt word combinations that balance hazard severity and test coverage, guiding a fine-tuned Stable Diffusion model to generate challenging scenes using two modes: from text prompts, or transforming existing images.

In addition to incorporating specific objectives, \textbf{contextual information} such as trajectory histories and map layouts is also used to \textbf{condition} the diffusion process, enabling the generation of more realistic and context-aware critical scenarios~\cite{P5, P50, P52}. Xie et al.~\cite{P4} proposed a framework (AdvDiffuser) that uses a guided diffusion model and a collision reward model to generate adversarial vehicle behaviors. The framework includes several key components: a context encoder that processes scenario context such as map and trajectory history, a reverse diffusion process that iteratively denoises to produce a realistic latent code representing future trajectories, and a reward model that guides this process toward generating adversarial behaviors. Finally, a trajectory decoder maps the latent code into full, explicit trajectories. Lin et al.~\cite{P50} proposed CCDiff, a diffusion-based framework that generates critical scenarios by balancing controllability (user-specified safety events) and realism (plausible agent behaviors). It uses spatial-temporal attention to capture agent interactions from the motion histories and builds a causal graph to identify which agents are most interactive and influential. During the diffusion sampling process, CCDiff focuses on these key agents while treating others as passive to maintain realism. Peng et al.~\cite{P52} proposed a latent diffusion model for generating safety-critical traffic scenarios. It uses a GNN encoder to embed past agent trajectories and map features into a latent space. Using the DDIM sampling strategy, the model denoises a latent variable conditioned on this context, and a decoder autoregressively generates future agent actions, refined by a kinematic bicycle model for realism. To guide generation, an objective function is used with three goals: ensure non-adversarial agents behave safely (no collisions and stay on road), keep adversarial agents realistic (with plausible behaviors), and induce collisions with the ego vehicle.

\par\vspace{2mm}

\textbf{(2.4) GAN-based models}, like diffusion models, can generate critical driving scenarios by \textbf{incorporating specific objectives} (e.g., collisions, style) during optimization. They are often \textbf{conditioned on context} (e.g., vehicle states, road layout) or \textbf{constrained} by rules to ensure the generated scenarios are both realistic and safety-critical~\cite{P15, P25, P76}. Yin et al.~\cite{P15} proposed a deep generative model (RouteGAN) to create diverse and critical driving scenarios for testing AV planners. The generator takes inputs such as the controlled vehicle’s initial position, ego vehicle’s state, goal location, and a bird’s-eye view of the road. A style variable (e.g., safe, aggressive, critical) controls the interaction behavior to adjust scenario criticality. The discriminator evaluates the generated scenarios based on realism, safety, and criticality. Zhao et al.~\cite{P25} proposed a model (BN-AM-SeqGAN) to generate dangerous lane-changing scenarios for testing ADS. The model includes a batch-normalized LSTM generator and a CNN-based discriminator with attention, trained with real data, to generate synthetic emergency lane-change trajectories. A collision-based constraint model is applied to compute the ego vehicle’s initial state, making collisions possible but avoidable. Similarly, Jing et al.~\cite{P76} proposed Traj-TimeGAN, a method for generating high-risk lane-changing scenarios using an improved TimeGAN trained on real emergency lane-changing data. The model includes four key components: an encoder to compress real trajectories, a decoder to reconstruct them, a generator to produce new latent trajectories from noise, and a discriminator to distinguish real from synthetic data. After generation, a safety distance constraint is applied to compute the ego vehicle’s initial position, speed, and lane, placing it just close enough to create a near-collision scenario.

\par\vspace{2mm}

\textbf{(2.5) Autoencoder-based models} are also used in several studies~\cite{P16, P54, P91} to generate critical driving scenarios. Like diffusion- and GAN-based models, the generation is either \textbf{guided by specific objectives} (e.g., adversarial goals like collisions) or \textbf{conditioned on contextual information} (e.g., maps or past trajectories). Rempe et al.~\cite{P54} introduced STRIVE, a generative method for creating realistic, accident-prone scenarios using a Conditional Variational Autoencoder (CVAE). It starts with a real-world scene and map, then perturbs the future trajectories of non-ego vehicles to cause a collision with the AV planner. The framework includes a conditional prior network, where each agent’s latent vector captures context from its past trajectory, and a decoder generates future motions conditioned on this. During adversarial optimization, past trajectories stay fixed while future ones are adjusted to match the  real planner’s predictions for realism and induce a collision. The loss function combines an adversarial term (to cause vehicle collisions), a prior term (to stay close to real distribution), and an initial term (to maintain consistency to original scene). Gao et al.~\cite{P16} also used CVAE, in which the model learns a prior distribution over the latent space from past trajectories and map context. Future trajectories are generated by sampling a latent variable and decoding it with an autoregressive GAT-based decoder. To create safety-critical cases, a DLOW-based sampler~\cite{yuan2020dlow} perturbs the latent codes in targeted ways, enabling the generation of diverse outcomes, including potential collisions. Additionally, Ding et al.~\cite{P91} proposed SAG, a semantically adversarial generative framework for creating controllable and challenging driving scenarios. It uses a Tree-Structured Variational Autoencoder (T-VAE), where encoders and decoders map structured graphs (of nodes and edges that represent scenarios) to and from a latent space. During generation, latent sampling is guided by functions representing node-level knowledge (e.g., object properties) and edge-level knowledge (e.g., relationships between objects), ensuring the generated scenarios are both adversarial and semantically meaningful.

\par\vspace{2mm}

\textbf{(2.6) Other models} are rarely used for critical scenario generation but offer unique mechanisms that \textbf{incorporate causality}. Specifically, Ding et al.~\cite{P8} proposed CausalAF, a framework for generating safety-critical traffic scenarios using a autoregressive flow model that integrates causality in the generation. Each scenario is represented as a behavior graph, where nodes are traffic participants and edges represent their interactions. The Autoregressive Flow (AF) model sequentially generates these nodes and edges, conditioned on existing ones. To ensure realism, it uses two causal masks: the Causal Order Mask (COM) enforces a meaningful generation order, and the Causal Visibility Mask (CVM) filters out irrelevant objects. Finally, reinforcement learning (REINFORCE algorithm~\cite{williams1992simple}) is applied to optimize the model toward generating collision-prone scenarios by rewarding risky configurations like small inter-vehicle distances.

\par\vspace{2mm}

\textbf{(2.7) Hybrid models} are used by two studies~\cite{P32, P94} for critical scenario generation, combining an LLM and a VLM to complete the task, using \textbf{prompt engineering}~\cite{sahoo2024systematic} techniques and \textbf{iterative refinement} strategy. Specifically, Li et al.~\cite{P32} first use an LLM to generate a behavior tree -- a logical scenario consisting of nodes that represent behaviors, control flows, and conditions of a scenario -- based on structured prompts (including e.g., node descriptions, function scenario description, targeted XML schema, and its thought process). Then, the VLM serves as an optimizer to fill in concrete values for each action node in the generated behavior tree, aiming to create scenarios that lead to accidents. The optimization is guided by a startup prompt (including e.g., task description, logical scenario, parameter description, expected output format, and feedback format) where the LLM provides initial parameter values. The VLM receives feedback such as key frames (images) and evaluation results to iteratively generate updated parameters until a crash occurs or a trial limit is reached. Additionally, Peng et al.~\cite{P94} proposed a framework (LD-Scene) that combines LDMs with LLM-based guidance to generate user-controlled adversarial driving scenarios from natural language. The LDM learns realistic traffic behaviors by encoding past and future trajectories into a compact latent space, applying a denoising diffusion process, and decoding them back into future trajectories of non-ego vehicles. To make scenarios adversarial, an LLM-based module turns user queries into a guidance loss function: a code generator with Chain-of-Thought reasoning determining adversarial levels and loss weights, while a code debugger tests and fixes errors iteratively to ensure valid implementation. This guidance loss steers the denoising process so that the adversarial vehicle collides with the ego vehicle as specified, while keeping other vehicles realistic.

\par\vspace{2mm}

\begin{table}[hbp]
\centering
\small
\caption{RQ1 -- Studies using Generative AI for scenario transformation.}
\begin{tabular}{|c|l|l|l|}
    \hline 
    \textbf{Test objective} & \textbf{Generative model} & \textbf{Highlighted mechanisms} &  \textbf{Studies} \\
    \hline
    \multirow{3}{*}{Scenario transformation} & Diffusion-based & Domain transformation, image-to-image translation & \cite{P23}, \cite{P19}, \cite{P73} \\
    \cline{2-4} & GAN-based & Domain transformation, image-to-image translation & \cite{P71}, \cite{P73}, \cite{P31} \\
    \cline{2-4} & Hybrid & Domain transformation, image-to-image translation & \cite{P7}, \cite{P55}, \cite{P70} \\
    \hline
\end{tabular}
\label{table:results:RQ1:models:testing:scenario_transformation}
\end{table}

\noindent\textbf{(3) Scenario transformation.} Several studies use generative models, including Diffusion-based, GAN-based, and Hybrid models, for scenario transformation through image-to-image translation, as shown in Table~\ref{table:results:RQ1:models:testing:scenario_transformation}. This involves \emph{converting an existing scenario from one domain to another (e.g., weather condition, object attributes, environment, synthetic-to-real, or viewpoint changes) while preserving its semantic structure}. While there may be some overlap with scenario augmentation, in our study, transformation refers to changes that preserve the original scenario’s semantic structure, whereas augmentation involves modifications that may add or remove elements within the scenario -- though both of them aim to generate new and diverse scenarios by introducing changes to existing ones.

\par\vspace{2mm}

\textbf{(3.1) Diffusion-based models} are applied for \textbf{domain transformation} through \textbf{image-to-image translation}~\cite{P23, P19, P73}. Xu et al.~\cite{P23} presents a framework that uses a fine-tuned latent diffusion model to create street-view images from Bird’s-Eye View (BEV) inputs. It first converts the BEV map into semantic street-view maps, then applies a Shape Refinement Network to enhance accuracy and resolution. These refined maps, along with a text prompt describing the scene, are used as conditions for the diffusion model to generate realistic street-view images that align with the correct camera viewpoints and styles described. Baresi et al.~\cite{P19} proposed an approach for editing driving scenario images to create new ones using diffusion models. It explored three editing methods using images from car cameras: instruction-editing, which modifies the image based on a textual instruction; inpainting, which combines an editing instruction with a mask to preserve the road; inpainting with refinement, which enhances visual consistency using an edge map. The method then evaluates the validity and semantic consistency between the original and modified images, discarding those below a certain threshold. To improve efficiency, a smaller model based on CycleGAN is trained to mimic the larger and complex model for translating images between different domains. Zhao et al.~\cite{P73} explored three image-to-image translation models, including two GAN-based models (Pix2pixHD~\cite{Wang_2018_CVPR} and OASIS~\cite{1360017285535413888}) and a diffusion-based model (ControlNet~\cite{Zhang_2023_ICCV}) to generate photo-realistic driving scenarios from semantic label maps of real or synthetic scenes, reproducing semantically consistent scenarios while reducing the simulation-to-reality gap.

\par\vspace{2mm}

\textbf{(3.2) GAN-based models} are, like other generative models used for scenario transformation, also applied by several studies~\cite{P71, P73, P31} for \textbf{domain transformation} through \textbf{image-to-image translation}. Specifically, Pan et al.~\cite{P71} proposed a framework to convert driving scenarios into foggy conditions with varying fog levels and directions for metamorphic testing of ADS. The metamorphic relation assumes that fog density and direction should not impact the ADS's steering angle. The framework takes a clear driving scene, applies a generator to synthesize a foggy version, and then verifies its realism with a validator. Two generators are used: OpenCV~\cite{culjak2012brief} and foggy-CycleGAN~\cite{foggy-cyclegan} -- a CycleGAN-based model that either adds fog to a clear image based on a specified fog rate or removes fog from a foggy image. Yu and Li~\cite{P31} developed a CycleGAN-based model to generate synthetic corner-case images for testing ADS. The approach transforms nominal images (e.g., scenarios recorded in normal environmental conditions) into corner-case images (e.g., with extreme conditions like high temperature). It uses a forward generator with a ResNet-based encoder to modify image features and a decoder to reconstruct the corner-case image. A discriminator judges whether the generated images with transformed conditions are real or fake. To avoid model collapse, a backward generator maps the corner-case image back to its original form, with a cycle-consistency loss ensuring the transformation is reversible.

\par\vspace{2mm}

\textbf{(3.3) Hybrid models} are used in several studies for scenario transformation, combining different generative models to perform \textbf{domain transformation} of existing scenarios via \textbf{image-to-image translation}~\cite{P7, P55, P70}. Specifically, Amini and Nejati~\cite{P7} explored three image-to-image translation methods -- CycleGAN~\cite{zhu2017unpaired}, Neural Style Transfer~\cite{gatys2016image}, and a proposed model called SAEVAE -- for translating between real and synthetic driving images. SAEVAE combines a Sparse Autoencoder (SAE)~\cite{le2013building} to reconstruct synthetic image features with a Variational Autoencoder (VAE)~\cite{kingma2022autoencodingvariationalbayes} to evaluate how closely translated images resemble real-world data. The study applied a comprehensive set of metrics to assess whether these translations help reduce the gap between training (real) and testing (synthetic) data, ultimately enhancing the robustness and realism of ADS testing. Zhang et al.~\cite{P55} proposed a metamorphic testing approach called DeepRoad, which uses a hybrid generative model (UNIT)~\cite{liu2017unsupervised} combining GAN and VAE. UNIT maps images from different weather domains into a shared latent space using a VAE, and transforms them between domains (e.g., sunny to rainy or snowy) using a GAN-based generator. The transformed images are then used in the DeepRoadMT framework to test whether ADS produce consistent steering predictions across weather conditions. Baresi et al.~\cite{P70} proposed a framework (DILLEMA) that combines a diffusion model and an LLM to transform given image senarios. The process starts by converting an input image into a text caption and identifying key elements of the image. An LLM then analyzes the caption to find keywords (e.g., weather, object color) that can be changed without affecting the semantics for intend tasks (e.g., semantic segmentation). It generates alternative keywords and updates the caption. Finally, a control-conditioned text-to-image diffusion model generates a new image based on the modified caption.

\par\vspace{2mm}

\begin{table}[tbp]
\centering
\small
\caption{RQ1 -- Studies using Generative AI for scenario augmentation.}
\begin{tabular}{|c|l|l|l|}
    \hline 
    \textbf{Test objective} & \textbf{Generative model} & \textbf{Highlighted mechanisms} &  \textbf{Studies} \\
    \hline
    \multirow{4}{*}{Scenario augmentation} & LLM & Prompt engineering, multi-agent collaboration & \cite{P74}, \cite{P75} \\
    \cline{2-4} & VLM & Prompt engineering & \cite{P77} \\
    \cline{2-4} & Diffusion-based & Object insertion, location conditioned & \cite{P12}, \cite{P20}, \cite{P30} \\
    \cline{2-4} & AE-based & Trajectory fusion & \cite{P11} \\
    \hline
\end{tabular}
\label{table:results:RQ1:models:testing:scenario_agumentation}
\end{table}

\noindent\textbf{(4) Scenario augmentation.} A few studies we reviewed applied generative AI for scenario augmentation by \emph{modifying existing scenarios to create new ones} for testing ADS. Unlike scenario transformation, augmentation introduces or edits elements in a way that can change the semantic structure of the scenario in our study. For example, turning an open-road driving scenario into a pedestrian-crossing one by adding a pedestrian to it. While one study used a VLM to add annotations to existing scenarios, enhancing their utility in training and testing ADS, most studies augment datasets by modifying existing scenarios through natural language instructions using LLMs or by inserting new elements (e.g., pedestrians, animals, or static objects) at specific locations using diffusion models, as shown in Table~\ref{table:results:RQ1:models:testing:scenario_agumentation}.

\par\vspace{2mm}

\textbf{(4.1) LLMs} are used to modify existing scenarios using \textbf{prompt engineering}~\cite{sahoo2024systematic} techniques, enabling effective scenario augmentation for ADS testing~\cite{P74, P75}. For instance, Li et al.~\cite{P74} introduced a framework (SeGPT) that integrates an LLM for scenario generation. Real-world driving scenarios are first preprocessed and stored in a library as a mathematical representation. The LLM then generates new scenarios based on these (e.g., modifying the vehicle trajectories, increasing the vehicle speed) using a carefully designed prompt, which includes a role setting, task description, reference text explaining scenario parameters, subtasks for analyzing input scenarios, a reasoning process with different generation strategies, and detailed generation instructions. Wei et al.~\cite{P75} presents a \textbf{multi-agent collaborative} system (ChatSim) for editing and generating photo-realistic driving scenes using multiple specialized LLM agents. Each agent has a distinct role with tailored prompts. A project manager agent orchestrates the process by breaking down user commands and assigning tasks to relevant agents. These include: a view adjustment agent for camera angle settings, a background rendering agent to generate the scene background, a vehicle deletion agent to remove specified vehicles, a 3D asset manager to select and customize digital assets (e.g., cars), a vehicle motion agent to create vehicle positions and movements, and a foreground rendering agent to integrate camera extrinsic information, 3D asset, and motion data. The foreground and background are then combined into a final video output for the user.

\par\vspace{2mm}

\textbf{(4.2) VLMs} are used in one study~\cite{P77} to augment the CODA~\cite{li2022coda} corner case dataset with textual pre-annotations, forming CODA-LM, for testing VLMs on three vision-language tasks for autonomous driving: general perception (understanding critical road entities in the driving scenarios), regional perception (recognizing corner-case objects), and driving suggestion (providing driving advice based on the driving scenarios). Using \textbf{prompt engineering}~\cite{sahoo2024systematic}, the VLM is guided to identify objects in each category, describe and explain them, and then convert the structured outputs into natural language as pre-annotations for general perception. For regional perception, visual prompts with either bounding boxes of corner case objects on the original image, or normalized coordinates are used. Then, the VLM takes the outputs from the previous two tasks and the original image to generate driving advices, serving as pre-annotations for driving suggestion. Finally, all pre-annotations are verified and refined by human annotators.

\par\vspace{2mm}

\textbf{(4.3) Diffusion-based models} are applied in several studies to augment scenarios by \textbf{inserting objects} at specific \textbf{locations of interest}~\cite{P12, P20, P30}. Guo et al.~\cite{P12} developed an approach to generate safety-critical scenarios in image form by inserting critical objects (e.g., cars, pedestrians) into real driving images at specified locations, guided by a location mask and text prompt, using a Latent Diffusion Model. A Mask-aware Adapter processes the masked image to produce clean feature maps, allowing natural blending of new objects. Spatial coordinates (bounding boxes) and object categories are encoded as embeddings through a Grounding Input Embedding module and refined by a Transformer encoder. A Region-Guided Cross Attention mechanism precisely controls what (objects) and where (locations) to generate during the diffusion process. Finally, A pre-trained diffusion model performs the actual denoising process, conditioned on three key inputs: cleaned feature maps from the Mask-aware Adapter, enhanced object embeddings from the Transformer encoder, and attention maps from the Region-Guided Cross Attention. Gannamaneni et al.~\cite{P20} proposed a diffusion-based pipeline to augment driving scenes by inserting pedestrians and creating safety-critical scenarios based on textual prompts (e.g., specifying pedestrian attributes like gender, skin color, and pose). The pipeline first selects a region of interest (a rectangular mask) using the LANG-SAM~\cite{langsam} model, then uses a latent diffusion model to inpaint the pedestrian into the image, guided by both the prompt and the selected region. To ensure realistic human poses, ControlNet~\cite{Zhang_2023_ICCV} with OpenPose~\cite{cao2019openpose} is integrated. Finally, a self-consistency check with CLIP~\cite{radford2021learning} is used to filter results -- only images that match the prompt are retained for the augmented dataset. Similarly, Yigit and Can~\cite{P30} presents an approach to create corner-case scenarios in infrared images by inserting objects (i.e., deer, dog, traffic cone, and trash bin) into background images. A diffusion model is guided with text prompts and masked regions indicating where each object should appear, allowing controlled placement. The final image is generated with the inserted objects, creating realistic and challenging infrared driving scenarios for testing.

\par\vspace{2mm}

\textbf{(4.4) Autoencoder-based models} are used in a study~\cite{P11} to \textbf{fuse trajectories} and synthesize safety-critical driving scenarios. Specifically, Ding et al.~\cite{P11} proposed CMTS, a generative framework for synthesizing safety-critical driving scenarios by blending real safe trajectories with artificial collision trajectories. Built on a Variational Autoencoder (VAE), CMTS includes three components: a GRU-based encoder for encoding safe and collision trajectories into latent space, a convolutional condition encoder that extracts road map features as style information, and a GRU-based decoder that reconstructs trajectories using both the latent code and map constraints. It uses style transfer (AdaIN~\cite{huang2017arbitrary}) to separate map features from driving behavior and applies latent space interpolation between safe and crash trajectories to synthesize near-collision scenarios. These synthesized scenarios consist of high-dimensional trajectory data enriched with road context, and can be used to augment datasets for improving AV trajectory prediction in risky situations.

\begin{table}[hbp]
\centering
\small
\caption{RQ1 -- Studies using Generative AI for scenario reconstruction.}
\begin{tabular}{|c|l|l|l|}
    \hline 
    \textbf{Test objective} & \textbf{Generative model} & \textbf{Highlighted mechanisms} &  \textbf{Studies} \\
    \hline
    \multirow{2}{*}{Scenario reconstruction} & LLM & Prompt engineering, accident reports & \cite{P22}, \cite{P33}, \cite{P45} \\
    \cline{2-4} & VLM & Prompt engineering, driving videos & \cite{P24} \\
    \hline
\end{tabular}
\label{table:results:RQ1:models:testing:scenario_reconstruction}
\end{table}

\noindent\textbf{(5) Scenario reconstruction.} Although relatively few in number, several generative models, including LLMs and VLMs, are used \emph{to reconstruct driving scenarios from sources like accident reports and driving videos for testing ADS} using natural language instructions with advanced prompt engineering techniques, as shown in Table~\ref{table:results:RQ1:models:testing:scenario_reconstruction}. Unlike LLMs, VLMs can process visual information, allowing users to include images or videos in the input.

\par\vspace{2mm}

\textbf{(5.1) LLMs} are used in several studies to reconstruct scenarios from \textbf{accident reports}, guided by natural language instructions and advanced \textbf{prompt engineering}~\cite{P22, P33, P45}. Specifically, Luo et al.~\cite{P22} developed an LLM-driven framework (TRACE) to reconstruct driving scenarios from multimodal accident reports (including summary descriptions and sketches). In the first stage, the framework uses an LLM to identify the road type involved and selects a matching prompt tailored to that geometry. A TrackMate module is developed to vehicle trajectories from crash sketches. The LLM then extracts information on the road network, environment, actors, and trajectories from the report, which is transformed into a standardized scenario representation via a pattern-matching algorithm. Prompts used are carefully crafted, incorporating role setting, task descriptions, chain-of-thought reasoning, demonstration examples, and self-validation. In the second stage, a scenario generation adapter converts this representation into executable code in simulation. Tang et al.~\cite{P33} presents LeGEND, an LLM-assisted approach for generating test scenarios from accident reports. It first uses an LLM to extract information and produce an intermediate representation, which outlines initial actions and interactions between entities. This is done using a prompt that includes the task description, output format, and adherence requirements, with the LLM repeating the process until the output is correct. A second LLM then converts this intermediate representation into a logical test scenario, guided by a task-specific prompt and examples. Finally, a multi-objective genetic algorithm searches for diverse and concrete critical scenarios based on the logical scenario. Guo et al.~\cite{P45} developed SoVAR, a tool that reconstructs scenarios from accident reports. SoVAR uses an LLM to extract key elements, such as the environmental conditions, road network, and dynamic objects, from the report's descriptive text. Using the extracted driving actions and road information, SoVAR computes trajectory waypoints, which, along with environmental conditions, are used to simulate and reconstruct the crash scenario detailed in the original report.

\par\vspace{2mm}

\textbf{(5.2) VLMs} are used in one selected study~\cite{P24} to reconstruct scenarios from real-world \textbf{driving videos} with carefully designed prompts. The proposed VLM-enabled framework converts natural driving videos, e.g., dashcam recordings, into Scenic~\cite{fremont2019scenic} code for simulation in Carla~\cite{dosovitskiy2017carla}. First, a VLM generates a Scenic script from the video using prompts with example pairs of simulation videos and corresponding scripts. This script is simulated in Carla~\cite{dosovitskiy2017carla}, and the resulting video is compared to the original one using a similarity check based on key scenario features extracted by another VLM. If differences exceed a set threshold, the script is refined using the discrepancies and the process is repeated until the difference falls below the threshold.

\par\vspace{2mm}

\begin{table}[hbp]
\centering
\small
\caption{RQ1 -- Studies using Generative AI for scenario understanding.}
\begin{tabular}{|c|l|l|l|}
    \hline 
    \textbf{Test objective} & \textbf{Generative model} & \textbf{Highlighted mechanisms} &  \textbf{Studies} \\
    \hline
    Scenario understanding & LLM & Prompt engineering & \cite{P9}, \cite{P79} \\
    \hline
\end{tabular}
\label{table:results:RQ1:models:testing:scenario_understanding}
\end{table}

\noindent\textbf{(6) Scenario understanding.} Utilizing generative AI for scenario understanding, which aims \emph{to interpret a scenario by providing or deriving key characteristics of interest}, has been reported to a very limited extent. Only two studies employed LLMs for this purpose, for example, to extract classification information or analyze the criticality of a given driving scenario (see Table~\ref{table:results:RQ1:models:testing:scenario_understanding}), facilitating understanding of the scenario from specific lens.

\par\vspace{2mm}

\textbf{(6.1) LLMs} are the only generative models reported for scenario understanding, used in two studies~\cite{P9, P79} with \textbf{prompt engineering} techniques~\cite{sahoo2024systematic}. Specifically, Zhao et al.~\cite{P9} proposed a framework (Chat2Scenario) to generate structured scenario classification information (e.g., vehicle maneuvers) from naturalistic driving data using structured prompts consisting of five segments: a scenario classification model, scenario description, intended output structure, template, and an example pairing of description and classification. The resulting classification is used for critical analysis and retrieving relevant scenarios from naturalistic data. Gao et al.~\cite{P79} proposed a framework to assess the safety criticality of ADS scenarios and provide feedback for transforming non-critical scenarios into safety-critical ones. Real-world datasets in CommonRoad format~\cite{althoff2017commonroad} are first converted into a structured 6-layer~\cite{scholtes20216} JSON representation (road, traffic signs, temporary changes, participants, environment, and digital map). A specially designed prompt is then used to guide an LLM in evaluating scenario criticality, incorporating role setting, task description, risk level definitions, evaluation metrics (e.g., TTC and MDC), and a reasoning process. The prompt also includes detailed context, such as Cartesian and Frenet coordinates, and additional safety-critical metrics like Distance and Temporal Safety Scores.

\par\vspace{2mm}

\faHandPointRight \ Overall, \emph{a wide range of generative AI models have been applied to support scenario-based testing of ADS}, including generating new and challenging scenarios, reconstructing scenarios from existing resources, transforming or augmenting scenarios, and understanding scenario content or characteristics. Among these, scenario generation, including critical scenario generation, receives the most attention (70 of 91, or 77\% studies), reflecting \emph{a strong need for diverse and relevant test scenarios to assess ADS} performance and safety. Mechanisms like prompt engineering (for LLMs and VLMs), goal- or context-conditioned generation (for diffusion, GAN-based, and AE-based models), and hybrid architectures that combine complementary models \emph{represent a shift from, and an extension for, traditional scenario generation methods} such as knowledge-based, data-driven, and search-based approaches~\cite{riedmaier2020survey, song2024empirically}. These generative methods have demonstrated promising results, positioning \emph{generative AI as a valuable direction for advancing ADS testing and warranting further research}.  

\subsection{RQ2 -- How Effective is Generative AI for Testing ADS}
\label{sec:results:RQ2}

Most of the selected studies evaluated their proposed generative approaches against baseline methods (e.g., TrafficGen~\cite{P47}, LCTGen~\cite{P83}, STRIVE~\cite{P54}, AdvSim~\cite{wang2021advsim}, L2C~\cite{ding2020learning}, BITS~\cite{xu2022bits}) using various metrics such as realism, diversity, and criticality, demonstrating their effectiveness in producing useful scenarios for testing ADS. Several studies also included human evaluation or visual inspection to assess scenario quality. As nearly all studies reported improved performance over baselines, we focus our analysis on the datasets, simulators, ADS systems, evaluation metrics, and baseline methods used for evaluation, rather than the performance results themselves. 

\begin{table}[hbp]
\centering
\small
\caption{RQ2 -- Datasets used for evaluation in selected studies.}
\begin{tabular}{|c|l|l|}
    \hline 
    \textbf{Evaluation setup} & \textbf{Instances} & \textbf{Studies} \\
    \hline
    \multirow{17}{*}{Dataset} & Waymo Open Motion Dataset & \cite{P18}, \cite{P43}, \cite{P57}, \cite{P58}, \cite{P69}, \cite{P81}, \cite{P82}, \cite{P84}, \cite{P86}, \cite{P87}, \cite{P90} \\
    \cline{2-3} & Waymo Open Dataset & \cite{P47}, \cite{P63}, \cite{P75}, \cite{P83}, \cite{P88} \\
    \cline{2-3} & nuScenes Dataset & \cite{P4}, \cite{P16}, \cite{P23}, \cite{P27}, \cite{P38}, \cite{P50}, \cite{P52}, \cite{P53}, \cite{P54}, \cite{P60}, \cite{P64}, \cite{P94} \\
    \cline{2-3} & highD Dataset & \cite{P9}, \cite{P13}, \cite{P35}, \cite{P40}, \cite{P49}, \cite{P76} \\
    \cline{2-3} & nuPlan Dataset & \cite{P64}, \cite{P68}, \cite{P69}, \cite{P86}, \cite{P90} \\
    \cline{2-3} & Argoverse Dataset & \cite{P15}, \cite{P44}, \cite{P67} \\
    \cline{2-3} & Argoverse 2 Dataset & \cite{P17}, \cite{P41}, \cite{P42} \\
    \cline{2-3} & NHTSA Crash Reports & \cite{P39}, \cite{P45}, \cite{P93}\\
    \cline{2-3} & Udacity self-driving car & \cite{P55}, \cite{P71} \\
    \cline{2-3} & Cityscapes Dataset & \cite{P20}, \cite{P73} \\
    \cline{2-3} & INTERACTION Dataset & \cite{P15}, \cite{P74} \\
    \cline{2-3} & \multicolumn{2}{c|}{NTHSA CIREN Database (\cite{P22}), NHTSA Crashworthiness Data System (CDS) Dataset (\cite{P32})} \\
    & \multicolumn{2}{c|}{NHTSA Pre-crash scenario descriptions (\cite{P46}), KITTI (\cite{P7}), Udacity Jungle Dataset (\cite{P7}), Lines (\cite{P11})} \\
    & \multicolumn{2}{c|}{MNIST (\cite{P11}), USPS (\cite{P11}), CityPersons (\cite{P12}), BDD100K (\cite{P12}), OpenStreetMap (\cite{P14}), SHIFT (\cite{P70})} \\
    & \multicolumn{2}{c|}{US Highway 101 Dataset (\cite{P25}), Honda Research Institute Driving Dataset (HDD) (\cite{P36})} \\
    & \multicolumn{2}{c|}{Car Crash Dataset (CCD) (\cite{P24}), ATG4D (\cite{P44}), C-NCAP (\cite{P46}), rounD (\cite{P51}), SPMD (\cite{P65}), UMTRI (\cite{P66})} \\
    & \multicolumn{2}{c|}{ImageNet1K (\cite{P70}), DAWN (\cite{P78}), INTERDRIVE (\cite{P82}), ProSim-Instruct-520k (\cite{P82}), Lyft (\cite{P89})} \\
    \hline
\end{tabular}
\label{table:results:RQ2:datasets}
\end{table}

\subsubsection{Datasets}
\label{sec:results:RQ2:datasets}

A number of public datasets are used for evaluation, with a few dominating the field -- most notably Waymo open data~\cite{waymoopen, ettinger2021large, chen2024womd}, nuScenes~\cite{caesar2020nuscenes}, highD~\cite{krajewski2018highd}, nuPlan~\cite{karnchanachari2024towards}, Agroverse~\cite{chang2019argoverse, wilson2023argoverse}, Udacity~\cite{udacity}, Cityscape~\cite{cordts2016cityscapes}, INTERACTION~\cite{zhan2019interaction}, and NTHSA Crash Reports~\cite{NHTSA}. Although not all datasets are strictly open-source -- often requiring free registration, acceptance of terms, and limiting use to non-commercial purposes -- they are all accessible for academic use. These datasets served various purposes, including \emph{model training and testing, providing ground-truth data for evaluating generated scenarios} (e.g., measuring discrepancies), \emph{assessing realism and diversity} (e.g., via distance or distribution metrics), and \emph{acting as benchmarks to test the effectiveness of scenarios in revealing ADS failures} (e.g., collisions) \emph{or improving their performance} (e.g., robustness, detection accuracy). Due to the large number of datasets used across studies (see Table~\ref{table:results:RQ2:datasets}), we focus on the most commonly used ones and briefly illustrate how they are used in evaluation. It is also worth to highlight that not all selected studies used existing datasets for experimentation and evaluation. Several studies relied on customized datasets that were specifically collected or curated for their research, while others used datasets without clearly specifying their names and sources.

\begin{enumerate}
    \item \textbf{Waymo Open Dataset} is a large-scale public dataset for autonomous driving~\cite{waymoopen}, comprising multiple sub-datasets: a motion dataset with object trajectories and corresponding 3D maps, a perception dataset with high-resolution sensor data and annotations, and an end-to-end driving dataset with camera feeds and high-level commands. It was used in several studies~\cite{P47, P63, P83, P88} to train and evaluate generative models, assess the quality of generated scenarios against ground-truth using relevant metrics (detailed in Section~\ref{sec:results:RQ2:metrics}), and compare performance with baseline methods. Furthermore, generated scenarios have been used to enhance ADS components, e.g., an RL-based ADS controller~\cite{P47} and a vehicle detector~\cite{P63}. 

    \vspace{1mm}
    
    \item Substantial studies~\cite{P18, P43, P69, P81, P82, P84, P86, P87, P90} specifically used the \textbf{Waymo Open Motion Dataset}~\cite{ettinger2021large, chen2024womd} to train and evaluate their generative models, and to compare produced scenarios against ground truth to evaluate their realism. In addition, Mei et al.~\cite{P57} used scenarios from this dataset to identify risky behaviors of background vehicles and to generate corresponding safety-critical scenarios. In Mei et al.~\cite{P58}, selected scenarios served as evaluation cases for a method where a LLM-based identifier detects adversarial vehicles exhibiting risky behavior toward AVs. A pretrained model then predicts the trajectories of these adversarial agents, thereby generating adversarial driving scenarios.

    \vspace{1mm}
    
    \item The \textbf{nuScenes} dataset is a large-scale public dataset for autonomous driving, developed by the team at Motional (formerly nuTonomy), with driving scenes collected in Boston and Singapore~\cite{caesar2020nuscenes}. It is claimed the first large-scale dataset to include data from the full sensor suite of an autonomous vehicle, comprising six cameras, one LiDAR, five radars, GPS, and an IMU~\cite{caesar2020nuscenes}. Specifically, scenarios from nuScenes data were used as input for an LLM to perform scenario augmentation in Wei et al.~\cite{P75}, where generated scenarios help augment data for object detection tasks. Besides, nuScenes is commonly used to train and evaluate generative models for scenario generation~\cite{P4, P16, P38, P50, P52, P53, P54, P60, P64, P93}, e.g., predicting future trajectories based on past ones, with results compared against ground-truth trajectories and selected baselines. In Rempe et al.~\cite{P54}, the generated critical scenarios were also used to augment the original nuScenes dataset, enabling the fine-tuning of a rule-based planner to better handle challenging situations. Furthermore, in Aasi et al.~\cite{P27}, nuScenes data served as a baseline for evaluating the OOD-ness (Out-of-Distribution) and rarity of generated scenarios.

    \vspace{1mm}

    \item The \textbf{highD} dataset is a collection of naturalistic vehicle trajectories recorded on German highways at six different locations using drone footage, and was initially developed to support the safety validation of ADS/AVs~\cite{krajewski2018highd}. The dataset served as the training data for a model that generates new lane-changing (cut-in) scenarios, and was used for evaluation purposes, where real-world cut-in scenarios extracted from the highD dataset were used as references to assess the quality of the generated scenarios in Fang et al.~\cite{P13}. Similarly, lane-changing scenarios from the highD dataset were employed in several other studies~\cite{P40, P49, P76} for training and evaluating proposed models, serving as ground truth for comparison against various baseline methods. In Chang et al.~\cite{P35}, a collection of naturalistic risky scenario segments from the highD dataset was selected as prompts for an LLM, which then generates hundreds of new scenarios intended to match the complexity and challenge of the original scenarios. In Zhao et al.~\cite{P9}, The highD dataset was used for evaluating and demonstrating proposed method for scenario understanding (i.e., scenario classification information extraction).

    \vspace{1mm}

    \item \textbf{nuPlan} is the world’s first large-scale planning benchmark for autonomous driving~\cite{karnchanachari2024towards}. The dataset comprises extensive driving data collected from four cities: Boston, Pittsburgh, Las Vegas, and Singapore. After collection, the data is enriched with detailed metadata, including semantic maps, object annotations, traffic light states, and observed scenario types~\cite{karnchanachari2024towards}. nuPlan data was used for training and evaluating the proposed models in several studies~\cite{P68, P69, P86, P90}, where the generated scenarios were assessed against established benchmarks based on specific quality attributes, such as realism, which we introduce in Section~\ref{sec:results:RQ2:metrics}.

    \vspace{1mm}

    \item \textbf{Argoverse} consists of publicly available datasets designed to support perception and prediction research in autonomous driving, complemented by detailed high-definition (HD) maps. The Argoverse 1~\cite{chang2019argoverse} release includes two main datasets: a 3D Tracking Dataset with fully annotated 3D scenes, and a Motion Forecasting Dataset containing driving scenarios. Argoverse 2~\cite{wilson2023argoverse} builds upon the original release, offering expanded data from six U.S. cities: Austin, Detroit, Miami, Pittsburgh, Palo Alto, and Washington, D.C. It includes: Argoverse 2 Sensor Dataset, which provides 3D-annotated scenarios with LiDAR, stereo imagery, and ring camera data, and Argoverse 2 Motion Forecasting Dataset, containing trajectory data across various object types. In Yin et al.~\cite{P15}, interaction data from straight road and intersection scenarios in the Argoverse dataset was used to evaluate a generative model designed to produce scenarios with critical interactions. In Tan et al.~\cite{P44}, the dataset's HD maps were used to generate new scenarios, which were then compared to real-world scenarios using various metrics to assess alignment with real data. In Mi et al.~\cite{P67}, Argoverse data was employed to train and evaluate a model for HD map generation, with the fidelity of the generated maps assessed based on topology, geometry, and urban planning attributes. Argoverse 2, specifically the Motion Forecasting~\cite{P17, P41} and Sensor~\cite{P42} datasets, were also used to train and test proposed generative models and baseline approaches, enabling comparisons of their effectiveness in scenario generation. Notably, in Zhang et al.~\cite{P17}, Argoverse 2 data was used to train a driving policy, which was subsequently fine-tuned using generated scenarios to evaluate performance improvements.

    \vspace{1mm}

    \item\textbf{NHTSA Crash Reports}, published by the National Highway Traffic Safety Administration (NHTSA), document detailed information about motor vehicle crashes in the United States~\cite{NHTSA}. These reports were utilized in two studies~\cite{P39, P45} to reconstruct real-world crash scenarios for experimental purposes and to demonstrate the effectiveness of the proposed approaches for scenario reconstruction. In Li et al.~\cite{P93}, scenario elements are extracted from accident report sketches and used as the basis for generating crash scenarios. 

    \vspace{1mm}

    \item The \textbf{Udacity self-driving car} dataset is a publicly available collection that includes multiple driving datasets, featuring unlabeled LiDAR data, camera frames, and more, captured in locations such as El Camino, the Udacity Office, and other areas~\cite{udacity}. It also includes annotated driving data collected in Mountain View, California, and neighboring cities under daylight conditions. In Zhang et al.~\cite{P55}, the Udacity dataset was used to train an image transformation model and evaluate ADS on both the original and transformed data to assess model consistency. Similarly, Pan et al.~\cite{P71} also employed this dataset to evaluate ADS model consistency under the original and transformed scenarios, further demonstrating the effectiveness of its approach for testing ADS.

    \vspace{1mm}

    \item\textbf{Cityscapes} is a large-scale dataset featuring a diverse collection of stereo video sequences captured in urban street scenes across 50 different cities. It includes high-quality, pixel-level semantic annotations, making it a valuable resource for scene understanding in urban driving environments~\cite{cordts2016cityscapes}. Like many similar datasets, Cityscapes is freely available for academic and non-commercial use, including research, teaching, scientific publications, and personal experimentation. In Gannamaneni et al.~\cite{P20}, several pre-trained diffusion models are used to inpaint pedestrians with varying attributes into existing Cityscapes scenarios, effectively augmenting the dataset and demonstrating the feasibility of the proposed approach. Similarly, Zhao et al.~\cite{P73} trained and evaluated multiple generative models to synthesize realistic images from Cityscapes semantic label maps, exploring how generative AI can help bridge the Sim2Real domain gap in autonomous driving.

    \vspace{1mm}

    \item\textbf{INTERACTION} is a publicly available dataset containing multi-agent interactive traffic scenarios from diverse geographical locations, including intersections, roundabouts, and merges across several countries such as the United States, Germany, China, and others~\cite{zhan2019interaction}. In Yin et al.~\cite{P15}, roundabout scenarios from the INTERACTION dataset were used to evaluate and demonstrate RouteGAN, a generative model designed to produce scenarios involving critical interactions. Additionally, in Li et al.~\cite{P74}, the dataset was used in conjunction with an LLM to generate safety-critical synthetic scenarios using carefully crafted prompts. Both the original INTERACTION scenarios and the generated ones were then used to train and test various ADS, demonstrating the utility of such synthetic scenarios in enhancing ADS testing.

\end{enumerate}

\faHandPointRight \ Overall, a variety of datasets were used for evaluation across the selected studies, with a few -- such as the Waymo Open Dataset -- being used more frequently and \emph{they are all available for academic research}. Rather than using entire datasets, studies often \emph{focus on specific subsets or scenario types (e.g., lane-changing, roundabout) that align with their research goals}. These selected scenarios are then used to \emph{evaluate the effectiveness of proposed generative models or the quality of generated scenarios for ADS testing}, with both qualitative assessments and quantitative measurements. 

\subsubsection{Simulators}
\label{sec:results:RQ2:simulators}

\begin{table}[hbp]
\centering
\small
\caption{RQ2 -- Simulators used for evaluation in selected studies. Simulators shown in bold (i.e., VTD, IPG CarMaker, BeamNG, and Unity Simulation Platform) are commercial, while the others are open-source.}
\begin{tabular}{|c|l|l|}
    \hline 
    \textbf{Evaluation setup} & \textbf{Instances} & \textbf{Studies} \\
    \hline
    \multirow{7}{*}{Simulator} & Carla & \cite{P10}, \cite{P14}, \cite{P19}, \cite{P21}, \cite{P24}, \cite{P27}, \cite{P32}, \cite{P34}, \cite{P39}, \cite{P46}, \cite{P56}, \cite{P61}, \cite{P62}, \cite{P91} \\
    \cline{2-3} & LGSVL & \cite{P33}, \cite{P36}, \cite{P45} \\
    \cline{2-3} & MetaDrive & \cite{P22}, \cite{P47}, \cite{P58} \\
    \cline{2-3} & \textbf{VTD} & \cite{P14}, \cite{P59}\\
    \cline{2-3} & \multicolumn{2}{c|}{\textbf{IPG CarMaker} (\cite{P9}), \textbf{BeamNG} (\cite{P22}), \textbf{Unity Simulation Platform} (\cite{P25})} \\
     & \multicolumn{2}{c|}{esmini (\cite{P9}), Udacity Simulator (\cite{P19}), ACAS Xu (\cite{P21}), CommonRoad (\cite{P79})} \\
     & \multicolumn{2}{c|}{BipedalWalker (\cite{P21}), CoopNavi (\cite{P21}), tbsim Simulator (\cite{P50}), SLEDGE (\cite{P68})} \\
     & \multicolumn{2}{c|}{Highway Environment Simulator (highway-env) (\cite{P85}), GUAM Simulator (\cite{P61})} \\
    \hline
\end{tabular}
\label{table:results:RQ2:simulators}
\end{table}

Several simulators (see Table~\ref{table:results:RQ2:simulators}) are used for evaluation purposes across the selected studies, including both open-source and commercial platforms. However, not every individual study employed a simulator (also referred to as a simulation platform) for experimentation or evaluation. Among the papers that explicitly mention the simulators used, Carla~\cite{dosovitskiy2017carla}, LGSVL~\cite{rong2020lgsvl}, MetaDrive~\cite{li2021metadrive}, and VTD~\cite{von2009virtual} are the most commonly adopted. These simulators were utilized for \emph{executing, visualizing, and analyzing the generated scenarios}, as well as for \emph{demonstrating the effectiveness of the proposed generative models and the generated scenarios in testing and improving ADS}. Again, we primarily focus on the four most commonly used simulators, given the large number of simulators employed across the selected studies.

\begin{enumerate}
    \item \textbf{Carla}~\cite{dosovitskiy2017carla} is an open-source urban driving simulator designed specifically for autonomous driving research and is widely used in this field~\cite{tang2023survey, lou2022testing}. Carla is the dominant simulator used in the selected studies and serving multiple purposes in the evaluation of generative models and ADS. Specifically, Carla was employed to execute generated scenarios -- including those that were synthesized~\cite{P10, P14, P21, P61}, transformed~\cite{P19}, or reconstructed~\cite{P32} -- for testing selected ADS, further demonstrating the effectiveness of generative models and the generated scenarios in revealing ADS failures, such as collisions. Additionally, Carla was used for simulating and analyzing generated scenarios, for instance, comparing original and reconstructed scenarios~\cite{P24}, or evaluating alignment between user-provided text instructions and the resulting scenarios~\cite{P39, P46, P56}. It was also utilized for visualization purposes, where generated scenarios were showcased for demonstration~\cite{P27, P34}, and for collecting data to support the generation of safety-critical scenarios for testing ADS~\cite{P62, P91}.

    \vspace{1mm}

    \item\textbf{LGSVL}~\cite{rong2020lgsvl} is a high-fidelity simulator for autonomous driving, used for evaluation in several studies~\cite{P33, P36, P45}, often alongside the Baidu Apollo ADS~\cite{Apollo}. Although LG announced the suspension of active development for the SVL Simulator as of January 1, 2022~\cite{lgsvl-github}, and the current version requires a cloud service (providing information such as assetGuid and download endpoints) that is no longer supported, efforts have been made to maintain its usability in research. In particular, SORA-SVL~\cite{SORA-SVL}, a locally hosted cloud environment compatible with SVL, has been adopted in several studies~\cite{P36, P45} to keep the simulator operational. LGSVL was employed to execute generated or reconstructed scenarios for various purposes: subsequent optimization of safety-critical scenarios using search algorithms~\cite{P33}; testing selected ADS~\cite{P36, P45}, for analyzing generated scenarios~\cite{P36} or demonstrating the effectiveness of proposed approaches for accident reconstruction~\cite{P45}.

    \vspace{1mm}
    \item\textbf{MetaDrive}~\cite{li2021metadrive} is a lightweight autonomous driving simulator developed by UCLA, designed to support customizable and realistic road scenarios and multiple types of ADS. It was used in several studies for diverse evaluation purposes. Luo et al.~\cite{P22} employed MetaDrive to execute reconstructed scenarios for ADS testing. In Feng et al.~\cite{P47}, it was used to train ADS, demonstrating the effectiveness of generated scenarios in improving driving performance. Additionally, in Mei et al.~\cite{P58}, MetaDrive was utilized to analyze generated scenarios and provide feedback (modification suggestions) for refining LLM-generated attacker identification code.

    \vspace{1mm}

    \item \textbf{VTD}~\cite{von2009virtual} (Virtual Test Drive) is a comprehensive driving simulation toolchain developed by Hexagon, designed for the development and testing of ADAS and ADS. It was used in selected studies to execute generated scenarios for ADS testing~\cite{P14, P59}, analyze the generated scenarios~\cite{P59}, and demonstrate the effectiveness of generated scenarios, e.g., exposing potential failures such as collisions during ADS evaluation in Zhou et al.~\cite{P14}.

\end{enumerate}

\faHandPointRight \ Overall, relatively few of the studies we surveyed employed a simulator in their experimentation or evaluation. This suggests that the \emph{use of a simulator is not essential} for assessing a generative approach or the quality of the produced scenarios. However, simulators can play a valuable role by \emph{enabling visualization and analysis of the generated scenarios, executing them in a controlled environment, and evaluating ADS performance} -- particularly by revealing system failures such as collisions and improvement after re-training or fine-tuning. Therefore, we regard simulators as useful tools that facilitate and enhance the evaluation process, even if they are not strictly required.

\begin{table}[tbp]
\centering
\small
\caption{RQ2 -- ADS used for evaluation in selected studies.}
\begin{tabular}{|c|l|l|c|}
    \hline 
    \textbf{Evaluation setup} & \textbf{Level} & \textbf{Category} & \textbf{Instance (Studies)} \\
    \hline
    \multirow{22}{*}{ADS} & \multirow{6}{*}{System} & \multirow{2}{*}{DNN-based} & Autumn (\cite{P71}, \cite{P55}), Chauffeur (\cite{P71}, \cite{P55}) \\
    & & & Rambo (\cite{P71}), Rwightman (\cite{P55})\\
    \cline{3-4} & & \multirow{2}{*}{RL-based} & SAC (\cite{P8}, \cite{P10}, \cite{P61}), PPO (\cite{P8}, \cite{P10}, \cite{P47}, \cite{P61}, \cite{P69})\\
    & & & TD3 (\cite{P10}, \cite{P58}, \cite{P61}), DDPG (\cite{P8}), MBRL (\cite{P8}) \\
    \cline{3-4} & & \multirow{2}{*}{General} & Baidu Apollo (\cite{P33}, \cite{P36}, \cite{P45}, \cite{P46}), InterFuser (\cite{P19}, \cite{P46})\\
    & & & Autoware (\cite{P46}), Dora-RS (\cite{P46}), IDM (\cite{P22}), PPO (\cite{P22}), Auto (\cite{P22})\\
    \cline{2-4} & \multirow{5}{*}{Module} & Decision-making & Data Planner (\cite{P15}), IDM Planner (\cite{P15}), Astar Planner (\cite{P15}) \\
    \cline{3-4} & & Planning & Replay Planner (\cite{P54}), Rule-based Planner (\cite{P54}, \cite{P94}) \\
    \cline{3-4} & & \multirow{2}{*}{Driving algorithm} & VI (\cite{P5}), RVI (\cite{P5}), D3QN  (\cite{P5}), GAIL (\cite{P17}), SAC (\cite{P17}) \\
    & & & DNN (\cite{P21}), TQC (\cite{P21}), RL/IL (\cite{P21}), CoopNavi (\cite{P21}) \\
    \cline{3-4} & & VLM & GPT-4 (\cite{P6}), Qwen2-VL (\cite{P6}) \\
    \cline{2-4} & \multirow{11}{*}{Function} & \multirow{2}{*}{Lane-keeping} & Dave2 (\cite{P7}, \cite{P19}), Epoch (\cite{P7}, \cite{P19}), Chauffeur (\cite{P7}, \cite{P19}) \\
    & & & Autumn (\cite{P7}), Vision-Transformer-based ADS (\cite{P19}), ALKS (\cite{P59}) \\
    \cline{3-4} & & \multirow{4}{*}{Trajectory prediction} & Vanilla-LSTM (\cite{P11}), Social-LSTM (\cite{P11}), CS-LSTM (\cite{P11}) \\
    & & & AutoBots (\cite{P38}, \cite{P74}), MATP (\cite{P74}), Rule-based Planner (\cite{P64})\\
    & & & Hybrid Planner (\cite{P64}), Learning-based Behavior Cloning Planner (\cite{P64}) \\
    & & & PDM-closed (\cite{P64}), IDM Planner (\cite{P64})\\
    \cline{3-4} & & Object detection & YOLO (\cite{P7}, \cite{P30}), Lift-Splat (\cite{P75}), FasterRCNN (\cite{P78}) \\
    \cline{3-4} & & \multirow{3}{*}{Classification} & ResNet (\cite{P70}, \cite{P78}), VGG (\cite{P78}), Xception (\cite{P78})\\
    & & & ConvNeXtSmall (\cite{P78}), InceptionV3 (\cite{P78}), MobileNet (\cite{P78}) \\ 
    & & & DenseNet (\cite{P78}), EfficientNetV2S (\cite{P78}) \\
    \cline{3-4} & & Segmentation & DeepLabV3 (\cite{P70}), OneFormer (\cite{P73}) \\
    \hline
\end{tabular}
\label{table:results:RQ2:ads}
\end{table}

\subsubsection{ADS}
\label{sec:results:RQ2:ads}

A wide range of ADS, including full systems, individual modules, and specific functions, were used as the system under test (SUT) across the selected studies, as shown in Table~\ref{table:results:RQ2:ads}. These systems were employed \emph{to evaluate and demonstrate the effectiveness of the proposed generative models or the generated scenarios by revealing failures and defects in ADS} behavior (e.g., collisions, off-roads, and route completion) and \emph{improvement of performance after retraining or finetuning with generated scenarios}. Given the large number of ADS utilized across the literature, we deliberately focus on a subset of systems that are more frequently used for evaluation in these studies. Our categorization of the ADS used (e.g., DNN-based, RL-based, for system-level ADS) is primarily based on the descriptions provided in the original studies, although there may be overlaps between categories and alternative grouping strategies may also be valid.

\begin{enumerate}
    \item Several \textbf{DNN-based ADS} were used as the SUT in the selected studies in evaluation. Autumn, Chauffeur, Rambo, Rwightman -- pretrained DNN-based regression models provided by the Udacity Self-Driving Car Challenge~\cite{udacitychallenge} -- are used to predict steering angles under different scenarios with transformed visual conditions. Specifically, in Pan et al.~\cite{P71}, original sunny images are transformed into foggy scenes with varying fog levels and orientations, while in Zhang et al.~\cite{P55}, original sunny images are transformed into snowy and rainy conditions. These models were evaluated by measuring inconsistencies in steering angle predictions between original and transformed images; deviations beyond a threshold were used to detect faults in the ADS.

    \vspace{1mm}
    
    \item Quite a few studies used \textbf{Reinforcement learning (RL)-based agents} as the SUT in evaluation. In Ding et al.~\cite{P8}, four RL driving agents were trained on normal scenarios and then evaluated on generated scenarios. The study further trained the same RL agents on generated scenarios and test them on various scenario types, demonstrating improved driving performance of agents trained with generated scenarios. Zhang et al.~\cite{P10} trained ADS using three RL algorithms (PPO~\cite{schulman2017proximal}, SAC~\cite{haarnoja2018soft}, and TD3~\cite{fujimoto2018addressing}) and evaluated their performance under safety-critical scenarios generated by the proposed method. Results indicate that adversarial training using these generated scenarios substantially reduced collision rates and significantly improved the overall performance and safety. Similarly, in Feng et al.~\cite{P47}, an RL controller was trained using PPO~\cite{schulman2017proximal} across three different datasets: real-world scenarios, generated scenarios, and augmented datasets combining both. Agents trained with the generated scenarios show a lower rate of safety violations and a higher success rate in reaching their destinations. In Xu et al.~\cite{P61}, three AV models were trained using RL algorithms (SAC, PPO~\cite{schulman2017proximal}, and TD3~\cite{fujimoto2018addressing}) on benign (normal) driving scenarios, and then evaluated on safety-critical scenarios generated. Additionally, the downstream utility of the generated scenarios was assessed by fine-tuning AV models on them. In Rowe et al.~\cite{P69}, an RL planner was trained with PPO~\cite{schulman2017proximal} on real-world scenes from the Waymo dataset and evaluated both on original Waymo data and on generated scenarios with respect to collision rate, off-road rate, and route completion. Lastly, in Mei et al.~\cite{P58}, an RL driving policy was trained using TD3~\cite{fujimoto2018addressing} and tested in generated adversarial scenarios to measure the frequency of ADS failure. These generated scenarios were further used for adversarial training, and evaluation shows that this process improved the ADS’s collision rate and route completion, further supporting the value of generated safety-critical scenarios for enhancing ADS robustness.

    \vspace{1mm}
    
     \item Several \textbf{general ADS} were used in the selected studies, with InterFuser~\cite{shao2023safety} and Apollo~\cite{Apollo} being the most frequently adopted. InterFuser is an open-source end-to-end ADS for urban driving, developed by OpenDILab~\cite{shao2023safety}. Specifically, Baresi et al.~\cite{P19} evaluated driving quality and system failures of it using various metrics such as driving score and route completion rate, showing that performance significantly degraded when tested with the augmented scenarios produced in their work. Apollo, an open-source industry-grade modular ADS developed by Baidu~\cite{Apollo}, supports a broad range of functionalities including perception, localization, prediction, routing, planning, and control. In Cai et al.~\cite{P46}, Apollo, along with Autoware, InterFuser, and Dora-RS, were tested on generated scenarios to record rule violations, collisions, and timeouts -- demonstrating the effectiveness of the test scenarios in testing ADS. In Tang et al.~\cite{P33}, Apollo (version 7.0) served as the SUT to evaluate the LeGEND framework, which successfully identified more types of critical scenarios indicating potential system defects. Similarly, Tian et al.~\cite{P36} run Apollo (7.0) in generated scenarios to detect a wide range of safety violations, showing that the approach (LEADE) can generate both abstract and executable concrete scenarios from real traffic videos. The study also demonstrates LEADE can efficiently create safety-critical, semantically equivalent scenarios for ADS testing. In Guo et al.~\cite{P45}, generated scenarios from SoVAR were used to test Apollo (6.0), helping to identify safety-violation behaviors and evaluate the system’s robustness.

    \vspace{1mm}

    \item Several ADS were used in the selected studies for \textbf{specific functions} of interest. Autumn, Chauffeur, and Epoch -- all open-source models contributed by the community through the Udacity Self-Driving Car Challenge~\cite{udacitychallenge}, and Dave2 (NVIDIA)~\cite{bojarski2016end}, are end-to-end deep neural network (DNN)-based regression models that predict steering angles to control the ego vehicle, and were used for lane-keeping tasks~\cite{P7, P19}. For object detection and classification, YOLO was widely adopted. Specifically, YOLOv5~\cite{yolov5} was used in Amini and Nejati~\cite{P7}, while YOLOv8~\cite{yolov8} was used in Yigit and Can~\cite{P30}, in which multiple versions of YOLOv8 (Nano, Small, and Medium) were trained on a generated dataset and evaluated both on that dataset and on real-world images to assess object detection performance. Autobots~\cite{girgis2021latent}, an open-source trajectory prediction model that employs multi-head self-attention to capture spatial and temporal relationships between traffic elements, was used for motion prediction in ADS and for evaluating generated scenarios~\cite{P38, P74}. Specifically, in Ding et al.~\cite{P38}, Autobots was trained on various dataset configurations—including the original nuScenes dataset~\cite{qian2024nuscenes}, a version augmented with random noise, and one augmented with generated scenarios -- and tested for performance metrics such as collision rate, demonstrating the utility of scenario augmentation for improving prediction robustness.
\end{enumerate}

\faHandPointRight \ Overall, incorporating ADS enables \emph{empirical evaluation of the effectiveness of the proposed approaches and produced scenarios for testing} -- such as their ability to trigger failures, uncover system defects, or performance degradation, for selected ADS. While the results may be influenced by the specific implementation and training of the ADS, and can vary across different systems, \emph{comparing performance of the proposed approaches across different baseline methods on the same ADS, or comparing performance of the system with and without scenarios produced in those studies} offers a consistent and informative perspective on the relative effectiveness of each approach for testing purposes. 

\subsubsection{Evaluation Metrics}
\label{sec:results:RQ2:metrics}

\begin{figure}[hbp]
    \centering
    \includegraphics[width=0.9\textwidth]{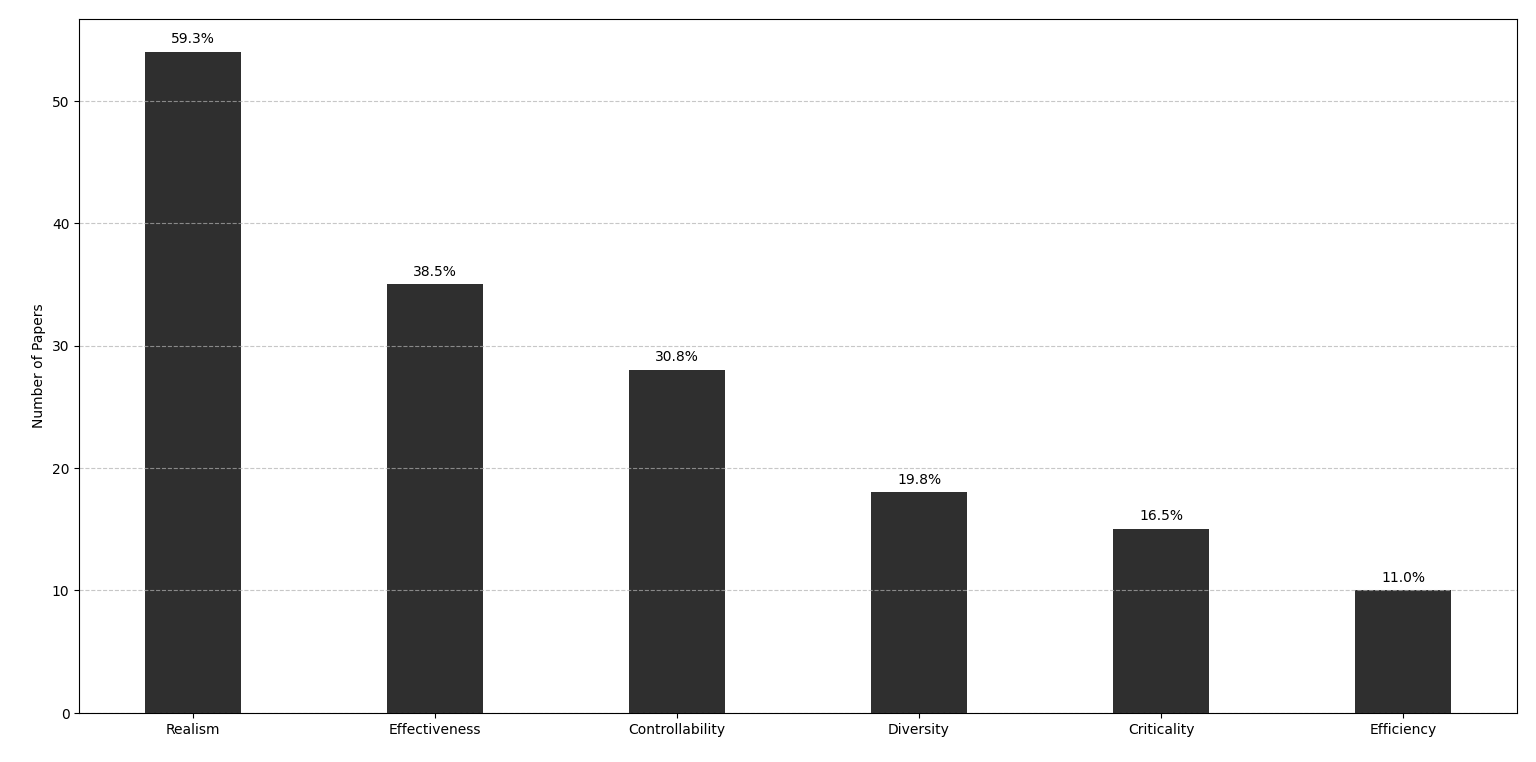}
    \caption{Number and proportion of papers employing metrics related to each quality attribute in the evaluation.}
    \label{fig:quality_attributes}
\end{figure}

More than 160 metrics were used across the surveyed studies \emph{to evaluate and quantify the quality of produced scenarios and to demonstrate the performance of the proposed approaches} in terms of effectiveness and efficiency. These metrics primarily concern six key quality aspects: Realism (e.g., similarity-based, common sense-based), Diversity (e.g., coverage-based, variation-based), Effectiveness (e.g., failure event-based, performance-based), Controllability (e.g., accuracy-based, correctness-based, alignment-based), Criticality (e.g., risk/severity-based), and Efficiency (e.g., time-based and resource-based). The primary goal of these metrics is \emph{to assess whether the proposed approaches can efficiently and effectively generate scenarios that meet desired properties} -- such as being realistic, diverse, adversarial, or user instruction aligned -- for testing and improving ADS. Given the large number of metrics reported, our analysis focuses on those that are most frequently used across studies. It is also worth noting that several studies used some evaluation metrics in the results without sufficient detail on how they are computed, what they measure, or whether they are intended to serve as evaluation metrics; therefore, they are not included in our analysis.

\par\vspace{2mm}

\begin{table}[tbp]
\centering
\small
\caption{RQ2 -- Similarity-based realism metrics used for evaluation in selected studies.}
\begin{tabular}{|l|l|l|}
    \hline 
    \textbf{Category} & \textbf{Metric} & \textbf{Studies} \\
    \hline
    \multirow{25}{*}{Similarity-based} &  \multirow{2}{*}{minimum Average Displacement Error (minADE)} & \cite{P16}, \cite{P17}, \cite{P18}, \cite{P38}, \cite{P47}, \cite{P52}, \cite{P74}, \cite{P81} \\
    & & \cite{P83}, \cite{P88}, \cite{P90} \\
    \cline{2-3} & \multirow{2}{*}{minimum Final Displacement Error (minFDE)} & \cite{P16}, \cite{P17}, \cite{P18}, \cite{P38}, \cite{P47}, \cite{P50}, \cite{P52}, \cite{P74} \\
    & & \cite{P81}, \cite{P83}, \cite{P88}, \cite{P90} \\
    \cline{2-3} & \multirow{2}{*}{Maximum Mean Discrepancy (MMD)} & \cite{P18}, \cite{P38}, \cite{P41}, \cite{P44}, \cite{P47}, \cite{P67}, \cite{P81}, \cite{P83} \\
    & & \cite{P86}, \cite{P87}, \cite{P89} \\
    \cline{2-3} & Jensen-Shannon Divergence (JSD) & \cite{P4}, \cite{P14}, \cite{P40}, \cite{P42}, \cite{P69}, \cite{P81} \\
    \cline{2-3} & Root Mean Square Error (RMSE) & \cite{P25}, \cite{P65}, \cite{P76}, \cite{P78} \\
    \cline{2-3} & Wasserstein distance (WD) &  \cite{P53}, \cite{P60}, \cite{P61}, \cite{P64} \\
    \cline{2-3} & Frechet Distance &  \cite{P61}, \cite{P67}, \cite{P68}, \cite{P69} \\
    \cline{2-3} & Hausdorff Distance & \cite{P13}, \cite{P14}, \cite{P90} \\
    \cline{2-3} & Scenario Collision Rate (SCR) & \cite{P18}, \cite{P81}, \cite{P90} \\
    \cline{2-3} & mean Intersection over Union (mIOU) & \cite{P23}, \cite{P73} \\
    \cline{2-3} & Matching distance & \cite{P28}, \cite{P80} \\
    \cline{2-3} & Kullback–Leibler Divergence (KL) & \cite{P61}, \cite{P65} \\
    \cline{2-3} & Peak signal-to-noise ratio (PSNR) & \cite{P71}, \cite{P78} \\
    \cline{2-3} & \multicolumn{2}{c|}{ Road Smoothness (\cite{P14}), Distributional realism (\cite{P25}, \cite{P51}, \cite{P82}), BLEU (\cite{P26}), ROUGE-L (\cite{P26})} \\
    & \multicolumn{2}{c|}{ Scenario Bahavior Coverage (\cite{P32}), Discriminative score (\cite{P40}), Predictive score (\cite{P40})} \\
    & \multicolumn{2}{c|}{ Negative log-likelihood (NLL) (\cite{P44}), Mean Absolute Error (MAE) (\cite{P49})} \\
    & \multicolumn{2}{c|}{ Symmetric Segment-Path Distance (SSPD) (\cite{P61}), Pixel-level L1 distance error (\cite{P63})} \\
    & \multicolumn{2}{c|}{ F1 score (\cite{P68}), Chamfer distance (\cite{P68}), Route length (\cite{P68}), Longest Route Length (\cite{P69})} \\
    & \multicolumn{2}{c|}{Endpoint Distance (\cite{P69}), Structural Similarity Index (SSIM) (\cite{P71})} \\
    & \multicolumn{2}{c|}{Frechet Video Distance (FVD) (\cite{P72}), Average Precision (AP) (\cite{P73}), Panoptic quality (PQ) (\cite{P73})} \\
    & \multicolumn{2}{c|}{Feature similarity index (FSIM) (\cite{P78}), Spectral Angle Mapper (SAM) (\cite{P78})} \\
    & \multicolumn{2}{c|}{Signal to Reconstruction Error Ratio (SRE) (\cite{P78}), Sim Agent Benchmark (\cite{P84})} \\
    \hline
\end{tabular}
\label{table:results:RQ2:metrics:realism:similarity}
\end{table}

\noindent\textbf{(1) Realism.} Realism primarily concerns whether the produced scenarios -- whether generated, transformed, or reconstructed -- closely resemble real-world scenarios. It is typically evaluated using similarity-based and common sense-based metrics. Similarity-based metrics were widely used in the selected studies to assess how realistic the produced scenarios were by comparing them to reference or ground-truth data, often drawn from real-world datasets. These evaluations were conducted from various perspectives, including spatial distance, structural similarity, and distributional similarity of scenario features. Among the most frequently used similarity-based metrics are minimum Average Displacement Error (minADE)~\cite{P16}, minimum Final Displacement Error (minFDE)~\cite{P16}, Maximum Mean Discrepancy (MMD)~\cite{P18}, Jensen-Shannon Divergence (JSD)~\cite{P4}, Root Mean Square Error (RMSE)~\cite{P25}, Wasserstein Distance~\cite{P53}, Fréchet Distance~\cite{P61}, and Hausdorff Distance~\cite{P13}, as shown in Table~\ref{table:results:RQ2:metrics:realism:similarity}. In addition, studies used metrics such as Scenario Collision Rate (SCR)~\cite{P18}, mean Intersection over Union (mIoU)~\cite{P23}, Fréchet Inception Distance (FID)~\cite{P18}, Matching Distance ~\cite{P28}, Kullback–Leibler Divergence (KL)~\cite{P61}, and Peak Signal-to-Noise Ratio (PSNR)~\cite{P71}. These metrics are designed to measure the degree of similarity between generated and real intended scenarios by comparing distance or distributional gaps of their underlying features (e.g., distance between the predicted trajectories and ground truth trajectories, or distribution gaps in vehicle velocity, acceleration, and time to collision between real and generated scenarios), thereby serving as indicators of how realistic the generated scenarios are.

\begin{table}[hbp]
\centering
\small
\caption{RQ2 -- Common sense-based similarity metrics used for evaluation in selected studies.}
\begin{tabular}{|c|l|l|}
    \hline 
    \textbf{Category} & \textbf{Metric} & \textbf{Studies} \\
    \hline
    \multirow{8}{*}{Common sense-based} &  Collision Rate & \cite{P4}, \cite{P38}, \cite{P42}, \cite{P47}, \cite{P52}, \cite{P60}, \cite{P64}, \cite{P69}, \cite{P81}, \cite{P83}, \cite{P87}, \cite{P88}, \cite{P94} \\
    \cline{2-3} & Off-road Rate &  \cite{P38}, \cite{P42}, \cite{P50}, \cite{P52}, \cite{P60}, \cite{P64}, \cite{P81}, \cite{P94} \\
    \cline{2-3} & Solution Rate & \cite{P16}, \cite{P54} \\
    \cline{2-3} & Classification accuracy & \cite{P29}, \cite{P48} \\
    \cline{2-3} & \multicolumn{2}{c|}{ Confusion matrix-based metrics (\cite{P30}), mean average precision (mAP) (\cite{P30})} \\
    & \multicolumn{2}{c|}{ Drivability (\cite{P41}), Semantic fidelity (\cite{P46}), Driving rationality (\cite{P46})} \\
    & \multicolumn{2}{c|}{ Discriminative score (\cite{P49}), Comfort distance (CFD) (\cite{P50}), Trajectory smoothness (\cite{P66})} \\
    & \multicolumn{2}{c|}{ Agreement rate on understandability and validity(\cite{P70})} \\
    \hline
\end{tabular}
\label{table:results:RQ2:metrics:realism:commonsense}
\end{table}

Common sense-based metrics evaluate the realism of generated scenarios based on generally accepted characteristics that real-world driving scenarios are expected to exhibit. These include the absence of unintended vehicle collisions or off-road events, as well as properties such as plausibility, drivability, comfort, and smoothness. Such metrics aim to capture whether a scenario aligns with what would be considered reasonable or realistic by human standards. Among the most commonly used metrics are Scenario Collision Rate~\cite{P4, P38, P42, P47, P52, P60, P64, P69, P81, P83, P87, P88, P94} and Off-road Rate~\cite{ P38, P42, P50, P52, P60, P64, P81, P94}, which respectively quantify the proportion of generated scenarios that result in unintended collisions or vehicles leaving the drivable area. Solution Rate, as used in Gao et al.~\cite{P16} and Rempe et al.~\cite{P54}, measured whether a plausible driving solution exists for the generated scenario. Another important indicator is whether the generated scenarios expose similar performance in the SUT, such as classification accuracy, when compared to real data, as examined in several studies~\cite{P29, P30, P48}. Pronovost et al.~\cite{P41} measured the Drivability of generated scenarios -- the fraction of waypoints that fall within the drivable area. To evaluates smoothness of generated trajectories, Ding et al.~\cite{P66}  measured the angle between consecutive movement directions, and Lin et al.~\cite{P50} computed a Comfort Distance metric. In addition to automated metrics, several studies incorporated human evaluation to assess qualities such as Semantic Realism and Rationality~\cite{P46}, and Understandability and Validity~\cite{P70} of the generated scenarios, as shown in Table~\ref{table:results:RQ2:metrics:realism:commonsense}. These subjective assessments provide valuable insight into how realistic and acceptable the scenarios are from the perspective of human users or domain experts.

\par\vspace{2mm}

\begin{table}[hbp]
\centering
\small
\caption{RQ2 -- Effectiveness metrics used for evaluation in selected studies.}
\begin{tabular}{|c|l|l|l|l|}
    \hline 
    \textbf{Quality attribute} & \textbf{Category 1} & \textbf{Category 2} & \textbf{Metric} & \textbf{Studies} \\
    \hline
    \multirow{20}{*}{Effectiveness} & \multirow{11}{*}{Failure events} & \multirow{5}{*}{Ratio-based} & \multirow{2}{*}{Collision rate} & \cite{P8}, \cite{P10}, \cite{P15}, \cite{P16}, \cite{P38}, \cite{P54}, \\
    & & & & \cite{P56}, \cite{P57}, \cite{P58}, \cite{P61} \\
    \cline{4-5} & & & Failure rate & \cite{P21}, \cite{P53}, \cite{P58}, \cite{P68}, \cite{P71} \\
    \cline{4-5} & & & \multicolumn{2}{c|}{Effective vulnerability rate (\cite{P5}), Driving Capability (\cite{P17})} \\
    & & & \multicolumn{2}{c|}{Critical scenario rate (\cite{P32}), Incomplete route rate (\cite{P61})} \\
    
    \cline{3-5} & & \multirow{6}{*}{Quantity-based}& Number of collisions & \cite{P22}, \cite{P45}, \cite{P46} \\
    \cline{4-5} & & & Number of failures & \cite{P21}, \cite{P71}, \cite{P7}, \cite{P55} \\
    \cline{4-5} & & & Number of safety violations  & \cite{P36}, \cite{P47} \\
    \cline{4-5} & & & Types of safety violations  & \cite{P36}, \cite{P45} \\
    \cline{4-5} & & & \multicolumn{2}{c|}{Out-of-Bound (OOB) distance (\cite{P7}), OOB counts (\cite{P7})} \\
    & & & \multicolumn{2}{c|}{Number of timeouts (\cite{P46}), Number of traffic rule violations (\cite{P46})} \\

    \cline{2-5} & \multirow{9}{*}{Performance} & \multirow{3}{*}{Driving-related} & Overall score & \cite{P10}, \cite{P56}\\
    \cline{4-5} & & & \multicolumn{2}{c|}{Driving comfort (\cite{P17}), Success rate (\cite{P47})} \\
    & & &  \multicolumn{2}{c|}{Route completion (\cite{P58}), Speed satisfaction (\cite{P61})} \\
    
    \cline{3-5} & & \multirow{6}{*}{Accuracy-related} & Model accuracy & \cite{P6}, \cite{P70}, \cite{P75}, \cite{P78}) \\
    \cline{4-5} & & & mean Average Precision (mAP) & \cite{P7}, \cite{P62} \\
    \cline{4-5} & & & \multicolumn{2}{c|}{mean Absolute Error (mAE)~\cite{P7}, Prediction error (\cite{P11})} \\
    & & & \multicolumn{2}{c|}{Predictive score (\cite{P13}), Discriminative score (\cite{P13})} \\
    & & & \multicolumn{2}{c|}{Confusion matrix-based metrics (\cite{P19}), Average Precision (AP) (\cite{P63})} \\
    & & & \multicolumn{2}{c|}{Recall (\cite{P63}), mean Intersection over Union (mIoU) (\cite{P70}), IoU (\cite{P91})} \\
    \hline
\end{tabular}
\label{table:results:RQ2:metrics:effectiveness}
\end{table}

\noindent\textbf{(2) Effectiveness.} The effectiveness of generative approaches and the resulting scenarios for testing ADS is commonly measured using two types of metrics: failure events-based and ADS performance-based metrics, as shown in Table~\ref{table:results:RQ2:metrics:effectiveness}. While there is often overlap among metrics -- e.g., collisions, off-road events, and incomplete routes are all typically considered failures -- we follow the original descriptions provided in the studies and avoid merging these categories to preserve important contextual details.

Failure events-based metrics focus on quantifying whether the SUT experiences specific types of failures when exposed to the produced scenarios. These metrics are further categorized into ratio-based and quantity-based metrics. Ratio-based metrics primarily compute the proportion of specific failure events in the generated scenarios, such as collision~\cite{P8, P10, P15, P16, P38, P54, P56, P57, P58, P61}, failure in a more general context (e.g., collisions, off-road events, and incomplete routes)~\cite{P21, P53, P58, P68, P71}, vulnerability (e.g., off-road events)~\cite{P5}, critical scenario (e.g., scenarios falling below a safe distance threshold)~\cite{P32}, and incomplete route incidents~\cite{P61}. In contrast, quantity-based metrics count the absolute number of specific failure events in the generated scenarios. These include number of collisions~\cite{P22, P45, P46}, failures in a general context~\cite{P21, P71}, types of safety violations~\cite{P36, P45}, number of safety violations~\cite{P36, P47}, Out-of-bound (OOB) incidents~\cite{P7}, misbehavior~\cite{P19}, traffic rule violations~\cite{P46}, timeouts (i.e., failure to complete a task within the allotted time)~\cite{P46}, and inconsistent behaviors, such as differing outputs under transformed inputs~\cite{P55}. These failure-based metrics are straightforward in quantifying how well the generative models produce scenarios that challenge the ADS, ultimately supporting their use for testing and improvement.

Performance-based metrics focus on evaluating the behavior and performance of SUT, which may refer to an ADS, or a specific function or module of ADS, within the generated scenarios. Performance-based metrics are further categorized into two groups: driving quality-related metrics and accuracy-related metrics. Driving quality-related metrics assess the overall driving performance of the SUT. Common examples include: overall driving score~\cite{P10, P56}, driving comfort~\cite{P17}, success rate of reaching the destination~\cite{P47, P58}, and adherence to regular or desired driving speed~\cite{P61}. Accuracy-based metrics evaluate the correctness or precision of specific ADS functions or modules. These include: question-answering accuracy of VLMs~\cite{P6}, lane-keeping function (using mAE)~\cite{P7}, trajectory prediction accuracy (measured by prediction error)~\cite{P11}, object detection (using metrics like accuracy and mAP)~\cite{P7, P62, P75, P78}, classification (using accuracy and mIoU)~\cite{P70, P78}, and segmentation performance, measured via IoU and accuracy~\cite{P70, P91}. Additionally, performance metrics may be used to evaluate specific components within the generative model or framework itself. For example, in Fang et al.~\cite{P13}, the discriminator and generator components were assessed using predictive and discriminative scores. In Baresi et al.~\cite{P19}, a semantic validator to assess the validity and semantic consistency between original and augmented images was evaluated using confusion matrix-based metrics.

\par\vspace{2mm}

\begin{table}[hbp]
\centering
\small
\caption{RQ2 -- Controllability metrics used for evaluation in selected studies.}
\begin{tabular}{|c|l|l|l|}
    \hline 
    \textbf{Quality attribute} & \textbf{Category} & \textbf{Metric} & \textbf{Studies} \\
    \hline
    \multirow{10}{*}{Controllability} & \multirow{1}{*}{Accuracy-based} & Information extraction accuracy & \cite{P22}, \cite{P36}, \cite{P45}, \cite{P46}, \cite{P93} \\
    \cline{2-4} & \multirow{2}{*}{Correctness-based} & Success rate & \cite{P37}, \cite{P39}, \cite{P24}, \cite{P26}, \cite{P46} \\
    \cline{3-4} & & Error rate & \cite{P34} \\
    \cline{2-4} & \multirow{8}{*}{Alignment-based} & Preference rate & \cite{P38}, \cite{P90} \\
    \cline{3-4} & & Matching rate & \cite{P75} \\
    \cline{3-4} & & Rule/Constraint satisfaction & \cite{P42}, \cite{P53}, \cite{P60}, \cite{P65} \\
    \cline{3-4} & & \multicolumn{2}{c|}{Clarity score (\cite{P6}), YOLO score (\cite{P12}), Confusion matrix-based metrics (\cite{P20})} \\
    & & \multicolumn{2}{c|}{Correctness (\cite{P36}), Conformity of command (\cite{P37}, \cite{P39}),  Collision rate (\cite{P50}, \cite{P51})} \\
    & & \multicolumn{2}{c|}{Action prediction loss (\cite{P72}), Validity (\cite{P76}), Text-Score (\cite{P77})} \\
    & & \multicolumn{2}{c|}{minimum Average Displacement Error(minADE) (\cite{P82}), Agreement matrices (\cite{P85})} \\
    & &  \multicolumn{2}{c|}{Reconstruction rate (\cite{P45}), Reconstruction error (\cite{P91})} \\
    \hline
\end{tabular}
\label{table:results:RQ2:metrics:controllability}
\end{table}

\noindent\textbf{(3) Controllability.} Controllability refers to the extent to which generative models can produce desired outputs in response to specified inputs or constraints. It is evaluated across several dimensions, including information extraction accuracy, output correctness, and alignment with user intent. Accuracy-based metrics assess the model's ability to correctly interpret and extract relevant information (e.g., road, environment, dynamic objects) from input data or user instructions, as seen in studies~\cite{P22, P36, P45, P46, P93}. Correctness-based metrics evaluate whether the generated outputs -- such as code or scenarios -- are valid, meaningful, and executable without errors~\cite{P24, P26, P37, P39, P46}. Complementing this, error rate is used in Petrovic et al.~\cite{P34} to quantify the percentage of incorrect outputs produced by the model.

A majority of studies evaluated controllability through alignment-based metrics, which assess how well the generated outputs match user-specified attributes, constraints, rules, formats, or intentions. Several studies incorporated human evaluation to assess alignment of the generated scenarios using metrics like preference rate and matching rate~\cite{P38, P75, P90}. Others verified whether the generated scenarios adhere to user-defined rules or constraints~\cite{P42, P53, P60, P65}. Numerous studies also assessed alignment of produced scenarios with control attributes or intended behaviors using various metrics, as shown in Table~\ref{table:results:RQ2:metrics:controllability}. Examples include: control attributes such as specified locations and entities (based on a YOLO score)~\cite{P12}, pedestrian characteristics (using Confusion matrix-based metrics)~\cite{P20}, number of lanes or vehicles~\cite{P37, P39}, risk levels (via collision rate)~\cite{P51}, or user-specified actions (measured by action prediction loss)~\cite{P72} in the generated scenarios. Intended behaviors derived from original video data~\cite{P36}, desired collisions~\cite{P50}, user-specified behaviors~\cite{P82, P85}. Desired quality attributes like clarity (graded by GPT4)~\cite{P6}, validity~\cite{P76}, accuracy, consistency, rationality, and relevance measured by Text-Score~\cite{P77}. Finally, two studies evaluated controllability through reconstruction accuracy, using metrics such as reconstruction rate or reconstruction error to measure how accurately the model can regenerate scenarios from input information -- such as accident reports~\cite{P45} or image scenes~\cite{P91}.

\par\vspace{2mm}

\begin{table}[hbp]
\centering
\small
\caption{RQ2 -- Diversity metrics used for evaluation in selected studies.}
\begin{tabular}{|c|l|c|}
    \hline 
    \textbf{Quality attribute} & \textbf{Category} & \textbf{Metric (Studies)} \\
    \hline
    \multirow{10}{*}{Diversity} & \multirow{2}{*}{Coverage-based} & Neuron coverage (NC) (\cite{P7}, \cite{P62}), Surprise adequacy (SA) (\cite{P7}, \cite{P62}) \\
    & & Intersection over Union (IoU) (\cite{P13}, \cite{P40}) \\
    \cline{2-3} & \multirow{3}{*}{Uniqueness-based} & Number of distinct vulnerability types (\cite{P5}), Cluster numbers (\cite{P11}) \\
    & & Number of distinct failures (\cite{P21}), Distinct types of critical scenarios (\cite{P33}) \\
    & & Agent Diversity (\cite{P56}), Road Diversity (\cite{P56}) \\
    \cline{2-3} & \multirow{5}{*}{Variation-based} & Final Displacement Diversity (FDD) (\cite{P4}, \cite{P52}), Geometric Diversity (GD) (\cite{P7}) \\
    & & minimum Scenarios Final Distance Error (minSFDE) (\cite{P4}), Distance (\cite{P66}) \\
    & & Average Displacement Error (ADE) (\cite{P10}), Average Feature Distance (\cite{P18}) \\
    & & Diversity score (\cite{P27}), Collision diversity (\cite{P64}), Chamfer distance (\cite{P67}) \\
    & & Kendall Tau correlation (\cite{P85}), Jensen-Shannon Divergence (JSD) (\cite{P85}) \\
    \hline
\end{tabular}
\label{table:results:RQ2:metrics:diversity}
\end{table}

\noindent\textbf{(4) Diversity.} Diversity refers to the variety or variability of generated scenarios and was evaluated across multiple dimensions using various metrics (see Table~\ref{table:results:RQ2:metrics:diversity}) in the selected studies. It plays a key role in assessing whether generative approaches can produce a broad and meaningful range of test scenarios for ADS. A few studies evaluated diversity through coverage-based metrics, such as neuron coverage and surprise adequacy~\cite{P7, P62}, which measures how many neural pathways in the ADS are activated by the generated scenarios, and scenario coverage~\cite{P13, P40}, which assesses how closely the generated scenarios resemble or intersect with real-world data. Other works measured diversity by counting distinct types of generated outputs. For example, Qiu et al.~\cite{P5} count the number of unique vulnerability types discovered, while Xu et al.~\cite{P21} and Tang et al.~\cite{P33} quantified diversity by identifying different types of critical or failure scenarios. Similarly, Ding et al.~\cite{P11} analyzes the number of clusters among generated scenarios to infer diversity, and Ruan et al.~\cite{P56} computed the rate of unique objects or road types appearing in the generated dataset.

Several studies also analyzed diversity through variation in scenario attributes or characteristics. Xie et al.~\cite{P4} and Peng et al.~\cite{P52} computed the variability in final vehicle positions across scenarios, while Xie et al.~\cite{P4} further measured the distance from a generated endpoint to the closest real scenario. Zhang et al.~\cite{P10} examined how the positions of adversarial objects vary across different scenes within the same scenario over time. In feature space, Amini and Nejati~\cite{P7} and Li et al.~\cite{P49} assessed the standard deviation and variation of generated scenarios, and Sun et al.~\cite{P18} calculated Average Feature Distance between feature vectors of the generated scenarios. Aasi et al.~\cite{P27} proposed a diversity score based on self-similarity within the dataset. Chang et al.~\cite{P64} analyzed variation in collision angles, relative speeds, and collision points to assess diversity in collision scenarios. Similarly, Ding et al.~\cite{P66} considered variations in inter-vehicle distances and vehicle speeds as indicators of scenario diversity. For map generation, Mi et al.~\cite{P67} used Chamfer Distance between generated and real-world maps as a geometric measure of diversity. Lastly, Nguyen et al.~\cite{P85} applied Jensen-Shannon Divergence (JSD) to evaluate behavioral diversity by comparing the distribution of actions or behaviors across different generated scenarios.

\par\vspace{2mm}

\begin{table}[hbp]
\centering
\small
\caption{RQ2 -- Criticality metrics used for evaluation in selected studies.}
\begin{tabular}{|c|l|l|l|}
    \hline 
    \textbf{Quality attribute} & \textbf{Category} & \textbf{Metric} & \textbf{Studies} \\
    \hline
    \multirow{13}{*}{Criticality} & \multirow{13}{*}{Rik/Severity-based} & Time to collision (TTC) & \cite{P25}, \cite{P57}, \cite{P76}, \cite{P79} \\
    \cline{3-4} & & Collision rate & \cite{P4}, \cite{P39}, \cite{P52}, \cite{P94} \\
    \cline{3-4} & & Acceleration rate & \cite{P4}, \cite{P52}, \cite{P54}, \cite{P94} \\
    \cline{3-4} & & Collision velocity & \cite{P4}, \cite{P54} \\
    \cline{3-4} & & Complexity & \cite{P37}, \cite{P39}, \cite{P68} \\
    \cline{3-4} & & Route Completion score & \cite{P37}, \cite{P39} \\
    \cline{3-4} & & Driving score & \cite{P37}, \cite{P39} \\
    \cline{3-4} & & Total score & \cite{P37}, \cite{P39} \\
    \cline{3-4} & & Success rate & \cite{P37}, \cite{P39} \\
    \cline{3-4} & & Minimal Distance & \cite{P11}, \cite{P79} \\
    \cline{3-4} & & \multicolumn{2}{c|}{Collision Severity (\cite{P4}), Maneuver Comfort (\cite{P4})} \\
    & & \multicolumn{2}{c|}{Mean distance (\cite{P11}), Mean Square Error (MSE) (\cite{P11}), OOD-ness (\cite{P27})} \\
    & & \multicolumn{2}{c|}{Rarity Score (\cite{P35}), Lane-Changing Risk Index (LCRI) (\cite{P49})} \\
    \hline
\end{tabular}
\label{table:results:RQ2:metrics:criticality}
\end{table}

\noindent\textbf{(5) Criticality.} Criticality often refers to how adversarial or challenging a scenario is for ADS, and was commonly evaluated using risk- or severity-based metrics across the selected studies. These metrics assess the severity of scenarios from various dimensions. Among the most frequently used metrics was Time-To-Collision (TTC)~\cite{P25, P57, P76, P79}, which quantifies the remaining time before a potential collision between the AV and a targeted vehicle -- typically a vehicle that poses a challenge to the AV, such as a lane-changing vehicle~\cite{P25, P76}. Collision Rate~\cite{P4, P39, P52, P94} was also widely used to measure the proportion of generated scenarios that result in a collision. Additional severity indicators include the acceleration rate of the targeted vehicle~\cite{P4, P52, P54, P94}, its velocity at the time of collision~\cite{P4, P54}, the minimum distance between vehicles~\cite{P11, P79}, and the Mean Distance between Neighbor waypoints (MDN)~\cite{P11}, which captures the smoothness of generated trajectories and indicates how adversarial they are.

Beyond direct measures of collision or collision-related metrics, some studies used performance-based indicators as proxies for criticality. For instance, Lu et al.~\cite{P37, P39} used route completion rate, driving performance, success rate, and a total score as metrics to reflect the difficulty level of scenarios. Rarity~\cite{P27, P35} was used to assess how uncommon or rare a generated scenario is, while Li et al.~\cite{P49} introduced LCRI (Lane Change Risk Index) to quantify the risk level in lane-changing situations and identify rare adversarial cases. In Ding et al.~\cite{P11}, Mean Square Error (MSE) between generated and original trajectories was used to assess how significantly a generated scenario deviates. Some studies also analyzed scenario complexity, e.g., in Lu et al.~\cite{P37, P39}, complexity was evaluated by computing the mean and standard deviation of features such as the number of lanes, map edges, route length, and vehicle count. Similarly, Chitta et al.~\cite{P68} used the number of turns and agents in a scenario as indicators of its difficulty.

\par\vspace{2mm}

\begin{table}[tbp]
\centering
\small
\caption{RQ2 -- Efficiency metrics used for evaluation in selected studies.}
\begin{tabular}{|c|l|l|l|}
    \hline 
    \textbf{Quality attribute} & \textbf{Category} & \textbf{Metric} & \textbf{Studies} \\
    \hline
    \multirow{5}{*}{Efficiency} & \multirow{2}{*}{Time-based} & Computational time & \cite{P7}, \cite{P19}, \cite{P22}, \cite{P24}, \cite{P46}, \cite{P33}, \cite{P34}, \cite{P52} \\
    \cline{3-4} & & Simulation time & \cite{P94} \\
    \cline{2-4} & \multirow{3}{*}{Resource-based} & Number of tokens & \cite{P34} \\
    \cline{3-4} & & Number of simulations & \cite{P33} \\
    \cline{3-4} & & Number of videos & \cite{P36} \\
    \hline
\end{tabular}
\label{table:results:RQ2:metrics:efficiency}
\end{table}

\noindent\textbf{(6) Efficiency.} The efficiency of the proposed approaches or generative models is primarily evaluated using two types of metrics: time-based and resource-based. Time-based metrics quantify the computational time required either for the entire scenario generation, transformation, or augmentation process~\cite{P7, P19, P22, P24, P33, P46, P52}, or for a single step within the generation pipeline, such as generating OCL rules, creating model instances, producing JSON specifications, or constructing feedback from test results in Petrovic et al.~\cite{P34}. Specifically, in Tang et al.~\cite{P33}, the reported time cost encompasses scenario generation, execution, and analysis phases. In contrast, resource-based metrics assess the amount of resources consumed during the process. These include the average number of tokens needed to perform a single generative step~\cite{P34}, the number of simulations required to identify the first and all types of critical scenarios~\cite{P33}, or the number of traffic videos needed to find safety violations for the SUT~\cite{P36}. Together, these metrics provide insights into the computational and practical efficiency of the scenario generation approaches.

\par\vspace{2mm}

\faHandPointRight \ In summary, a wide range of metrics were employed \emph{to evaluate various quality aspects of the generated scenarios and the efficiency of the proposed generative approaches}. While some studies incorporated human ratings in their assessments -- introducing potential bias and subjectivity -- many others relied on \emph{quantitative metrics that offer objective insights into the realism, diversity, effectiveness, controllability, and criticality} of the resulting scenarios. These evaluations collectively demonstrate the value of the generated scenarios for testing and improving ADS. Notably, we observe that significantly more studies focus on evaluating realism, suggesting that \emph{generating realistic driving scenarios remains a primary goal and the minimum required for practical testing} in current research. Lastly, it is important to note that metrics lacking clear definitions or explanations were excluded from our analysis to maintain consistency and clarity. 

\subsubsection{Benchmarks}
\label{sec:results:RQ2:benchmarks}

\begin{table}[tbp]
\centering
\small
\caption{RQ2 -- Baseline methods used for evaluation in selected studies.}
\begin{tabular}{|c|l|l|}
    \hline 
     \textbf{Category} & \textbf{Baseline Methods} & \textbf{Studies} \\
    \hline
    \multirow{20}{*}{(Critical) Scenario generation} & TrafficGen & \cite{P17}, \cite{P41}, \cite{P81}, \cite{P83}, \cite{P86}, \cite{P87}, \cite{P88}, \cite{P89}, \cite{P90} \\
    \cline{2-3} & LCTGen &  \cite{P17}, \cite{P22}, \cite{P38}, \cite{P81}, \cite{P86}, \cite{P87}, \cite{P89}, \cite{P90} \\
    \cline{2-3} & STRIVE &  \cite{P4}, \cite{P8}, \cite{P16}, \cite{P50}, \cite{P52}, \cite{P61}, \cite{P64}, \cite{P94}\\
    \cline{2-3} & AdvSim &  \cite{P4}, \cite{P10}, \cite{P52}, \cite{P56}, \cite{P61}, \cite{P94} \\
    \cline{2-3} & L2C &  \cite{P8}, \cite{P10}, \cite{P56}, \cite{P61}, \cite{P91} \\
    \cline{2-3} & BITS &  \cite{P17}, \cite{P50}, \cite{P53}, \cite{P60} \\
    \cline{2-3} & Random &  \cite{P21}, \cite{P33}, \cite{P58}, \cite{P85} \\
    \cline{2-3} & SceneGen &  \cite{P42}, \cite{P47}, \cite{P86} \\
    \cline{2-3} & Adversarial Trajectory Optimization &  \cite{P10}, \cite{P56}, \cite{P61} \\
    \cline{2-3} & TimeGAN &  \cite{P13}, \cite{P40}, \cite{P76} \\
    \cline{2-3} & HDMapGen &  \cite{P18}, \cite{P68} \\
    \cline{2-3} & RandomTrip &  \cite{P37}, \cite{P39} \\
    \cline{2-3} & SimNet &  \cite{P50}, \cite{P53} \\
    \cline{2-3} & TrafficSim &  \cite{P50}, \cite{P53} \\
    \cline{2-3} & CTG & \cite{P50}, \cite{P60} \\
    \cline{2-3} & Carla Scenario Generator  &  \cite{P10}, \cite{P61} \\
    \cline{2-3} & VAE &  \cite{P14}, \cite{P91} \\
    \cline{2-3} & SeqGAN &  \cite{P25}, \cite{P76} \\
    \cline{2-3} & RankGAN &  \cite{P25}, \cite{P76} \\
    \cline{1-3} \multicolumn{3}{|c|}{Replay (\cite{P4}), Adv-RL (\cite{P4}), VDARS (\cite{P5}), MMG	(\cite{P8}), SAC (\cite{P8}), MTG	(\cite{P11}), Perturbed	(\cite{P11}), GLIGEN (\cite{P12})} \\
    \multicolumn{3}{|c|}{Blended Diffusion (\cite{P12}), COSCI-GAN	(\cite{P13}), GAN (\cite{P14}), SceneTransformer (\cite{P17}), SETR (\cite{P10}), ICNet (\cite{P20})} \\
    \multicolumn{3}{|c|}{MDPFuzz (\cite{P21}), BEVGen	(\cite{P23}), Grid Search (\cite{P32}), AV-Fuzzer (\cite{P33}), Manual Creation (\cite{P34}), M-CPS (\cite{P36})} \\
    \multicolumn{3}{|c|}{LEADE	(\cite{P36}), CRISCO (\cite{P36}), AE	(\cite{P38}), Contrastive AE (\cite{P38}), Masked AE	(\cite{P38}), AEGAN	(\cite{P40}), AC3R (\cite{P45})} \\
    \multicolumn{3}{|c|}{Random Log Selection (\cite{P41}), ATISS (\cite{P42}), Procedural (\cite{P44}), Probabilistic Grammar	(\cite{P44}), MetaSim (\cite{P44})} \\
    \multicolumn{3}{|c|}{Lane Graph	(\cite{P44}), LayoutVAE (\cite{P44}), SoVAR (\cite{P45}), Bicycle (\cite{P54}), ChatScene (\cite{P56}), LLM-F (\cite{P57}), Min TTC (\cite{P58})} \\
    \multicolumn{3}{|c|}{A Transformer-based motion prediction model (\cite{P57}), Kinetic Field (\cite{P58}), Adversarial RL (\cite{P61}), DiffScene (\cite{P64}, \cite{P94})} \\
    \multicolumn{3}{|c|}{VAE (\cite{P65}), $\beta$-VAE (\cite{P66}), InfoGAN (\cite{P66}), SketchRNN	(\cite{P67}), PlainGen (\cite{P67}), DiT (RSI) (\cite{P68}), SLEDGE (\cite{P69})} \\
    \multicolumn{3}{|c|}{DriveSceneGen (\cite{P69}), Rule-based IDM (\cite{P69}), Data-driven Trajeglish (\cite{P69}), DeepTest (\cite{P71}), DeepRoad (\cite{P71})} \\
    \multicolumn{3}{|c|}{Action-RNN (\cite{P72}), SAVP (\cite{P72}), GameGAN (\cite{P72}), World Model (\cite{P72}), DVGO (\cite{P75}), Mis-NeRF360 (\cite{P75}), CAD (\cite{P84})} \\
    \multicolumn{3}{|c|}{S-NeRF (\cite{P75}), F2NeRF (\cite{P75}), BN-AM-SeqGAN (\cite{P76}), Actions-Only (\cite{P81}), Decision Transformer (\cite{P81}), CTG++ (\cite{P81})} \\
    \multicolumn{3}{|c|}{CtRL-Sim  (\cite{P81}), MotionCLIP (\cite{P83}), UniGen (\cite{P86}), SceneDMF (\cite{P84}), MTR\_E (\cite{P84}), Multipath (\cite{P84}), MTR++ (\cite{P84})} \\
    \multicolumn{3}{|c|}{QCNeXt (\cite{P84}), sim\_agents\_tutorial (\cite{P84}), DIAYN (\cite{P85}), Direct Search (\cite{P91}), SPIRAL (\cite{P91}), Grammar-VAE (\cite{P91})} \\
    \hline
\end{tabular}
\label{table:results:RQ2:benchmarks}
\end{table}

The majority of the included studies (62 out of 91) \emph{compared their proposed generative approaches against established baseline methods from existing work}. As shown in  Table~\ref{table:results:RQ2:benchmarks}, more than 100 baseline methods have been identified and recorded. These comparisons were typically made in terms of the generated scenarios, evaluated using various metrics such as realism, diversity, and criticality, as outlined in Section~\ref{sec:results:RQ2:metrics}, to demonstrate the effectiveness of the proposed methods in producing desired scenarios for testing ADS. However, some studies employed customized or modified baseline methods, which were not always clearly defined and are therefore not reported here. Other studies also incorporated \emph{human evaluation, particularly visual inspection, to analyze the quality of the resulting scenarios}. 

Given the large number of baseline methods identified, we focus on the most commonly used ones and briefly outline their objectives and underlying mechanisms. Specifically, TrafficGen~\cite{P47}, LCTGen~\cite{P83}, STRIVE~\cite{P54}, SceneGen~\cite{P44}, HDMapGen~\cite{P67}, CTG~\cite{P53}, TimeGAN~\cite{NEURIPS2019_c9efe5f2}, and SeqGAN~\cite{Yu_Zhang_Wang_Yu_2017} are included in this survey and already described in Section~\ref{sec:results:RQ1:testing}. Several other methods are also utilized across different studies. Except for TrafficGen~\cite{P47} and the Random approach, which were used as baseline methods for scenario reconstruction in two studies~\cite{P22, P33}, other methods were employed as benchmark methods in studies focused on scenario generation and critical scenario generation.

\begin{enumerate}
    \item \textbf{AdvSim}~\cite{wang2021advsim} is a framework for generating safety-critical scenarios using search algorithms, such as Bayesian optimization and genetic algorithms, to manipulate vehicle trajectories and induce failures. AdvSim has been employed as a baseline method in several studies on critical scenario generation~\cite{P4, P10, P52, P56, P61}.

    \vspace{1mm}
    
    \item \textbf{L2C} (short for Learning to Collide)~\cite{ding2020learning} is a framework for generating safety-critical scenarios by employing reinforcement learning method (i.e., policy ingredient) to optimize the sampling of scenario parameters (within a probabilistic structure designed by human knowledge) toward safety-critical regions. L2C has been commonly used as a baseline method for comparison in several studies focusing on critical scenario generation~\cite{P8, P10, P61, P91}, with the exception of one study~\cite{P56} that is categorized as scenario generation.

    \vspace{1mm}
    
    \item \textbf{BITS}~\cite{xu2022bits} is a bi-level imitation learning model that generates traffic behaviors by exploiting both high-level intent inference and low-level driving behavior imitation. Similar to AdvSim\cite{wang2021advsim} and L2C\cite{ding2020learning}, BITS was also commonly used as a baseline method for critical scenario generation~\cite{P17, P53, P60}, with one exception where it was used as a benchmark in one included study for scenario generation~\cite{P50}.

    \vspace{1mm}

    \item The \textbf{random} generation approach refers to a general method that employs random mechanisms to produce scenarios for comparison and has been used as a baseline method in several studies. Specifically, Xu et al.~\cite{P21} employed randomly sampled scenarios for evaluation, Tang et al.~\cite{P33} used scenarios with randomly generated NPC (Non-Player Character) vehicles and their driving actions, Mei et al.~\cite{P58} used a baseline approach that randomly identified adversarial vehicles from all background vehicles for critical scenario generation, and Nguyen et al.~\cite{P85} utilized a baseline with random behaviors for comparing diversity of generated trajectories.

    \vspace{1mm}
    
    \item \textbf{Adversarial Trajectory Optimization}~\cite{zhang2022adversarial} is a scenario optimization approach that introduces adversarial perturbations -- small, carefully crafted changes to input trajectories -- while respecting real-world traffic rules and physical constraints to create adversarial scenarios. Adversarial Trajectory Optimization was used as a benchmark in studies on scenario generation~\cite{P56} and critical scenario generation~\cite{P10, P61}.

    \vspace{1mm}
    
    \item \textbf{RandomTrip}~\cite{randomtrips}, provided by the SUMO~\cite{krajzewicz2002sumo} tool, creates random scenarios within a given network by randomly selecting vehicle origin and destination edges, based on a specified arrival rate. Lu et al.~\cite{P37} adopted it as a baseline for scenario generation due to its ability to provide broad coverage of potential scenarios, while Lu et al.~\cite{P39} applied it as a baseline method for critical scenario generation.

    \vspace{1mm}
    
    \item \textbf{SimNet}~\cite{bergamini2021simnet} is a machine learning approach that simulates driving scenes using a deterministic behavior-cloning model. Similarly, \textbf{TrafficSim}~\cite{suo2021trafficsim} is a multi-agent behavior model designed to simulate realistic and reactive driving behaviors. Both of them were used for benchmarking in the same studies, with one applied to scenario generation~\cite{P53} and the other to critical scenario generation~\cite{P50}.

    \vspace{1mm}
    
    \item \textbf{Carla Scenario Generator}~\cite{dosovitskiy2017carla} employs rule-based methods and grid search to find optimal scenarios across three predefined settings. It served as a baseline method for comparison in two studies focusing on critical scenario generation~\cite{P10, P61}.
    
\end{enumerate}

In addition to comparisons with existing methods, human evaluation, particularly visual inspection, was frequently employed to assess the quality of the generated scenarios. Specifically, human evaluators were used to confirm that generated questions were clear and answerable~\cite{P6}, to assess the semantic validity of generated scenarios~\cite{P19}, to evaluate the quality of augmented data~\cite{P20}, to evaluate the controllability of generated scenarios~\cite{P38}, to qualitatively assess the alignment between descriptions and generated scenarios~\cite{P90}, to judge the accuracy of extracted trajectories from accidents as well as the reconstructed scenarios~\cite{P22}, and to verify the validity and interpretability of the augmented images~\cite{P70}. Additionally, Cai et al.~\cite{P46} and Gao et al.~\cite{P79} used human experts as baselines for comparison. Other than that, visual inspection of selected examples is also frequently used by several studies to qualitatively analyze the quality of the generated scenarios. This includes qualitative comparisons between scenarios produced by different generative approaches~\cite{P73}, assessment of the alignment between user descriptions and generated scenarios~\cite{P3}, comparisons between real and generated scenarios~\cite{P13, P40, P80}, as well as showcasing specific examples via visualization accompanied by qualitative analysis and discussion of results~\cite{P15, P16, P24, P25, P29, P44, P83}.

\par\vspace{2mm}

\faHandPointRight \ Overall, the included studies (62 of 91) \emph{predominantly compared their proposed approaches with generative models or methods from existing work}. These baseline approaches are not necessarily machine learning–based and may involve conventional techniques such as search algorithms, optimization, adversarial perturbations, or random generation. By comparing outcomes, such as the number of critical scenarios generated or the alignment with user intentions, the studies aimed to demonstrate the effectiveness of their proposed methods. However, not all studies relied on established models or approaches as baselines. A few instead \emph{incorporated human assessments, used human subjects as a baseline}, or showcased their evaluation results through visualizations accompanied by qualitative analysis. 

\subsection{RQ3 -- What Limitations Exist for Generative AI}
\label{sec:results:RQ3}

Although generative AI has demonstrated promising results and potential in supporting ADS testing, including tasks such as scenario generation, transformation, reconstruction, augmentation, and understanding, as described in Section~\ref{sec:results:RQ1}, numerous limitations have been identified across the reviewed studies. To provide a comprehensive view, we have compiled \emph{not only the practical challenges encountered in individual study but also the general limitations of generative AI as discussed} in these works, as shown in Table~\ref{table:results:RQ3:limitations}. While several studies have proposed mitigation strategies or potential solutions to address specific issues, many limitations remain unresolved and highlight important research directions for improving the reliability, generalizability, and efficiency of generative models in ADS testing.

\begin{table}[tbp]
\centering
\small
\caption{RQ3 -- Limitations reported or discussed in selected studies.}
\begin{tabular}{|c|l|}
    \hline 
     \textbf{Models} & \textbf{Limitations} \\
    \hline
    \multirow{11}{*}{LLMs} & \textbullet\quad Exhibit hallucination issues in their outputs; \\
    & \textbullet\quad Struggle with complex user commands; \\
    & \textbullet\quad Fail to recognize inputs that are semantically equivalent but phrased differently; \\
    & \textbullet\quad Exhibit biases toward common patterns; \\
    & \textbullet\quad Limitations in handling subtle and domain-specific tasks; \\
    & \textbullet\quad Inability to produce precise outputs when given minimal context or information; \\
    & \textbullet\quad Subject to a knowledge cut-off date; \\
    & \textbullet\quad Suffer from catastrophic forgetting; \\
    & \textbullet\quad Requires a deep understanding of both the SUT and its operating environment; \\
    & \textbullet\quad Lack depth and complexity for generated scenarios; \\
    & \textbullet\quad Different LLMs used within the same framework may yield inconsistent results; \\
    \hline
    \multirow{3}{*}{VLMs} & \textbullet\quad Exhibit hallucination issues in their outputs; \\
    & \textbullet\quad Inability to reliably generalize to unseen or underrepresented data; \\
    & \textbullet\quad Lack robust multi-modal in-context learning capabilities;\\
    \hline
    \multirow{3}{*}{Diffusion-based models} & \textbullet\quad Produce unrealistic or flawed outputs; \\
    & \textbullet\quad Inability to reliably generalize to unseen or underrepresented data; \\
    & \textbullet\quad Demands extensive computation resources;\\
    \hline
    \multirow{3}{*}{GAN-based models} & \textbullet\quad Produce unrealistic or flawed outputs; \\
    & \textbullet\quad Inability to reliably generalize to unseen or underrepresented data; \\
    & \textbullet\quad Lack of diversity in their outputs;\\
    \hline
    \multirow{2}{*}{AE-based models} & \textbullet\quad Produce unrealistic or flawed outputs; \\
    & \textbullet\quad Limitations in modeling multi-agent interactions in scenario generation;\\
    \hline
    \multirow{2}{*}{Other models} & \textbullet\quad Inability to reliably generalize to unseen or underrepresented data; \\
    & \textbullet\quad Struggle to capture causal relationships in pre-crash scenarios;\\
    \hline
    \multirow{3}{*}{Hybrid models} & \textbullet\quad Exhibit realism, consistency, and controllability issues in scenario generation; \\
    & \textbullet\quad Inability to reliably generalize to unseen or underrepresented data; \\
    & \textbullet\quad Demands extensive computation resources;\\
    \hline
\end{tabular}
\label{table:results:RQ3:limitations}
\end{table}

\par\vspace{2mm}

\textbf{LLMs}, being the most frequently used generative models in the included studies, have been associated with several limitations. These limitations primarily concern the generation of suboptimal or incorrect outputs, as well as the inability to effectively handle complex, domain-specific, or alternatively phrased user prompts.

\begin{enumerate}
    \item \textbf{Hallucination issues} in LLM outputs are commonly discussed and reported in the literature. For example, Güzay et al.~\cite{P3} reports that LLMs can produce incorrect simulation files, such as those containing wrong road geometries, lanes, or vehicle placements, especially in complex scenarios. Zhang et al.~\cite{P10} discusses that, while using generative models (e.g., LLMs) to directly generate test scenarios in Scenic code~\cite{fremont2019scenic} appears promising, it frequently results in invalid outputs such as non-compatible code or references to non-existent APIs. These issues are likely due to the limited availability of Scenic examples in the training data and the inherent complexity of scenario code generation. Besides, Rubavicius et al.~\cite{P26} identified a variety of errors emerging during scenario generation, including execution errors, where syntax issues cause the generated code to be non-executable, and pass errors, where code executes but does not behave correctly in simulation. Similarly, Petrovic et al.~\cite{P34} encountered hallucination-related issues when generated content failed to conform to a strict schema, resulting in incorrect or unusable outputs. In Lu et al.~\cite{P37}, the evaluation revealed hallucinations of the LLM-powered tool (OmniTester) such as generating an incorrect number of lanes and using improperly formatted keywords. Lu et al.~\cite{P39} also experienced failures with incorrect formatting of road network generation keywords, vehicle model names, and vehicle distance estimations, often stemming from hallucinated or misunderstood concepts. In Guo et al.~\cite{P45}, errors arose when extracting information from accident reports; many inaccuracies were attributed to implicit actions described in natural language in the reports. Additional examples of hallucinations include LLMs introducing traffic lights that were not specified in the prompt~\cite{P56}, and producing semantically incorrect loss functions when generating code for learning tasks, as seen in Zhong et al.~\cite{P60}. These findings highlight the ongoing challenge of hallucinations in LLM-driven scenario generation and underscore the need for robust validation, schema enforcement, and possibly domain-specific fine-tuning. Specifically, studies~\cite{P3, P45} proposed intentionally training or fine-tuning a pretrained model using labeled data as a potential solution to mitigate hallucinations. In contrast, Zhong et al.~\cite{P60} suggested an interactive approach, where simulation results are provided as feedback to guide the LLM in detecting and correcting semantic errors in its responses.

    \vspace{1mm}

    \item LLMs tend to \textbf{struggle with complex user commands}. For example, Lu et al.~\cite{P39} observed that LLMs often fail to accurately extract road networks when faced with complex scenes, particularly when interpreting scenes involving numerous buildings and vehicles, resulting in significant deviations from the intended layout. These difficulties are compounded when interpreting complex or subtle vehicle behaviors and multi-agent interactions, particularly when dealing with implicit expressions~\cite{P46}. This limitation is frequently attributed to error propagation in few-shot learning, where LLMs rely heavily on prior examples and perform best when given closely related input-output pairs. In Zhong et al.~\cite{P60}, it is reported that the proposed model CTG++ does not support complex user commands involving intensive interactions with the map, and it can fail to understand some basic driving concepts (e.g., "cut in") properly. To address these limitations, Zhong et al.~\cite{P60} proposed that providing more relevant examples and offering feedback during generation may help LLMs better understand and execute complex commands.

    \vspace{1mm}

    \item LLMs often fail to recognize \textbf{semantically equivalent but differently phrased inputs}, which can lead to inconsistent or incorrect scenario generation. Specifically, Cai et al.~\cite{P46} reports that variations in scenario descriptions, despite being semantically identical, can result in disparate scenario representations generated by the LLM. Similarly, Ruan et al.~\cite{P56} observed that LLMs frequently overlook alternative phrasings in user prompts. For example, a phrase like "a car in front does not move" may not be correctly interpreted as a blocking behavior, leading to failures in understanding user intent. These findings highlight the sensitivity of LLMs to surface-level linguistic variation and underscore the need for better semantic generalization in prompt interpretation.

    \vspace{1mm}

    \item Xu et al.~\cite{P21} specifically reports a limitation wherein LLMs tend to exhibit \textbf{biases toward common patterns}, a consequence of their underlying probabilistic nature. This often results in outputs that lack precision and nuance, particularly when fine-grained control or adjustment is required. For example, LLMs struggle to make subtle distinctions in numeric values or capture minor variations between scenarios, which limits their suitability for local exploration -- that is, generating near-failure scenarios that lie close to critical thresholds. To address this limitation, the study introduced a Potential Analysis method to evaluate the proximity of generated scenarios to known critical scenarios. Based on this analysis, the authors adopted an adaptive strategy, dynamically switching between LLM-driven generation and random mutation depending on whether the potential exceeds a predefined threshold.

    \vspace{1mm}

    \item LLMs exhibit notable limitations in handling \textbf{subtle and domain-specific tasks}, such as accurately filling in parameter ranges (e.g., vehicle speed or cut-in distance) required for scenario generation, as reported in Tang et al.~\cite{P33}. To address this issue, Tang et al.~\cite{P33} proposed an adaptive strategy, where the LLM is used to generate parameter ranges only when it demonstrates sufficient confidence; otherwise, default ranges are applied to ensure reliability. Additionally, Lu et al.~\cite{P37} highlights the difficulty LLMs face in generating domain-specific and complex topological structures that adhere to precise spatial relationships, further underscoring the challenges in using LLMs for structured, geometry-aware scenario generation.

    \vspace{1mm}

    \item A number of general limitations of LLMs have been discussed across several studies focused on scenario generation. One commonly noted challenge is that LLMs often struggle to interpret scenario generation requests and produce precise outputs when \textbf{given minimal context or information}, largely due to being trained on broad, general-purpose corpora rather than domain-specific data~\cite{P37}. Additionally, LLMs like GPT-4 have a \textbf{knowledge cut-off date}, meaning they lack access to any information or developments that occurred after that point~\cite{P3}. LLMs are also known to suffer from \textbf{catastrophic forgetting}, especially when processing overly long input sequences or performing complex extraction tasks, which results in reduced accuracy~\cite{P22}. Further, Xu et al.~\cite{P21} highlights several additional concerns. First, effective prompt engineering requires \textbf{a deep understanding of both the SUT and its operating environment}; without this, blindly applying prompts may lead to suboptimal or irrelevant outputs. Second, LLM-generated scenarios often \textbf{lack depth and complexity}, in part due to the absence of robust benchmarks, which limits their ability to produce intricate or realistic testing environments. Finally, it notes that different LLMs used within the same framework may yield \textbf{inconsistent results}, introducing variability that can affect reproducibility and reliability.
\end{enumerate}

\par\vspace{2mm}

\textbf{VLMs} have been observed to produce incorrect or unrealistic outputs. Additionally, these models often exhibit limited generalization capabilities in scenarios involving underrepresented data from the training set and struggle with in-context learning in complex scenarios when sufficient or relevant contextual information is not available.

\begin{enumerate}
    \item VLMs have also been reported to suffer from \textbf{hallucination issues} in their generated outputs. For instance, Miao et al.~\cite{P24} observed that some scenarios generated in Scenic using VLMs contained syntax errors, making them invalid for execution. In Tian et al.~\cite{P36}, the authors found that the baseline implementation (\textit{LEADE$_D$}) was infeasible for directly generating concrete scenarios from abstract representations on given road networks, with the correctness of the generated scenarios considered low. Further, Marathe et al.~\cite{P78} highlights several cases where VLMs produced unrealistic images. These include scenes with hurricanes depicted in implausible ways, tornadoes appearing on top of cars, entities with missing body parts, and even extra terrestrial creatures crossing roads. The study notes that, as the prompts shift to more out-of-distribution settings, the model’s outputs increasingly favor unrealistic or fantastical imagery. This is likely due to the scarcity of realistic training images captured under such rare conditions. Additionally, the study observed frequent spatial anomalies, such as incorrect placement, orientation, and interaction between objects, which compromise the spatial coherence of the generated scenes. Besides, as discussed in Li et al.~\cite{P93}, VLMs cannot reliably capture details about roads and users that depend on spatial understanding of road accident sketch images, and they remain prone to hallucinations, sometimes producing outcomes that are logically inconsistent or physically implausible, especially in complex scenes. These findings collectively highlight the challenges VLMs face in maintaining realism and structural validity in complex or low-frequency scenarios.

    \vspace{1mm}

    \item Similar to previously described hallucination issues, a broader limitation of VLMs is their \textbf{inability to generalize well to unseen or underrepresented data}, which can result in biases and performance degradation in those circumstances~\cite{P24}. As a consequence, the effectiveness of VLM-based frameworks often depends heavily on the careful and manual selection of diverse, high-quality examples for prompt engineering. 

    \vspace{1mm}
    
    \item In Chen et al.~\cite{P77}, VLMs, specifically GPT-4V, is reported to \textbf{lack robust multimodal in-context learning capabilities}. This limitation makes few-shot evaluation indispensable when dealing with complex and highly variable autonomous driving scenarios. As a result, when VLMs were used as evaluators or "judges" in such tasks, the observed human consistency rate remains relatively low, at approximately 70\% across all three evaluation tasks, indicating challenges in achieving reliable and human-aligned judgment.

\end{enumerate}

\par\vspace{2mm}

\textbf{Diffusion-based models} also encounter challenges such as producing incorrect or unrealistic outputs and exhibiting limited generalization on underrepresented data. In addition, they are reported to be computationally expensive, with high inference times that may limit their scalability for real-time or large-scale applications.

\begin{enumerate}
    \item Several studies have reported \textbf{unrealistic or flawed outputs} produced by diffusion-based generative models, particularly in the context of autonomous driving scenario generation. In Chang et al.~\cite{P64}, certain adversarial scenarios generated by the model were found to be unrealistic, such as instances where the adversarial agent collided prematurely with non-adversarial agents before reaching the ego vehicle. Additionally, some scenarios resulted in collisions for which the ADS (ego vehicle) was not at fault, undermining the validity of the evaluation. In Baresi et al.~\cite{P19}, issues with temporal consistency were reported, where generated scenarios were incoherent across sequential frames. These inconsistencies, stemming from the lack of rendering continuity, led to drastic visual or behavioral changes between consecutive frames, which are problematic for maintaining temporal coherence in simulation-based testing. To address this issue, the study integrates the diffusion model with a CycleGAN to enhance domain consistency during generation. Similarly, Gannamaneni et al.~\cite{P20} highlighted a misalignment between the intended input conditions for the safety-critical objects (i.e., pedestrians) and the generated scenarios. The study observed that diffusion models occasionally failed to adhere to specified attributes, such as clothing color, resulting in significant deviations. For example, while the input prompt specified a red shirt, the generated pedestrian scenario sometimes featured red shoes or pants instead, and recall for certain colors (e.g., brown and grey) was notably poor. These failures suggest that textual conditioning mechanisms in diffusion models may be unreliable in certain contexts, particularly when precise object-level attributes are crucial for scenario validity.

    \vspace{1mm}

    \item Like LLMs and VLMs, diffusion-based generative models also face significant limitations in \textbf{generalizing to unseen or underrepresented data}, often leading to biases or performance degradation in certain scenarios. For instance, Xu et al.~\cite{P23} highlights that most large-scale image diffusion models available today are fundamentally trained to generate standard, normal-shaped images, making them well-suited for typical visual outputs but ill-equipped for specialized tasks. When applied to multi-view street image generation, these models struggle to produce high-quality results, particularly due to the demands of generating images with wide or panoramic fields of view. This limitation becomes evident in their inability to render scenes with the spatial coverage required for realistic multi-view or street-level imagery. Furthermore, Xu et al.~\cite{P23} also reports an overfitting risk when diffusion models are trained from scratch on limited datasets and can't generalize well on unseen situations, causing the generation of low-diversity and low-quality images. To address these limitations, Xu et al.~\cite{P23} proposed retraining or fine-tuning models specifically for the target task and output format using specialized datasets, thereby improving their ability to generate relevant and high-quality outputs in underrepresented domains. Similarly, Wang et al.~\cite{P88} finds that the performance of the proposed model DragTraffic degrades when generating scenarios involving cyclists, a group that is underrepresented in the training data. This suggests that the quality and diversity of the training dataset play a crucial role in the model’s ability to produce realistic and inclusive outputs. Across these studies, the inability of diffusion models to generalize beyond their training distributions remains a key limitation, especially when high-fidelity, diverse, or specific scenarios are required.

    \vspace{1mm}

    \item Diffusion-based models are reported to be \textbf{computationally expensive} and exhibit high inference times, as noted in Baresi et al.~\cite{P19}. To address this limitation, the study employed knowledge distillation by integrating the diffusion model with a CycleGAN, aiming to maintain domain consistency in generation while significantly improving inference efficiency and throughput.
\end{enumerate}

\par\vspace{2mm}

\textbf{GAN-based models}, like other generative models, are reported to generating suboptimal results and exhibit poor generalization on underrepresented data in their training sets. Additionally, they often suffer from limited diversity in their outputs due to overfitting, as the generator tends to learn patterns that merely fool the discriminator, rather than capturing the full range of variation present in real-world data.

\begin{enumerate}
    \item GAN-based models have also been reported to \textbf{generate unrealistic or flawed scenarios}. Specifically, Spooner et al.~\cite{P29} observed that some generated pose sequences, particularly those involving pedestrians crossing from the left, appeared unrealistic or erratic, resulting in unrealistic pedestrian-crossing scenarios. This issue was attributed to the presence of noisy or low-quality samples in the original training dataset, which misled the generator and resulted in the model learning incorrect patterns. As a potential mitigation strategy, Spooner et al.~\cite{P29} proposed improving dataset quality by identifying and removing poor-quality or inconsistent sequences, thereby reducing the likelihood of the model incorporating undesirable behaviors during training.

    \vspace{1mm}

    \item Like any other generative models, GAN-based models also exhibit limitations in \textbf{generalizing to unseen or underrepresented data}, which can lead to biases or degraded performance in certain contexts. For instance, Spooner et al.~\cite{P48} reports that when the model (Ped-Cross GAN) is trained on an imbalanced dataset, it inherits the biases of the dominant class, predominantly generating samples from well-represented categories while failing to adequately capture minority classes, such as pedestrian jogging scenarios. By adding additional data to improve representation for underrepresented classes, the model’s ability to generate more realistic and diverse samples that better reflect real-world distributions significantly improves. In a related observation, Li et al.~\cite{P49} shows that although the generated scenarios are diverse, the distribution of generated data points (meaning scenarios) is uneven, with noticeable clustering in some areas and sparsity in others. This imbalance implies that the model may be overfocused on generating specific scenario types, possibly favoring rare or critical cases, while under-representing more typical lane-changing behaviors. While such a focus might be intentional, particularly to highlight critical or adversarial scenarios, it also reveals a limitation in the model’s ability to comprehensively capture the full range of real-world lane-changing behavior.

    \vspace{1mm}

    \item Continuing on the issue of generalization, GAN-based models have also been reported to suffer from a \textbf{lack of diversity} in their outputs, often due to overfitting in attempts to fool the discriminator. In Spooner et al.~\cite{P29}, the Ped-Cross GAN tend to generate similar or repetitive pedestrian crossing sequences, which, while realistic enough to deceive the discriminator, lack the variety observed in real-world behavior. This suggests the model has overfit to a small set of representative patterns that suffice for the adversarial objective but fall short in reflecting the broader distribution. To address this, the authors propose incorporating a diversity constraint or adding explicit loss terms that penalize the model when it focuses solely on fooling the discriminator, thereby encouraging greater variability while maintaining realism. A similar issue is reported in Yu and Li~\cite{P31}, where the model exhibits mode collapse, a common failure mode in GAN-based models. In this case, the generator maps different nominal inputs to the same corner-case image, resulting in reduced diversity. To mitigate this, the study introduced a backward generator along with a cycle-consistency loss, enforcing a form of reversibility such that outputs must map back to their corresponding inputs. This mechanism discourages the generator from collapsing multiple distinct inputs into a single output, thereby preserving output diversity.
\end{enumerate}

\par\vspace{2mm}

\textbf{Autoencoder-based models}, though used less frequently in the reviewed studies, have been observed to produce unrealistic results and exhibit limitations in modeling agents separately in multi-agent scenarios. 

\begin{enumerate}
    \item Autoencoder-based models have also been reported to produce \textbf{unrealistic results} in scenario generation, particularly due to the absence of physical or kinematic constraints (e.g., acceleration limits or realistic vehicle dynamics). As a result, some synthesized scenarios in Ding et al.~\cite{P11}, though visually plausible, may violate physical feasibility, making them less suitable for reliable autonomous driving testing. The authors acknowledge this limitation and identify the integration of physical constraints into the model architecture as an important direction for future improvement.

    \vspace{1mm}

    \item VAEs have been noted to face limitations in \textbf{modeling multi-agent interactions}, particularly in vehicle trajectory generation. Specifically, Gong et al.~\cite{P65} reports that traditional VAE models, which operate within a single unified framework, are unable to separately process the trajectories of leading and following vehicles. This design results in limited modeling of vehicle-to-vehicle interactions and fails to capture the dynamic relationships between multiple agents in a traffic scenario. To address these issues, Gong et al.~\cite{P65} proposed Dual-VAE, an enhanced architecture that runs two VAEs in parallel -- one dedicated to the leading vehicle and the other to the following vehicle -- enabling better representation of individual trajectories and more realistic modeling of their interactions.

\end{enumerate}

\par\vspace{2mm}

\textbf{Other generative models} that do not fall into the previously discussed categories also face two commonly reported issues: performance degradation on underrepresented data and the inability to adequately capture causal relationships in pre-crash scenarios.

\begin{enumerate}
    \item Other deep generative models also face generalization challenges, leading to \textbf{biases or performance degradation when handling unseen or underrepresented data}. For example, the framework proposed in Tan et al.~\cite{P44} (SceneGen) occasionally generates near-collision scenarios that, while statistically plausible based on the training data, are highly unlikely in real-world contexts. This highlights a gap between statistical realism and practical safety considerations, suggesting that realism based solely on data distribution may not be sufficient for safety-critical applications. Additionally, the incorporation of traffic light data into the generation process was found to slightly degrade performance during evaluation. This decline is likely due to the sparse or inconsistent representation of traffic light information in the dataset, which introduces ambiguity and confuses the model during training. Furthermore, the study notes that pure sampling from deep autoregressive models can result in degenerate or unrealistic examples, primarily due to the models’ unreliable long-tail behavior. To mitigate this, the authors employ a modified sampling strategy where actors involved in collisions are discarded during sampling. This approach helps preserve diversity while avoiding implausible scenarios.

    \vspace{1mm}

    \item Ding et al.~\cite{P8} also highlights that common generative models struggle to \textbf{capture causal relationships} in pre-crash scenarios. While humans can intuitively recognize causal chains in traffic situations such as a pedestrian crossing causing the ego vehicle to brake, deep generative models lacking an explicit causal structure often fail to model such dependencies accurately. To address this, Ding et al.~\cite{P8} introduced CausalAF, which incorporates explicit causal modeling to better represent the cause-effect relationships inherent in complex driving situations. The study underscores that causal reasoning is essential for reliably simulating scenarios that reflect real-world safety dynamics.
\end{enumerate}

\par\vspace{2mm}

\textbf{Hybrid models}, although designed to leverage the strengths of multiple generative models or architectures, still suffer from common limitations observed in their individual components. These include the generation of suboptimal results, performance degradation on underrepresented data, and high computational costs.

\begin{enumerate}
    \item Hybrid generative models, which integrate multiple generative models such as LLMs, VLMs, and diffusion models, also encounter challenges related to \textbf{realism, consistency, and controllability}. In Zhang et al.~\cite{P17}, which combines an LLM with a VLM, the authors observe that generating scenarios based on diverse goals selected by the VLM can lead to a slight loss of consistency across the generated outputs. Similarly, Jiang et al.~\cite{P43}, which integrates an LLM with a diffusion model, reports realism reduction when constraints are applied only after the generation process. In such cases, late-stage constraint enforcement can result in unrealistic layouts, and if the constraints influence only a small portion of the scene, the model may simply ignore them in subsequent steps of generation. Additionally, Pronovost et al.~\cite{P41} highlights limited precision in controllability in their hybrid model, Scenario Diffusion, which uses agent- and scene-level tokens to guide generation. While the model is capable of modifying high-level scene or agent properties, the granularity of control is restricted -- complex or highly detailed specifications are not always accurately reflected in the generated scenarios. These findings suggest that, despite their potential, hybrid generative models still face significant challenges in maintaining consistency, respecting constraints, and supporting fine-grained user control.

    \vspace{1mm}

    \item In Pronovost et al.~\cite{P41}, the hybrid model combining VAE and diffusion architectures exhibits \textbf{generalization issues when applied to unseen or underrepresented data}. Specifically, the model struggles when the scene distributions differ significantly, such as between geographically distinct regions (e.g., suburban vs. urban environments). This behavior suggests that the model is strongly conditioned on regional structural patterns, limiting its ability to generalize effectively to out-of-distribution scenarios. Such a constraint presents a notable limitation for deploying the model in diverse real-world contexts where driving environments vary widely.

    \vspace{1mm}

    \item In Zhong et al.~\cite{P60}, the hybrid model integrating an LLM and a diffusion model is reported to be \textbf{computationally expensive}, with the trajectory generation process taking approximately one minute per simulated scenario. This highlights a limitation of the approach -- it is both computationally intensive and time-consuming.
\end{enumerate}

\par\vspace{2mm}

\faHandPointRight \ Overall, \emph{each type of generative model presents certain limitations}, as observed or discussed across the included studies. Despite their architectural differences, these models tend to exhibit similar challenges, including the \emph{generation of suboptimal outputs} that may be incorrect, inaccurate, or unrealistic; \emph{difficulty in handling complex or domain-specific tasks}; \emph{poor generalization to underrepresented data} in the training set; and being \emph{computationally expensive}. These limitations highlight that \emph{improving current generative models to produce accurate and realistic results} remains a primary concern in order to use them for realistic testing of ADS. This observation aligns with our findings in Section~\ref{sec:results:RQ2:metrics}, where a majority of studies focused on evaluating and demonstrating the realism of the generated scenarios. In addition, enhancing generative models to better \emph{handle uncommon or complex scenarios, particularly in domain-specific contexts}, represents another critical challenge for future research and requires substantial improvement.  

\section{Discussion}
\label{sec:discussion}

The objective of this survey is to analyze the relevant literature and investigate how generative AI is applied in testing ADS (RQ1), how effective these approaches are (RQ2), and what limitations they present (RQ3). In the following sections, we present our findings addressing these research questions and discuss their broader implications.

\subsection{RQ1 -- How Generative AI is used for Testing ADS} 

We observed a growing trend in publications that apply generative AI to testing ADS, particularly from 2023 onward, reflecting the increasing attention and perceived potential of this approach. A wide variety of generative models are employed across studies, with LLMs, especially those from the GPT~\cite{openai2024gpt4technicalreport, hurst2024gpt} family, being the most commonly used. While some works explicitly target specific ADS modules or functionalities, the majority describe testing objectives in a more general context, without focusing on individual components. Generative AI~\cite{feuerriegel2024generative, sengar2024generative}, with its primary goal of creating novel and diverse content, is widely used to generate new scenarios, especially critical scenarios that are rare and hazardous, for testing ADS. Other applications include scenario reconstruction, transformation, and augmentation -- each aimed at creating diverse scenarios under varying conditions and inputs, despite differences in underlying models and techniques. A few studies explored the use of generative AI for scenario analysis and understanding, still within the broader context of scenario-based testing of ADS. These findings align with recent survey studies by Tang et al.~\cite{tang2023survey} and Lou et al.~\cite{lou2022testing}, both of which highlight efficient scenario generation as an urgent need and ongoing challenge in testing ADS, which eventually validates the functionality and safety of ADS under diverse environmental and driving conditions. Our research contributes practical insights into which generative AI models are used, what ADS functions or modules are tested, and the approaches and mechanisms employed. Overall, this study offers a comprehensive overview of how generative AI is used for ADS testing and serves as a valuable reference for future research in this area.

\subsection{RQ2 -- How Effective is Generative AI for Testing ADS}

In general, generative AI has demonstrated promising results in fulfilling the objective generative tasks for testing ADS, as outlined in previous sections, with the proposed generative models or approaches often outperforming selected baseline methods in evaluation across various studies. In the evaluation, most studies relied on publicly available datasets for training and testing models, open-source simulators for executing produced scenarios, and open-source ADS implementations to demonstrate the effectiveness of their approaches in testing and improving ADS performance. Their evaluation typically focused on multiple quality dimensions of the generated scenarios and proposed methods, including realism, diversity, effectiveness for testing, controllability, criticality, and efficiency. Among these, realism emerged as the most frequently emphasized attribute, reflecting its fundamental importance in ensuring test scenarios are realistic and it aligns with existing studies~\cite{song2024industry, song2024empirically, song2025synthetic, stocco2022mind, beringhoff2022thirty} in which the realism gap has been consistently reported as a significant challenge for ADS test scenarios. In this survey, we provide practical insights into the effectiveness of generative AI for ADS testing, detailing the datasets, simulators, systems under test, and evaluation metrics used, particularly those most commonly adopted in the selected studies. Overall, evaluations indicate that generative AI offers a promising and effective mechanism for ADS testing, with significant potential. These findings are consistent with recent reviews~\cite{feuerriegel2024generative, sengar2024generative, bengesi2024advancements, cao2023comprehensive} that examine the applications and usefulness of generative AI in broader contexts.

\subsection{RQ3 -- What Limitations Exist for Generative AI} 

Although generative AI has proven effective and holds significant potential for facilitating ADS testing, several common limitations remain evident across current generative models. A recurring issue is the generation of suboptimal outputs (e.g., syntactically incorrect, semantically inaccurate, or unrealistic scenarios), which has been reported across nearly all types of generative models and still requires substantial improvement to achieve the desired quality. Beyond this, generative models often exhibit generalization limitations. These models can display biases and struggle with underrepresented data, reducing their effectiveness for rare or unseen input patterns. Moreover, current generative AI models often face challenges in handling complex, domain-specific tasks. This is largely due to their training on general-purpose corpora, necessitating retraining or fine-tuning to achieve reliable performance in specialized autonomous driving contexts. In addition, computational cost is frequently cited as a practical challenge, raising the need for more resource-efficient generative AI approaches. Collectively, these limitations reflect the weaknesses of current generative models despite their wide adoption and impressive results, and highlight critical gaps and opportunities for future research in improving generative AI for ADS testing.

\section{Related Work}
\label{sec:related_work}

In general, very few literature surveys have specifically focused on the use of generative AI for testing ADS. While there are several existing surveys on ADS testing, they do not typically examine or report findings related to generative AI. Conversely, although a number of surveys explore generative AI in the context of autonomous driving, their focus is often on reviewing such techniques for enabling autonomous driving, rather than on testing, making them not directly relevant to our scope. Nevertheless, we briefly describe these studies to provide a more comprehensive view, while focusing on the few relevant literature surveys that closely align with the scope of this work.

\subsection{ADS testing} 

Several recent survey studies have been conducted on ADS testing, although generative AI was not reported in these works. Notably, Tang et al.~\cite{tang2023survey} provide a comprehensive overview of current testing techniques at both the module level and the system level. Their study also compiles key testing resources, including datasets, tools, open-source ADS platforms, and scenario description languages, and synthesizes major challenges in ADS testing. In a separate study, Lou et al.~\cite{lou2022testing} investigated industry practices and emerging needs in ADS testing through interviews and online surveys with practitioners. Based on their findings, they conducted a literature review and identified gaps between current research and real-world industrial needs. Several studies have focused on surveying scenario generation approaches. Specifically, Schütt et al.~\cite{schutt20231001} and Riedmaier et al.~\cite{riedmaier2020survey} developed taxonomies to classify scenario generation methods used in the selected literature. Zhang et al.~\cite{zhang2022finding} and Ding et al.~\cite{ding2023survey} conducted literature reviews centered on the generation of safety-critical scenarios. Beyond that, Cai et al.~\cite{cai2022survey} focused on data-driven approaches -- a key category that leverages collected or synthetic data for scenario creation. Additionally, Ren et al.~\cite{ren2022survey} reviewed a broad scope of research on autonomous driving scenarios and scenario databases, systematically organizing a large body of relevant publications.

\subsection{Generative AI in AD} 

Several studies have presented surveys on the use of generative AI in autonomous driving (AD). While some of these works discuss studies related to scenario generation and understanding, their primary focus is empowering ADS with generative AI instead of testing ADS, and the findings specifically related to ADS testing remain limited. Among these studies, Wang et al.~\cite{wang2025generative} conducted a comprehensive survey on the role of generative AI in AD, highlighting frontier applications in data generation as well as LLM-guided reasoning and decision-making. They categorized practical applications into areas such as end-to-end AD, digital twin systems, and intelligent transportation networks. In contrast, Nie et al.~\cite{nie2025exploring} provided a broader survey of how LLMs are reshaping modern transportation systems, covering topics such as predictive analytics, traffic forecasting, and autonomous driving. Both Zhu et al.~\cite{zhu2024will} and Li et al.~\cite{li2024large} provide surveys on recent advancements in leveraging LLMs for AD, with a focus on their applications in both individual ADS functional modules and end-to-end ADS. Wu et al.~\cite{wu2025multi} presents a survey specifically examining the integration of LLMs in multi-agent ADS. In a related effort, Cui et al.~\cite{cui2024survey} conducted a survey and present an overview of existing multi-modal LLM tools used for AD tasks. Similarly, Zhou et al.~\cite{zhou2024vision} offer a comprehensive and systematic review of advances in VLMs for enabling a wide range of AD capabilities, including perception and understanding, end-to-end AD, navigation, and planning. In addition, Yang et al.~\cite{yang2023llm4drive} systematically reviewed the research landscape concerning the use of both LLMs and VLMs in AD. Furthermore, Guan et al.~\cite{guan2024world} and Tu et al.~\cite{tu2025role} provide an in-depth overview of world models (aiming for interpreting sensor data and predicting future driving scenarios) for AD tasks.

\subsection{Generative AI for ADS testing}

Tian et al.~\cite{tian2024large} provide a comprehensive review of current research on the applications of LLMs and VLMs in AD, focusing on four key areas: modular integration, end-to-end integration, data generation, and evaluation platforms. In their study, they present a few works that utilize LLMs and world models to support simulation and scenario generation, which falls within the broader scope of testing. In contrast, our focus is on testing as a whole rather than solely on scenario generation. Moreover, we explore a wider range of generative models, including  diffusion-based models, GAN-based models, and other generative models apart from LLMs and VLMs. In a related study, Zhao et al.~\cite{zhao2025survey} surveyed the applications of LLMs for scenario-based testing and outlined open challenges and future research directions. In contrast, our work focuses more broadly on generative AI models, including but not limited to LLMs. Additionally, Gao et al.~\cite{gao2025foundation}, a recent survey published as a preprint on arXiv in June 2025, present work that closely aligns with the focus of our study. Their survey specifically investigates foundation models for scenario generation and analysis, covering papers published between October 2022 and May 2025, including many of the primary studies considered in our work. In contrast, our study is grounded in the broader context of ADS testing, encompassing -- but not limited to -- scenario generation and analysis, as we present in Section~\ref{sec:results:RQ1:testing}. We intentionally select studies that are explicitly framed within the context of ADS testing. While we also examine foundation models (i.e., large generative models trained on unlabeled general data), our scope further includes other deep generative models that do not fall under the category of foundation models. Beyond these differences in scope, our primary objective is to analyze the approaches and mechanisms by which generative AI supports ADS testing, assess their effectiveness, identify associated limitations, and highlight directions for future research in this field.

\section{Conclusion}
\label{sec:conclusion}

Generative AI is used to produce novel content in various formats to support diverse tasks, and has become a widely adopted technology across numerous application domains, including autonomous driving. While autonomous driving is also expected to contribute greatly to our societal advancement, testing ADS to validate their functionality and safety in all kinds of driving scenarios is a critical yet challenging endeavor, as the operational environment is open-ended and the range of potential driving scenarios is vast. In this study, we collected and analyzed 91 papers on the application of generative AI for testing ADS, examining their underlying mechanisms and identifying their limitations. Our findings show that generative AI is predominantly applied in scenario-based testing, particularly in scenario generation, where it has demonstrated notable effectiveness. Empirical evaluations of the proposed techniques further highlight the utility of generative AI in testing and improving ADS. However, several limitations persist, including the generation of suboptimal results and the relatively narrow scope of current applications. Consequently, while generative AI represents a promising approach with significant potential for ADS testing, substantial advancements are still needed to ensure the generation of realistic and accurate outputs across a wider range of tasks in this special domain.

\section*{Acknowledgement}
This work was supported in part by the Wallenberg AI, Autonomous Systems and Software Program (WASP). We thank the authors of the papers included in our survey who provided us with valuable feedback. 

\bibliographystyle{ACM-Reference-Format}
\bibliography{reference}


\end{document}